\documentclass[twocolumn,pre,twoside,floatfix]{revtex4-1}

\usepackage{adjustbox}
\usepackage{amsmath}
\usepackage{amssymb}
\usepackage{graphicx}
\usepackage{dcolumn}
\usepackage{bm}
\usepackage{color}
\usepackage[T1]{fontenc}

\def\epsilon{\varepsilon}

\def\theta{\vartheta}
\def\rho{\varrho}
\def\Int#1#2#3{\int_{#1}\!\mathrm{d}^{#2}{#3}\;}

\def\vec#1{\pmb{#1}}

\def\modifiedRed#1{\textcolor{black}{#1}}
\def\modifiedBlue#1{\textcolor{black}{#1}}
\def\modifiedGreen#1{\textcolor{black}{#1}}
\def\modifiedMagenta#1{\textcolor{black}{#1}}
\def\modifiedOrange#1{\textcolor{black}{#1}}
\def\MODIFIED#1{\textcolor{black}{#1}}
\def\MODIfied#1{\textcolor{black}{#1}}

\graphicspath{{Figs/}}

\begin{document}


\title{Smectic phases in ionic liquid crystals}

\author{Hendrik Bartsch}
\email{hbartsch@is.mpg.de}
\author{Markus Bier}
\email{bier@is.mpg.de}
\author{S.\ Dietrich}
\affiliation
{
   Max-Planck-Institut f\"ur Intelligente Systeme,\\ 
   Heisenbergstr.\ 3,
   70569 Stuttgart,
   Germany and\\
   Institut f\"ur Theoretische Physik IV,
   Universit\"at Stuttgart,
   Pfaffenwaldring 57,
   70569 Stuttgart,
   Germany
}

\date{ \today }

\begin{abstract}
Ionic liquid crystals (ILCs) \modifiedBlue{are} anisotropic mesogenic molecules
which carry charges and \modifiedBlue{therefore combine} properties of liquid crystals,
e.g., \modifiedBlue{the formation of} mesophases,
and of ionic liquids, \modifiedBlue{such as} low melting temperatures
and tiny triple-point pressures.
\modifiedBlue{
Previous density functional calculations have revealed
that the phase behavior of ILCs is strongly affected 
by their molecular properties, i.e., their aspect ratio,
the loci of the charges, and their interaction strengths.}
Here, we report \modifiedBlue{new} findings \modifiedBlue{concerning}
the phase behavior of ILCs \modifiedBlue{as obtained by density functional theory}
and Monte Carlo simulations.
The most \modifiedBlue{important} result is the occurrence of a novel,
wide smectic-A phase $S_{AW}$, at low temperature,
\modifiedBlue{the} layer spacing \modifiedBlue{of which} is larger than that of the 
ordinary high-temperature smectic-A phase $S_{A}$.
\MODIFIED{Unlike the ordinary smectic $S_A$ phase, 
the \MODIfied{structure of the} $S_{AW}$ phase 
\MODIfied{consists of} alternating layers of
particles oriented parallel to the layer normal and
oriented perpendicular to \MODIfied{it}.}
\end{abstract}

\maketitle

\section{Introduction}

Ionic liquid crystals (ILCs)
\modifiedBlue{\cite{Binnemans2005,Kondrat_et_al2010}
combine characteristics of liquid crystals and ionic liquids
such as anisotropic material properties and ionic conductivity, respectively.
They attract} steadily growing scientific and technological interest.
A common molecular structure \modifiedBlue{of ILCs} is that of charged imidazolium rings
with highly anisotropic alkyl chains attached.
\modifiedBlue{Varying} the length of the alkyl chains
as well as the number and \modifiedBlue{loci} of the charged groups offers the possibility
\modifiedBlue{to optimize and tune material properties upon synthesis~\cite{Binnemans2005}.}
\modifiedGreen{For instance, 
ILCs forming either columnar or smectic phases
can show a high conductivity in one dimension (parallel to the columnar stacks),
respectively in two dimensions (perpendicular to the smectic layer normal).
\modifiedMagenta{Therefore} they can potentially be used as anisotropic electrolytes
in batteries~\cite{Yoshio2004,Bruce2008,Kato2010}.
Moreover, ILCs can be synthesized such that they exhibit
high thermal as well as mechanical stability~\cite{Binnemans2005,Goossens2016}.
The combination of (low-dimensional) high conductivity and durability
renders ILCs promising candidates as electrolyte constituents, e.g.,
in solar cells~\cite{Yamanaka_et_al2005,Yamanaka_et_al2007}.
Additionally, since ILCs can be regarded as anisotropic solvents,
they can also be used as organized reaction media~\cite{Lee2000,Goossens2016}
\modifiedMagenta{which, due to their nanostructure, facilitate chemical reactions
or offer a higher degree of control over the reactions.}
}

\modifiedBlue{Another,
but \modifiedGreen{closely} related, class of liquids are room temperature ionic liquids
(RTILs) which at ambient pressure exhibit a melting temperature below room temperature. For these 
materials, as well as for ILCs, it is the combination of molecular shape-anisotropy
and the presence of charges which leads to a variety of astonishing properties of these fluids.
Besides the remarkable low melting temperature of RTILs,
caused by a suppression of crystallization at room temperature
(due to the underlying molecular shape-anisotropy),}
\modifiedGreen{RTILs show an almost negligible vapor pressure
which renders them candidates as solvents for ultrahigh vacuum
applications~\cite{Liu_et_al2002,Wang_et_al2004,Suzuki_et_al2007,Bermudez_et_al2009,Bier2010}.
The \modifiedMagenta{notion} \emph{room temperature ionic liquid}
emphasizes that this class of material remains liquid at standard conditions.
\modifiedMagenta{But, on one hand, RTILs may in addition} exhibit liquid crystalline phases
(below room temperature), which renders these room temperature
ionic liquids also \emph{ionic liquid crystals}.
On the other hand, if the molecular structure of RTILs is sufficiently asymmetric,
no \modifiedMagenta{liquid-crystalline} orientational ordering can be established and thus no mesophases will occur,
which distinguishes this kind of RTILs from ILCs.}

\modifiedMagenta{The} technological use of \modifiedGreen{ILCs and RTILs} 
\modifiedMagenta{requires} an in-depth understanding of the microscopical mechanisms,
\modifiedBlue{in particular the interplay of molecular shape-anisotropy and
the presence of charges}, leading to the
\modifiedBlue{remarkable behaviors} of those materials.
\modifiedBlue{Thus, theoretical studies,
which incorporate anisotropic charged particles and which allow
\modifiedGreen{one} to vary molecular properties
\modifiedGreen{(e.g., by} tuning the aspect-ratio or the charge distribution of the 
underlying particles\modifiedGreen{), might elucidate the role these 
microscopic properties play for the above mentioned
\modifiedMagenta{remarkable} macroscopic features of 
both classes of fluids, ILCs and RTILs}.}

\modifiedBlue{
\modifiedMagenta{On the one hand,}
previous theoretical studies mainly focused either on the effect of
molecular shape-anisotropy on thermodynamic properties or \modifiedGreen{on} 
ionic liquids within simplistic models.
For instance,} the \emph{restricted primitive model} (RPM),
\modifiedBlue{for which both ion species are considered to be uniformly charged hard spheres
of the same size and the same charge \modifiedGreen{strength},
has been studied intensively} in the past,
both in the continuum~\cite{Stell1976,Gillan1983}
as well as on lattices~\cite{Stell1999,Panagiotopoulos1999,Bartsch2015}.
However, \modifiedBlue{models incorporating spherically-shaped} ions are
\modifiedBlue{designed to study gross features such as the nature of criticality}
\cite{Fisher1994,Caillol1997,Orkoulas1999,Stell1999,Panagiotopoulos1999,Panagiotopoulos2002}.

On the other \modifiedMagenta{hand,} there \modifiedBlue{is} a vast number of theoretical studies concerning
ordinary (uncharged) liquid crystals, which are based on
\modifiedGreen{the elongated shapes} of the underlying molecules and 
\modifiedGreen{their} \modifiedBlue{anisotropic}
pair potentials~\cite{Onsager1949,Maier1958,Maier1959,Maier1960,
McMillan1971,McMillan1972,Straley1971,Straley1973,Straley1976,Sheng1976}.
Depending on the effective shape of the particles and \modifiedBlue{their} interaction potentials
one observes a huge diversity of mesophases
\modifiedBlue{
-- phases in between the isotropic liquid and the crystalline phase
distinguishable by their degree of positional and orientational ordering --}
occurring in systems of liquid crystals.
While plate-like particles (\emph{discotic mesogenes})
at high densities \MODIFIED{tend to} form columnar phases, in which the particles
form stack-like structures, in dense systems of elongated particles,
like thin rods or prolate ellipsoids (\emph{calamitic mesogens})
one \MODIFIED{typically} observes smectic structures,
in which the particles are located in layers\MODIFIED{~\cite{DeGennes1974}}.
\MODIFIED{(We note that \MODIfied{although} the formation of a columnar phase in a binary mixture of
hard spherocylinders has been reported~\cite{Stroobants1992},
this phase is not formed in a monodisperse system.)}
This distinct behavior due to the molecular anisotropy gives
rise to macroscopically measurable optical and mechanical anisotropies of liquid-crystalline materials
and drives phenomena like \modifiedBlue{self-assembly} or \modifiedBlue{nano-structuring}
\modifiedBlue{on microscopic scales}~\cite{Kato2006,Goodby2008,Tschierske2013}.

It is the \modifiedBlue{very} interplay of \modifiedBlue{shape-anisotropy}
and electrostatic interactions, which gives rise to the vast phenomenology
observed for ILC materials and at the same time \modifiedBlue{poses}
a particular challenge for theoretical studies.
Establishing a theoretical framework,
which is applicable to this kind of materials and allows one to
gain a deeper understanding of the origin of their properties, is an ongoing process.
Recently, Goossens et al.~\cite{Goossens2016} discussed in their review article the latest developments
in synthesis, characterization, and applications of ILCs. In particular, they \modifiedGreen{concluded}
that a deeper understanding of the role of the size, the shape,
and the charge distribution of the ILC molecules for the properties of those materials is required.
The aim of the present \modifiedGreen{contribution}
is to show that the considered molecular model of an ionic fluid,
incorporating orientational degrees of freedom as well as an anisotropic charge distribution,
gives rise to \modifiedBlue{a phenomenology concerning} the phase behavior
and \modifiedGreen{the} structural properties of the bulk phases,
\modifiedBlue{which is much richer than the one of}
simpler models of spherical ions or \modifiedBlue{of} ordinary liquid crystals.
The most \modifiedBlue{important new} result is the occurrence of a novel smectic phase $S_{AW}$
at low \modifiedGreen{temperatures},
\modifiedBlue{the layer spacing of which} is larger than that of the 
ordinary high-temperature smectic phase $S_{A}$.
The present findings stress the crucial role of the \modifiedBlue{loci}
of the charges on the ILC molecules and at the same time
emphasize the necessity of considering such kind of sophisticated model
in order to study \modifiedBlue{reliably} complex ionic \modifiedBlue{liquids}
such as room temperature ionic liquids.

The present \modifiedGreen{study} is structured as follows:
In Sec.~\ref{sec:theory} the employed model is presented as well as
\modifiedBlue{the outlines of} the methods
\modifiedBlue{which encompass density functional theory and Monte Carlo simulations.}
Our results for the phase behavior and \modifiedGreen{for} the structure of
various smectic phases of ILCs are discussed in Sec.~\ref{sec:discussion}.
Finally, in Sec.~\ref{sec:conclusions} we summarize the results and draw our conclusions.


\section{\label{sec:theory}Model and methods}

\modifiedBlue{This section presents in detail the}
molecular model of ILCs
\modifiedBlue{as employed here.}
In particular, we discuss the intermolecular pair potential,
which \modifiedBlue{can be applied} to a wide range of
ionic and liquid crystalline materials
due to its \modifiedBlue{flexibility provided by a large} set of parameters.

\modifiedBlue{This} model is studied by 
density functional theory (DFT) as well as
\modifiedBlue{by} grandcanonical Monte Carlo simulations.
The methodological and technical details
of both approaches are \modifiedBlue{described}
in Secs.~\ref{sec:theory:DFT} and \ref{sec:theory:GCMC},
\modifiedBlue{respectively}.
\begin{figure}[!t]
 \includegraphics[width=0.45\textwidth]{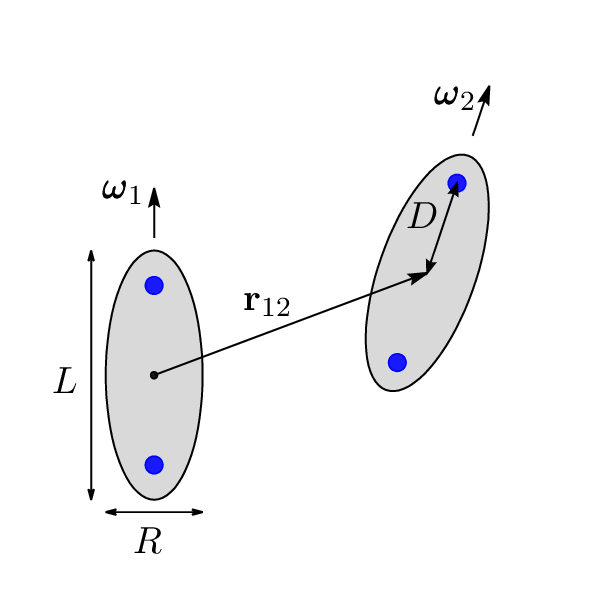}
  \caption{
  Cross-sectional view of two ILC molecules
  \modifiedBlue{
  in the plane spanned by the orientations $\vec\omega_i,~i=1,2$,
  of their long axis.}
  The particles are treated as rigid prolate ellipsoids,
  characterized by their length-to-breadth ratio $L/R$.
  Their orientations are fully described by the direction of
  their long axis $\vec\omega_i$; 
  $\vec{r}_{12}$ is the center-to-center distance vector.
  The \modifiedBlue{charges of the} ILC molecules (blue dots) are
  \modifiedBlue{located} on the long axis
  at \modifiedBlue{a} distance $D$ from their geometrical center.
  The counterions are not modeled explicitly,
  but they are implicitly accounted for in terms of a background,
  giving rise to the screening of the charges of the ILC molecules. 
  }
 \label{fig:ellipsoids}
\end{figure}
%

\subsection{\label{sec:theory:model} Molecular model and pair potential}
We consider a coarse-grained description of the ILC molecules
as rigid prolate ellipsoids of length-to-breadth ratio $L/R$
(see Fig.~\ref{fig:ellipsoids}).
Thus, the orientation of a molecule is fully described by the
direction \modifiedGreen{$\vec\omega(\phi,\theta)$ of its long axis,
where $\phi$ and $\theta$ denote the azimuthal and polar angle, respectively.}

The two-body interaction potential consists
of a hard core repulsive and an additional contribution
$U_\text{GB}+U_\text{es}$ beyond the
contact distance $R\sigma$, \modifiedBlue{the sum of} which can
\modifiedBlue{be attractive} or repulsive:
\begin{equation}
   U= 
   \begin{cases}
      \infty
      &, 
      |\vec{r}_{12}|    < R\sigma(\vec{\hat r}_{12},\vec\omega_1,\vec\omega_2) \\
      \begin{array}{l}
       U_\text{GB}(\vec{r}_{12},\vec\omega_1,\vec\omega_2)+\\
       U_\text{es}(\vec{r}_{12},\vec\omega_1,\vec\omega_2)
      \end{array}
      &,
      |\vec{r}_{12}| \geq R\sigma(\vec{\hat r}_{12},\vec\omega_1,\vec\omega_2),
   \end{cases}
\label{eq:Pairpot}
\end{equation}
where $\vec{r}_{12}:=\vec{r}_2-\vec{r}_1$ denotes the center-to-center distance vector
between \modifiedBlue{the} two particles labeled as 1 and 2,
and $\vec\omega_i$, $i=1,2$, are their orientations.
The contact distance $R\sigma(\vec{\hat r}_{12},\vec\omega_1,\vec\omega_2)$
depends on the orientations of both particles and their relative direction, 
which is expressed by the unit vector $\vec{\hat r}_{12}:=\vec{r}_{12}/|\vec{r}_{12}|$.
In Eq.~(\ref{eq:Pairpot}), we subdivided the contributions beyond the contact distance
$|\vec{r}_{12}|\geq R\sigma$ into two parts:
$U_\text{GB}(\vec{r}_{12},\vec\omega_1,\vec\omega_2)$
is the well-known Gay-Berne potential~\cite{Berne1972,Gay_Berne1981},
which incorporates an attractive \modifiedBlue{van der Waals-like} interaction between molecules
and which can be understood as a generalization of the Lennard-Jones pair potential
to ellipsoidal particles:
\begin{equation}
 \begin{split}
  &    U_\text{GB}(\vec     {r}_{12},\vec\omega_1,\vec\omega_2)
  =      4\epsilon(\vec{\hat r}_{12},\vec\omega_1,\vec\omega_2)\\
  & \times \left[  \left(1+\frac{|\vec{r}_{12}|}{R}-\sigma(\vec{\hat r}_{12},\vec\omega_1,\vec\omega_2) \right)^{-12} \right.\\
  &        \left.-~\left(1+\frac{|\vec{r}_{12}|}{R}-\sigma(\vec{\hat r}_{12},\vec\omega_1,\vec\omega_2) \right)^{-6}  \right]\\
 \end{split}
 \label{eq:Pairpot_GB}
\end{equation}
\modifiedBlue{with}
\begin{equation}
 \begin{split}
    \sigma(\vec{\hat r}_{12},\vec\omega_1,\vec\omega_2)
  & =\left[ 1-\frac{\chi}{2}\left(\frac{(\vec{\hat r}_{12}\cdot(\vec\omega_1+\vec\omega_2))^2}{1+\chi\vec\omega_1\cdot\vec\omega_2}\right.\right.\\
  &  \left.  +              \left.\frac{(\vec{\hat r}_{12}\cdot(\vec\omega_1-\vec\omega_2))^2}{1-\chi\vec\omega_1\cdot\vec\omega_2}\right)\right]\\
 \end{split}
\end{equation}
\modifiedBlue{and}
\begin{equation}
 \begin{split}
     \epsilon(\vec{\hat r}_{12},\vec\omega_1,\vec\omega_2)
  & =\epsilon_0\left(1-(\chi\vec\omega_1\cdot\vec\omega_2)^2\right)^{-1/2}\\
  & \times\left[ 1-\frac{\chi'}{2}\left(\frac{(\vec{\hat r}_{12}\cdot(\vec\omega_1+\vec\omega_2))^2}{1+\chi'\vec\omega_1\cdot\vec\omega_2}\right.\right.\\
  & \left.  +               \left.\frac{(\vec{\hat r}_{12}\cdot(\vec\omega_1-\vec\omega_2))^2}{1-\chi'\vec\omega_1\cdot\vec\omega_2}\right)\right].\\
 \end{split}
 \label{eq:Pairpot_GB_epsilon}
\end{equation}
The contact distance $R\sigma(\vec{\hat r}_{12},\vec\omega_1,\vec\omega_2)$ 
and the direction- and orientation-dependent
interaction strength $\epsilon(\vec{\hat r}_{12},\vec\omega_1,\vec\omega_2)$ are both
parametrically dependent on the length-to-breadth ratio
$L/R$ via the auxiliary function $\chi=((L/R)^2-1)/((L/R)^2+1)$.
Additionally, $\epsilon(\vec{\hat r}_{12},\vec\omega_1,\vec\omega_2)$ can be tuned via 
$\chi'=((\epsilon_R/\epsilon_L)^{1/2}-1)/((\epsilon_R/\epsilon_L)^{1/2}+1)$,
where $\epsilon_R/\epsilon_L$ is called the anisotropy parameter, defined 
\modifiedMagenta{in terms of} the ratio of
$\epsilon_R$, \modifiedGreen{which is} the depth of the potential minimum for parallel particles
positioned side by side
$(\vec{\hat r}_{12}\cdot\vec\omega_1=\vec{\hat r}_{12}\cdot\vec\omega_2=0)$\modifiedMagenta{, and}
$\epsilon_L$\modifiedGreen{, which} is the depth of the potential minimum for parallel particles
positioned end to end
$(\vec{\hat r}_{12}\cdot\vec\omega_1=\vec{\hat r}_{12}\cdot\vec\omega_2=1)$.
The energy scale of the Gay-Berne pair interaction is set by $\epsilon_0$.
\modifiedBlue{
Thus, the Gay-Berne pair potential has four independent free parameters:
$\epsilon_0, R, L/R$, and $\epsilon_R/\epsilon_L$.
Note that in the case of spherical particles, \modifiedGreen{i.e., for} $L=R$,
the Gay-Berne pair potential (Eq.~(\ref{eq:Pairpot_GB}))
reduces to the well-known isotropic Lennard-Jones pair potential
iff, additionally, the Gay-Berne anisotropy parameter 
equals unity, i.e., $\epsilon_R/\epsilon_L=1$, because \modifiedGreen{then}
$\sigma(\vec{\hat r}_{12},\vec\omega_1,\omega_2)=1$ and
$\epsilon(\vec{\hat r}_{12},\vec\omega_1,\omega_2)=\epsilon_0$.}

The second contribution 
\modifiedBlue{$U_\text{es}(\vec{r}_{12},\vec\omega_1,\vec\omega_2)$}
in Eq.~(\ref{eq:Pairpot}) is the \modifiedGreen{\emph{e}lectro\emph{s}tatic repulsion}
of ILC molecules. Within the scope of \modifiedBlue{the present study},
the counterions are not modeled explicitly,
but they will be considered to be much smaller in size
than the ILC molecules such that they can be treated as 
a continuous background. On \modifiedBlue{the level of} linear response,
this background gives rise to the screening of the pure Coulomb potential
between two \modifiedGreen{charged} sites
on a length scale given by the Debye screening length $\lambda_D$
such that the effective electrostatic interaction of the ILC molecules is given by
\begin{equation}
 \begin{split}
      U_\text{es}(\vec     {r}_{12},\vec\omega_1,\vec\omega_2)=~
  &   \gamma\left[\frac{\exp\left(-\frac{|\vec{r}_{12}+D(\vec\omega_1+\vec\omega_2)|}{\lambda_D}\right)}{|\vec{r}_{12}+D(\vec\omega_1+\vec\omega_2)|}\right.\\  
  &   +     \left.\frac{\exp\left(-\frac{|\vec{r}_{12}+D(\vec\omega_1-\vec\omega_2)|}{\lambda_D}\right)}{|\vec{r}_{12}+D(\vec\omega_1-\vec\omega_2)|}\right.\\  
  &   +     \left.\frac{\exp\left(-\frac{|\vec{r}_{12}-D(\vec\omega_1+\vec\omega_2)|}{\lambda_D}\right)}{|\vec{r}_{12}-D(\vec\omega_1+\vec\omega_2)|}\right.\\  
  &   +     \left.\frac{\exp\left(-\frac{|\vec{r}_{12}-D(\vec\omega_1-\vec\omega_2)|}{\lambda_D}\right)}{|\vec{r}_{12}-D(\vec\omega_1-\vec\omega_2)|}\right].  
 \end{split}
\label{eq:PairPot_ES}
\end{equation}
\modifiedBlue{The} charges $q$ are located symmetrically on their long axis at
\modifiedGreen{a} distance $D$
from the geometrical center of the particles (compare Fig.~\ref{fig:ellipsoids});
$\gamma=q^2/(4\pi\epsilon)$ characterizes the
electrostatic energy scale with permittivity $\epsilon$.
\modifiedBlue{
In principle, the Debye screening length 
\begin{equation}
 \lambda_D=\sqrt{\frac{kT}{q^2\rho_\text{c}}}
\label{eq:Debyelength}
\end{equation}
is a function of temperature $T$ and
of the number density $\rho_\text{c}$ of the counter ions. Thus,
it depends on the thermodynamic state of the fluid.
However, in the present model $\lambda_D$ is \modifiedGreen{taken} to be a constant parameter.
In order to compare results, obtained within this model, with
\modifiedGreen{data from} actual physical systems,
one could measure the value of the Debye screening length experimentally
and tune the model parameter $\lambda_D$ accordingly.
}

\begin{figure}[!t]
 \includegraphics[width=0.45\textwidth]{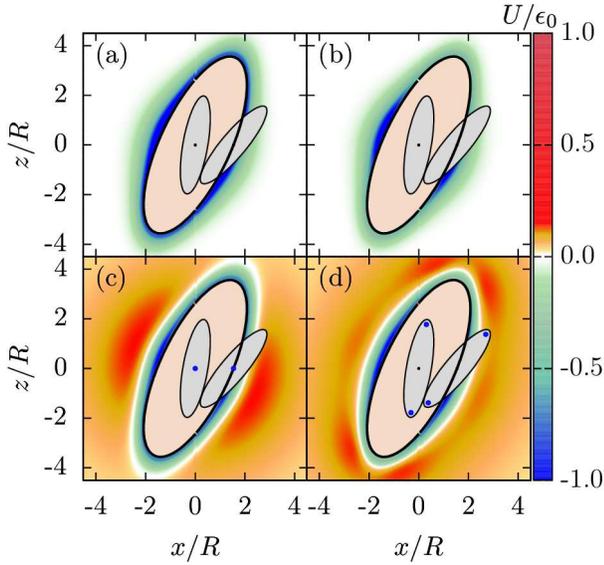}
  \caption{
  Contour-plots of the pair potential $U$ 
  for $|\vec{r}_{12}|\geq R\sigma$ in the $x$-$z$-plane
  for four \modifiedBlue{cases} of particles
  with \modifiedBlue{fixed} length-to-breadth ratio $L/R=4$ and fixed orientations.
  \modifiedGreen{In each panel the centers of both particles lie
  in the plane $y=0$.}
  \modifiedBlue{In order} to illustrate the orientations
  \modifiedBlue{of the ellipsoids},
  they have been included in the plots \modifiedBlue{at contact}
  \modifiedBlue{with} relative direction $\vec{\hat r}_{12}=\vec{\hat x}$.
  The \modifiedBlue{set} of points at contact
  in the $x$-$z$-plane is illustrated by the black curve and the 
  centers \modifiedBlue{of the particles} are shown by small black dots.
  Panel (a): uncharged liquid crystal with $\epsilon_R/\epsilon_L=2$.
  Panel (b): uncharged liquid crystal with $\epsilon_R/\epsilon_L=4$.
  The anisotropy of the potential is increased \modifiedBlue{slightly}.
  Panel (c): ILC with
  $\epsilon_R/\epsilon_L=2,D/R=0,\lambda_D/R=5,\gamma/(R\epsilon_0)=0.25$.
  Panel (d): ILC with
  $\epsilon_R/\epsilon_L=\modifiedBlue{2},D/R=1.8,\lambda_D/R=5,\gamma/(R\epsilon_0)=0.25$.
  \modifiedBlue{In (c) and (d) the loci of the charges are indicated as blue dots.}
  \modifiedGreen{The salmon-colored area is the excluded volume for given orientations of
  the two particles.}
  }
 \label{fig:pairpot}
\end{figure}
\modifiedGreen{In Fig.~\ref{fig:pairpot}}
we illustrate the full pair potential (Eq.~(\ref{eq:Pairpot}))
beyond the contact distance 
for certain choices of the parameters.
The two top panels, (a) and (b), show the pure Gay-Berne potential
(uncharged liquid crystals),
which is predominantly attractive in the space
outside the overlap volume (cream-colored area).
The shape of the overlap volume changes by varying the particle orientations
as well as by changing the length-to-breadth ratio $L/R$.
\modifiedBlue{However, these dependences are not apparent from} Fig.~\ref{fig:pairpot},
since $L/R=4$ and the particle orientations $\vec\omega_i$ are \modifiedBlue{kept fixed}
for all panels.
In panel (b) the anisotropy parameter $\epsilon_R/\epsilon_L=4$
is chosen to be  two times larger than for panel (a) ($\epsilon_R/\epsilon_L=2$).
Thus, the ratio of the well depth at the tails and at the sides is increased.
The two bottom panels, (c) and (d), show the same \modifiedGreen{choices} for the Gay-Berne parameters
\modifiedBlue{as for panel (a), but} the electrostatic repulsion of
the charged groups on the molecules, illustrated by blue dots,
is included ($\gamma/(R\epsilon_0)=0.25$).
In panel (c) the \modifiedBlue{loci of the two} charges
of \modifiedGreen{the particles} \modifiedBlue{coincide at its} center
\modifiedBlue{(i.e., $D/R=0$)}
while in panel (d) they are located near the \modifiedBlue{tips ($D/R=1.8$)}.
For both cases \modifiedBlue{with charge}, the effective interaction \modifiedBlue{range}
is significantly increased \modifiedBlue{compared with the uncharged case} and is
governed by the Debye screening length, chosen \modifiedGreen{as} $\lambda_D/R=5$.

\subsection{\label{sec:theory:DFT} Density functional theory}

\subsubsection{\label{sec:theory:DFT:Formalism} Formalism}

The degrees of freedom of the particles (compare Sec.~\ref{sec:theory:model})
are fully described by \modifiedBlue{the positions $\vec{r}$ of their centers} and
the orientations \modifiedBlue{$\vec\omega$ of} their long axes.
Thus, within density functional theory an appropriate
\modifiedBlue{variational} grand potential functional $\beta\Omega[\rho]$
of position- and orientation-dependent
\modifiedBlue{number density} profiles $\rho(\vec{r},\vec\omega)$ has to be found;
\modifiedBlue{its} minimum corresponds to the equilibrium density profile.
The grand potential functional for uniaxial particles,
in \modifiedBlue{the} absence of external fields,
can generically \modifiedBlue{be} expressed as
\begin{equation}
 \begin{split}
    \beta\Omega\left[\rho\right]=
  & \Int{\mathcal{V}}{3}{r}\Int{\mathcal{S}}{2}{\omega}
    \rho(\vec{r},\vec\omega)
    \left[\ln\left(4\pi\Lambda^3\rho(\vec{r},\vec\omega)\right)\right.\\
  &-\left.\left(1+\beta\mu\right)\right]
   +\beta\mathcal{F}\left[\rho\right],
 \end{split}
\label{eq:Omega}
\end{equation}
where the integration domains $\mathcal{V}$ and $\mathcal{S}$ denote the system volume
and the full solid angle, respectively.
The first term in Eq.~(\ref{eq:Omega}) is the purely entropic free energy contribution
of non-interacting uniaxial particles,
where \modifiedBlue{$\beta=1/(k_BT)$} denotes the inverse thermal energy,
$\mu$ the chemical potential,
and $\Lambda$ the thermal de Broglie wavelength.
\modifiedBlue{The last term is the excess free energy
$\beta\mathcal{F}\left[\rho\right]$ in units of $k_BT$,}
which incorporates the effects of the \modifiedBlue{inter-}particle interactions. 
\modifiedBlue{Minimizing} Eq.~(\ref{eq:Omega}) leads to the 
Euler-Lagrange equation, \modifiedBlue{which implicitly determines} the equilibrium density profile
\modifiedBlue{$\rho(\vec{r},\vec\omega)$}:
\begin{equation}
  \rho(\vec{r},\vec\omega)=
  \frac{e^{\beta\mu}}{4\pi\Lambda^3}
  \exp\left[c^{(1)}\left(\vec{r},\vec\omega,[\rho]\right)\right],
\label{eq:ELG}
\end{equation}
where
\begin{equation}
  c^{(1)}\left(\vec{r},\vec\omega,[\rho]\right)=
  -\frac{\delta\beta\mathcal{F}[\rho]}{\delta\rho}
\label{eq:Dir1CorrFunc}
\end{equation}
is the one-particle direct correlation function.
It is fully determined by the excess free energy functional $\beta\mathcal{F}[\rho]$.

Since the excess free energy functional is the characterizing quantity
of the underlying many-body problem,
in general it is not known exactly \modifiedBlue{so that}
one \modifiedBlue{has} to find appropriate approximations of \modifiedBlue{it}.
The starting point of the present \modifiedBlue{study} is
a weighted density formulation of $\beta\mathcal{F}[\rho]$
in the spirit of \modifiedBlue{Ref.}~\cite{Tarazona1985}:
\begin{equation}
  \beta\mathcal{F}[\rho]=
  \frac{1}{2}\Int{\mathcal{V}}{3}{r}\Int{\mathcal{S}}{2}{\omega}
  \rho(\vec{r},\vec\omega)
  \beta\psi\left(\vec{r},\vec\omega,[\bar\rho]\right),
\label{eq:F_WDA}
\end{equation}
which immediately leads to the following expression for 
the one-particle direct correlation function:
\begin{align}
  &c^{(1)}\left(\vec{r},\vec\omega,[\rho]\right)=-
  \frac{1}{2}\bigg[\beta\psi(\vec{r},\vec\omega,[\bar\rho])+
  \Int{\mathcal{V}}{3}{r'}\Int{\mathcal{S}}{2}{\omega'}\rho(\vec{r}',\vec\omega')\notag\\
  &\times
  \Int{\mathcal{V}}{3}{r''}\Int{\mathcal{S}}{2}{\omega''}
  \frac{\delta\beta\psi(\vec{r}',\vec\omega',[\bar\rho])}{\delta\bar\rho(\vec{r}'',\vec\omega'')}
  \,
  \frac{\delta\bar\rho (\vec{r}'',\vec\omega'',[\rho])}{\delta\rho(\vec{r},\vec\omega)}\bigg].
\label{eq:Calc_c1} 
\end{align}
In order to evaluate Eq.~(\ref{eq:Calc_c1}),
one needs to know the effective one-particle potential
$\beta\psi[\bar\rho]$ as a functional of the weighted density
$\bar\rho(\vec{r},\vec\omega,[\rho])$, which in the present case is chosen
as a projection of the full density profile $\rho(\vec{r},\vec\omega)$ onto
a certain functional subspace (see below).

The present work aims at studying the phase behavior of ILCs,
composed of uniaxial prolate particles.
Hence, one expects the occurrence of
isotropic (no positional and no orientational order),
nematic (no positional, but orientational order),
and smectic phases (one-dimensional positional \modifiedBlue{order in $z$-direction}
and orientational order).
\modifiedRed{At sufficiently low \modifiedGreen{temperatures and sufficiently large densities 
the homogenous phases mentioned above,} i.e., the isotropic and nematic phases,
or partially homogenous phases, i.e., the smectic phases, undergo transitions
to crystalline phases.}
\modifiedBlue{The first three}
types of phases can be represented
by spatially periodic density profiles $\rho(\vec{r},\vec\omega)$
with wavelength $d$ in $z$-direction and \modifiedGreen{spatially} 
\modifiedBlue{constant density} perpendicular to it.
\modifiedBlue{For a uniform density in $z$-direction $d$ is not uniquely defined and
can} be chosen arbitrarily,
whereas for smectic structures with layers perpendicular to the $z$-direction
$d$ is an integer multiple of the layer spacing.
\modifiedMagenta{
(Although there is no need to introduce $d$ for uniform phases, within the present
approach based on the projected density $\bar\rho(\vec{r},\vec\omega)$
(see Eqs.~(\ref{eq:WeightedDensity})-(\ref{eq:ExpansionCoeffs2}) below),
also a uniform density profile $\rho(\vec{r},\vec\omega)$ demands a value for $d$
entering into Eq.~(\ref{eq:ExpansionCoeffs2}).
However, the corresponding results do not depend
on such a choice of $d$; any value $d>0$ is valid.)}
This observation motivates the \modifiedBlue{approach to consider} a projected density
$\bar\rho(\vec{r},\vec\omega,[\rho])$, which is obtained by 
weighting the original density profile $\rho(\vec{r},\vec\omega)$
within a periodic cell of volume
$\mathcal{V}_d=A\times d$ around the position $\vec{r}$,
where $A$ is the cross-sectional area of the system.
\modifiedBlue{In order to express the orientational dependence of the projected density
$\bar\rho(\vec{r},\vec\omega)$ explicitly, in addition to the Fourier \modifiedGreen{series} expansion
of $\rho(\vec{r},\vec\omega)$
in terms of $\cos(2\pi iz/d)$ \modifiedMagenta{(with $i=0,1,2$)}
$\bar\rho(\vec{r},\vec\omega)$ (Eq.~(\ref{eq:WeightedDensity}))
is \modifiedGreen{determined} by performing
\modifiedMagenta{furthermore} an expansion of $\rho(\vec{r},\vec\omega)$
in \modifiedGreen{terms of} Legendre polynomials $P_l(y=\cos\theta)$
up to \modifiedGreen{and including} second order\modifiedOrange{, i.e.,} \modifiedMagenta{$l=0,2$}.
The \modifiedGreen{contribution} corresponding to $l=1$ vanishes due to the symmetry of
the underlying pair potential $U(\vec{r}_{12},\vec\omega_1,\vec\omega_2)$
(Eq.~(\ref{eq:Pairpot})):}
\begin{equation}
 \begin{split}
  & \bar\rho(\vec{r},\vec\omega,[\rho]) = \frac{1}{4\pi}\bigg[
    Q_0(\vec{r},[\rho])+
    Q_1(\vec{r},[\rho])\cos\left(2\pi z/d\right)\\
  &+Q_2(\vec{r},[\rho])\cos\left(4\pi z/d\right)+5P_2(\cos(\theta))
    \bigg(Q_3(\vec{r},[\rho])\\
  &+Q_4(\vec{r},[\rho])\cos\left(2\pi z/d\right)
   +Q_5(\vec{r},[\rho])\cos\left(4\pi z/d\right)\bigg)\bigg],
 \end{split}
\label{eq:WeightedDensity}
\end{equation}
where \modifiedGreen{$P_0(y)=1$}, $P_2(y)=(3y^2-1)/2$,
and \modifiedMagenta{with} coefficients $Q_i(\vec{r},[\rho])$ defined as
\begin{align}
 Q_i(\vec{r},[\rho])&=\frac{1}{\mathcal{V}_d}
 \Int{\mathcal{V}}{3}{r'}\Int{\mathcal{S}}{2}{\omega'}
 \rho(\vec{r}',\vec\omega')w_i(z,z',\theta')
 \label{eq:ExpansionCoeffs}
\end{align}
\modifiedBlue{with}
\begin{align}
 w_0&=\Theta(d/2-|z-z'|),\nonumber\\
 w_1&=2\Theta(d/2-|z-z'|)\cos\left(2\pi z'/d\right),\nonumber\\
 w_2&=2\Theta(d/2-|z-z'|)\cos\left(4\pi z'/d\right),\nonumber\\
 w_3&=\Theta(d/2-|z-z'|)P_2(\cos(\theta')),\nonumber\\
 w_4&=2\Theta(d/2-|z-z'|)P_2(\cos(\theta'))\cos\left(2\pi z'/d\right),\nonumber\\
 w_5&=2\Theta(d/2-|z-z'|)P_2(\cos(\theta'))\cos\left(4\pi z'/d\right).
 \label{eq:ExpansionCoeffs2}
\end{align}
Here $\Theta(x)$ denotes the Heaviside step function;
\modifiedGreen{concerning $\theta$ see below}.
\modifiedBlue{Without loss of generality}, for the three relevant bulk phases
one can consider the entire system being composed of \modifiedBlue{a set of} periodic
\modifiedGreen{macro-cells}.

Although \modifiedBlue{in general} the coefficients $Q_i(\vec{r},[\rho])$
depend on the position $\vec{r}$,
e.g., close to interfaces, for the scope of \modifiedBlue{the present}
\modifiedGreen{study} they are constant,
$Q_i(\vec{r},[\rho])=Q_i=\text{const}$,
\modifiedBlue{because} here we consider spatially periodic bulk profiles \modifiedBlue{only}.
\modifiedBlue{Thus, the \modifiedMagenta{coefficients} $Q_i$
in Eqs.~(\ref{eq:ExpansionCoeffs}) and (\ref{eq:ExpansionCoeffs2})
represent the first coefficients of a Fourier expansion
of the spatially periodic function $\rho(\vec{r},\vec\omega)$.
Note, that the factor $2$ for $w_1$, $w_2$, $w_4$, and $w_5$ in Eq.~(\ref{eq:ExpansionCoeffs2})
\modifiedGreen{is due to the definition of the first and second Fourier modes.}
Similarly, the factor $5$ in the \modifiedMagenta{last} term of Eq.~(\ref{eq:WeightedDensity})
is due to the \modifiedGreen{definition of the coefficient}
of the second order term of an expansion 
\modifiedMagenta{in terms of} Legendre polynomials.}
\modifiedBlue{The normal} of the smectic layers is chosen to be parallel to the $z$-axis.
\modifiedBlue{We restrict our analysis of smectic phases
to the case in} which the director field $\vec{\hat n}(\vec{r})=\vec{\hat z}$,
\modifiedBlue{describing the mean orientation of the particles}, is homogenous
and points along the $z$-direction as well (smectic-A ($S_A$)~\cite{DeGennes1974}).
Additionally, only distributions of orientations $\vec\omega$, 
\modifiedBlue{which are symmetric} around the director $\vec{\hat n}$, are considered,
\modifiedBlue{with} the polar angle $\theta$
between the director and the long-axis of one particle
is given by $\cos(\theta):=\vec{\hat n}\cdot\vec\omega$.
\MODIFIED{Thus, our description is restricted to uniaxial phases, like the isotropic, nematic,
and the smectic-A phase considered here. In order to study biaxial phases
(e.g., smectic-C phases where the director is tilted with respect to the layer normal)
within the present DFT-approach one would need to keep the full orientational
dependence of the projected density $\bar\rho(\vec{r},\vec\omega)$
on both the polar angle $\theta$ and the azimuthal angle $\phi$.}
\MODIfied{However, the present computer simulations did not reveal any evidence of the occurrence
of biaxial phases in the investigated systems. In particular, the smectic-A-type phases
were the only smectic phases that could be observed (see Sec.~\ref{sec:discussion}).
Therefore, the restriction to uniaxial structures seems to be adequate for the systems
studied here.}

In the final step of constructing the density functional,
the effective one-particle potential $\beta\psi[\bar\rho]$ needs to be \modifiedBlue{specified}.
Here, it \modifiedBlue{consists of} two contributions.
The first one \modifiedBlue{is due to} the hard-core interaction.
\modifiedBlue{For this contribution we adopt}
the well-studied Parsons-Lee approach~\cite{Parsons1979,Lee1987}
\begin{align}
  &\beta\psi_\text{PL}(\vec{r},\vec\omega,[\bar\rho]) = \modifiedBlue{-}
   \Int{\mathcal{V}}{3}{r'}\Int{\mathcal{S}}{2}{\omega'}
   \bar\rho(\vec{r}',\vec\omega')\nonumber\\
  &\times\frac{\mathcal{J}(Q_0(\vec{r}))+\mathcal{J}(Q_0(\vec{r}'))}{2}
   f_M(\vec{r}-\vec{r}',\vec\omega,\vec\omega'),
\label{eq:Eff1Pot_PL}
\end{align}
where $f_M(\vec{r}-\vec{r}',\vec\omega,\vec\omega')$ is the \MODIFIED{Mayer f-function~\cite{Hansen1976}}
of the hard core pair interaction potential and
\modifiedBlue{$\mathcal{J}(Q_0)$ modifies} the \modifiedBlue{corresponding}
original Onsager free energy functional
(\modifiedBlue{i.e.,} the second-order virial approximation) such that
the Carnahan-Starling equation of state~\cite{Lee1987} is reproduced
for spheres\modifiedBlue{, i.e.,} $L=R$~\cite{Onsager1949,VanRoij2005}:
\begin{equation}
 \mathcal{J}(Q_0)=\frac{1-\frac{3}{4}\eta_0(Q_0)}{(1-\eta_0(Q_0))^2},
\label{eq:ScalingFunctionJ}
\end{equation}
where $\eta_0=Q_0\,LR^2\pi/6$ denotes the mean packing fraction
within the volume $\mathcal{V}_d$.
It is proportional to the coefficient $Q_0$
which gives the mean density
within the volume $\mathcal{V}_d$.
\modifiedBlue{The original Onsager functional is recovered by replacing
$\mathcal{J}(Q_0)$ by $Q_0$ in Eq.~(\ref{eq:Eff1Pot_PL}).}

The second contribution to
the effective one-particle potential $\beta\psi[\bar\rho]$
\modifiedBlue{takes into account} the interactions beyond the contact distance
(\modifiedBlue{see} the case $|\vec{r}_{12}|\geq R\sigma$ in Eq.~(\ref{eq:Pairpot}))
within \modifiedBlue{the} modified mean-field approximation~\cite{Teixeira1991},
a variant of the extended random phase approximation (ERPA)~\cite{Evans1979}:
\begin{align}
  &\beta\psi_\text{ERPA}(\vec{r},\vec\omega,[\bar\rho]) =
   \Int{\mathcal{V}}{3}{r'}\Int{\mathcal{S}}{2}{\omega'}
   \bar\rho(\vec{r}',\vec\omega')\nonumber\\
  &\times\beta U(\vec{r}-\vec{r}',\vec\omega,\vec\omega')
     (1+f_M(\vec{r}-\vec{r}',\vec\omega,\vec\omega')).
\label{eq:Eff1Pot_ERPA}
\end{align}

For the sake of simplicity, instead of using the 
full \modifiedGreen{angular} expressions for the two contributions to the effective
one-particle potential, given by Eqs.~(\ref{eq:Eff1Pot_PL}) and (\ref{eq:Eff1Pot_ERPA}),
we utilize their expansions in \modifiedBlue{terms of} Legendre polynomials (up to second order)
\modifiedGreen{which provides an explicit expression for}
the orientational dependence of the effective one-particle potential:
\begin{align}
  \beta\psi(\vec{r},\vec\omega,[\bar\rho]) & =
  \zeta_{0}(\vec{r})+
  \zeta_{2}(\vec{r})P_2(\cos(\theta)),
  \notag\\
  \zeta_l(\vec{r}) & = \frac{1}{4\pi}
  \Int{\mathcal{S}}{2}{\omega'}(\beta\psi_\text{PL}(\vec{r},\vec\omega')+
                                \beta\psi_\text{ERPA}(\vec{r},\vec\omega')
                               )\nonumber\\
  & \times  
   \begin{cases}
     1 &, l=0 \\
     5P_2(\cos(\theta')) &, l=2.\\
   \end{cases}
\label{eq:Eff1Potential}
\end{align}

In order to \modifiedBlue{determine} the equilibrium density profile in Eq.~(\ref{eq:ELG}),
\modifiedBlue{one has} to \modifiedGreen{calculate} the one-particle direct correlation function
\modifiedBlue{(Eq.~(\ref{eq:Calc_c1}))}, using the definition of the weighted density
$\bar\rho(\vec{r},\vec\omega)$ (Eqs.~(\ref{eq:WeightedDensity})-(\ref{eq:ExpansionCoeffs2})),
and the effective one-particle potential $\beta\psi(\vec{r},\vec\omega,[\bar\rho])$
(Eqs.~(\ref{eq:Eff1Pot_PL})-(\ref{eq:Eff1Potential})).

For the particular case of bulk phases,
\modifiedBlue{in} which the coefficients $Q_i$ in Eq.~(\ref{eq:WeightedDensity})
do not depend on the position $\vec{r}$,
one finds the following expression for the equilibrium 
density profile (see \modifiedBlue{Appendix~\ref{sec:appendix:derivation}})
\begin{equation}
 \begin{split}
   \rho^{(0)}(\vec{r},\vec\omega):=\exp
  &\bigg[\sum_{i=0}^{2}A_i\cos(2\pi i z/d)\,+\\
  &   P_2(\cos(\theta))B_i\cos(2\pi i z/d)\bigg],
 \end{split}
 \label{eq:DensityGenericForm}
\end{equation}
where the \modifiedBlue{constant} coefficients $A_i$ and $B_i$ are to be determined by
evaluating Eqs.~(\ref{eq:ELG}) and (\ref{eq:Calc_c1})
\modifiedBlue{for this expression of $\rho^{(0)}(\vec{r},\vec\omega)$}.
As expected, the bulk density profile depends \modifiedBlue{only} on the $z$-coordinate
and the polar angle $\theta$.

It turns out, that the \modifiedBlue{precise} evaluation
of the coefficients $A_i$ \modifiedBlue{and} $B_i$ is very costly in terms of
computational resources and almost not feasible
with reasonable computational effort.
\modifiedBlue{In order to} circumvent those numerical difficulties, 
\modifiedBlue{from here on} we \modifiedBlue{shall} follow two different routes.
\modifiedBlue{Along the} first one,
instead of solving the \modifiedBlue{full} Euler-Lagrange equation \modifiedBlue{and}
using Eq.~(\ref{eq:Calc_c1}) \modifiedBlue{in order}
to evaluate Eqs.~(\ref{eq:ELG}) and (\ref{eq:Dir1CorrFunc}),
we modify the expression for the one-particle direct correlation function
\modifiedBlue{in} Eq.~(\ref{eq:Calc_c1}),
by replacing \modifiedBlue{in the integrand} the true density profile $\rho(\vec{r},\vec\omega)$
\modifiedBlue{by} the weighted density $\bar\rho(\vec{r},\vec\omega)$.
Consequently, Eq.~(\ref{eq:Calc_c1}) now reads 
\begin{equation}
 \begin{split}
   &\tilde c^{(1)}\left(\vec{r},\vec\omega,[\rho]\right)=-
    \frac{1}{2}\bigg[\beta\psi(\vec{r},\vec\omega,[\bar\rho])\\
   &+
    \Int{\mathcal{V}}{3}{r'}\Int{\mathcal{S}}{2}{\omega'}\bar\rho(\vec{r}',\vec\omega')
    \frac{\delta\beta\psi(\vec{r}',\vec\omega',[\bar\rho])}{\delta\bar\rho(\vec{r},\vec\omega)}\bigg]
 \end{split}
 \label{eq:Calc_c1_Approx_tent}
\end{equation}
where
\modifiedGreen{ 
$\frac{\delta\bar\rho (\vec{r}'',\,\,\vec\omega'')}{\delta\bar\rho(\vec{r},\,\,\vec\omega)}=\delta(\vec{r}''-\vec{r})\delta(\vec\omega''-\vec\omega)$} has been used.
\modifiedBlue{In Eq.~(\ref{eq:Calc_c1_Approx_tent}), evaluating}
the functional derivative of the effective one-particle potential
$\beta\psi[\bar\rho]$ w.r.t. the weighted density $\bar\rho$ \modifiedBlue{and}
using Eq.~(\ref{eq:Eff1Potential}) yields the
\modifiedBlue{following} final \modifiedBlue{expression}
for the modified one-particle direct correlation function
$\tilde c^{(1)}$:
\begin{equation}
 \begin{split}
  &\tilde c^{(1)}\left(\vec{r},\vec\omega,[\rho]\right)=
   -\beta\psi(\vec{r},\vec\omega,[\bar\rho])+
   \frac{\partial_{Q_0}\mathcal{J}(Q_0)}{2\mathcal{V}_d}\times\\
  &\Int{\mathcal{V}}{3}{r'}\Int{\mathcal{S}}{2}{\omega'}
   \bar\rho(\vec{r}',\vec\omega')\Theta(d/2-|z-z'|)\times\\
  &\Int{\mathcal{V}}{3}{r''}\Int{\mathcal{S}}{2}{\omega''}
   \bar\rho(\vec{r}'',\vec\omega'')f_M(|\vec{r}'-\vec{r}''|,\vec\omega',\vec\omega''),
 \end{split}
\label{eq:Calc_c1_Approx}
\end{equation}
where we used that
$\frac{\delta Q_0(\vec{r}',\,\,[\bar\rho])}{\delta\bar\rho(\vec{r},\,\,\vec\omega)}=
\Theta(d/2-|z-z'|)/\mathcal{V}_d$ holds for bulk phases.
\modifiedGreen{Due to the product rule of functional differentiation}\modifiedMagenta{, the}
\modifiedBlue{evaluation of the last term in Eq.~(\ref{eq:Calc_c1_Approx_tent})
\modifiedGreen{produces} a second term $-\frac{1}{2}\beta\psi(\vec{r},\vec\omega,[\bar\rho])$
and the latter term in Eq.~(\ref{eq:Calc_c1_Approx}).}
\modifiedBlue{As expected,} the solution of the modified Euler-Lagrange equation indeed 
differs from the exact \modifiedBlue{one}.
However, the solution \modifiedBlue{obtained from} the modified one-particle direct
correlation function $\tilde c^{(1)}\left(\vec{r},\vec\omega,[\rho]\right)$
\modifiedBlue{exhibits} the same \modifiedBlue{functional} form \modifiedBlue{as} the exact solution
\modifiedBlue{in} Eq.~(\ref{eq:DensityGenericForm}),
but with modified coefficients $A_i$ \modifiedBlue{and} $B_i$
(\modifiedBlue{see} the last paragraph \modifiedBlue{in Appendix~\ref{sec:appendix:derivation}}).

On the other hand, \modifiedBlue{one} could have followed,
\modifiedGreen{as mentioned above,}
a second route,
\modifiedBlue{which utilizes the} knowledge of
the functional form of the (exact) equilibrium density profile \modifiedBlue{in}
Eq.~(\ref{eq:DensityGenericForm}).
By plugging this generic form into the grand potential functional and 
\modifiedBlue{by} minimizing it w.r.t. the coefficients $A_i$ and $B_i$,
\modifiedGreen{$i=0,1,2$,}
\begin{equation}
  \frac{\partial\beta\Omega[\rho^{(0)}]}{\partial X_i}\bigg|_{X_j}=0,~X_i=A_i,B_i,~i\neq j,
\end{equation}
one obtains six equations, which determine the equilibrium values for the coefficients
$A_i$ \modifiedBlue{and} $B_i$ and therefore yield the exact equilibrium density profile 
for the considered excess free energy functional.
However, this generic form holds only for the bulk profiles, 
\modifiedGreen{because} the periodic structure is essential 
\modifiedBlue{for the validity of} this expression.
Therefore, this scheme cannot be extended to 
study interfacial problems, e.g., free interfaces,
\modifiedBlue{by} using coexisting bulk phases as boundary conditions.
\modifiedBlue{This is} unlike the first approach,
which is \modifiedBlue{applicable even for} non-periodic density profiles.

\modifiedBlue{However}, by comparing the two different approaches,
one can analyze, how the modification
leading to Eq.~(\ref{eq:Calc_c1_Approx})
quantitatively affects the exact bulk solution.
It turns out, that for all examined cases
the \modifiedGreen{form of the} bulk profiles,
obtained by the solution of the modified Euler-Lagrange equation,
can be assigned to \modifiedGreen{that of the corresponding} equivalent exact solution
\modifiedBlue{and the quantitative differences of both approaches are only minor}
(see \modifiedBlue{Appendix~\ref{sec:appendix:comparison}}).
\modifiedBlue{Although} for nematic and smectic phases the coefficients differ quantitatively,
the phase behaviors
\modifiedBlue{predicted by the two solutions do not differ qualitatively.}
It is worth mentioning, that for isotropic fluids both solutions are identical,
\modifiedBlue{because for isotropic phases $\bar\rho(\vec{r},\vec\omega)=\rho(\vec{r},\vec\omega)$.}

\subsubsection{\label{sec:theory:DFT:PhaseBehavior} Phase behavior}

In order to study the phase behavior of ionic liquid crystals
within the present DFT approach, we turn to the first
minimization scheme \modifiedBlue{discussed in} Sec.~\ref{sec:theory:DFT:Formalism},
which \modifiedBlue{is based on} a modified expression 
\modifiedBlue{(Eq.~(\ref{eq:Calc_c1_Approx})) for} the
one-particle direct correlation function $\tilde c^{(1)}\left(\vec{r},\vec\omega,[\rho]\right)$,
\modifiedBlue{in order} to evaluate the Euler-Lagrange equation \modifiedBlue{in} Eq.~(\ref{eq:ELG}).
For given \modifiedBlue{values of the} chemical potential $\mu$ and temperature $T$
the (bulk) solutions are described by a set of 
coefficients $Q_i$ (Eqs.~(\ref{eq:ExpansionCoeffs}) and (\ref{eq:ExpansionCoeffs2}))
which is obtained by numerically solving
Eqs.~(\ref{eq:ELG}) and (\ref{eq:Calc_c1_Approx}), \modifiedBlue{thereby}
using the definition of the projected density $\bar\rho(\vec{r},\vec\omega)$
\modifiedBlue{in} Eq.~(\ref{eq:WeightedDensity}).
The numerical evaluation is \modifiedBlue{carried out by}
employing a Picard algorithm with retardation.
Subsequently, the (approximate) equilibrium density profile $\rho^\text{eq}(\vec{r},\vec\omega)$
is obtained by evaluating Eq.~(\ref{eq:ELG}),
using the set of coefficients $Q_i$ of the solution.
\modifiedBlue{We note} that
\modifiedBlue{$\rho^\text{eq}(\vec{r},\vec\omega)$ exhibits the same functional form}
as the exact bulk solution \modifiedBlue{in} Eq.~(\ref{eq:DensityGenericForm})
\modifiedBlue{and that the exact and the approximate solution
of the Euler-Lagrange-equation yield only minor quantitative differences
(see Appendix~\ref{sec:appendix:comparison} and Table~\ref{tab:op_comparison}).}

In order to distinguish different types of bulk phases,
we define the following four order parameters:
\modifiedBlue{
\begin{align}
   n_0 &=\frac{1}{\mathcal{V}_d}\Int{\mathcal{V}_d}{3}{r'}
         n(\vec{r}'),
         \nonumber\\
   W_0 &=\frac{2}{\mathcal{V}_d}\Int{\mathcal{V}_d}{3}{r'}
         n(\vec{r}')\cos(2\pi z'/d),
         \nonumber\\
 S_{20}&=\frac{1}{\mathcal{V}_d}\Int{\mathcal{V}_d}{3}{r'}
         S_2(\vec{r}'),
         \nonumber\\
   W_2 &=\frac{2}{\mathcal{V}_d}\Int{\mathcal{V}_d}{3}{r'}
         S_2(\vec{r}')\cos(2\pi z'/d).
 \label{eq:OrderParameters}
\end{align}
}
\modifiedBlue{The mean density $n_0$} in a volume of size $\mathcal{V}_d$ and $W_0$ 
are the first two coefficients of a Fourier \modifiedGreen{series} expansion of the number density
$n(\vec{r}):=\Int{\mathcal{S}}{2}{\omega}\rho(\vec{r},\vec\omega)$,
while the \modifiedBlue{mean orientational order parameter $S_{20}$} and $W_2$
are the first two coefficients of a Fourier \modifiedGreen{series} expansion of
the (\modifiedGreen{spatially varying}) orientational order parameter
$S_2(\vec{r}):=\Int{\mathcal{S}}{2}{\omega}P_2(\cos(\theta))f(\vec{r},\vec\omega)$,
where $f(\vec{r},\vec\omega):=\rho(\vec{r},\vec\omega)/n(\vec{r})$
is the orientational distribution function.
\MODIFIED{For $S_2(\vec{r})=1$ the particles at position $\vec{r}$
are perfectly aligned with the director $\vec{\hat n}$,
while for $S_2(\vec{r})=-0.5$ they are perfectly perpendicular to the director
(recall $\vec{\hat n}\cdot\vec\omega=\cos\theta$).
In the case of \MODIfied{$|S_2(\vec{r})|\ll1$}
particles at $\vec{r}$ do not show orientational order.}
%
\modifiedBlue{In the case of the three relevant bulk phases,
$n(\vec{r})$ and $S_2(\vec{r})$ are periodic functions in $z$-direction
and can be expanded in \modifiedGreen{terms of the} Fourier series
\begin{equation}
 n(z)=a_0+\sum_{k=1}^{\infty}a_{2k}\cos(2\pi k z/d)
 \label{eq:Fourier_totnumberdens}
\end{equation}
and
\begin{equation}
 S_2(z)=b_0+\sum_{k=1}^{\infty}b_{2k}\cos(2\pi k z/d)
 \label{eq:Fourier_S2}
\end{equation}
where the first two non-\modifiedGreen{zero} expansion coefficients, $a_0$ and $a_2$ 
\modifiedGreen{and} $b_0$ and $b_2$\modifiedGreen{, follow from}
$n_0$ and $W_0$, \modifiedGreen{and from} $S_{20}$ and $W_2$,
\modifiedGreen{respectively} (see Eq.~(\ref{eq:OrderParameters})).
We note that antisymmetric terms proportional to $\sin(2\pi k z/d)$, $k\in\mathbb{N}$,
vanish, because $n(z)$ and $S_2(z)$ are even functions.
}

The four order parameters in Eq.~(\ref{eq:OrderParameters})
allow \modifiedBlue{one} to distinguish between the
\modifiedBlue{following distinct} bulk phases:
\begin{itemize}
 \item isotropic fluid: $n_0\neq0,S_{20}=W_0=W_2=0$,
 \item nematic fluid: $n_0\neq0,S_{20}\neq0,W_0=W_2=0$,
 \item smectic-A fluid: $n_0\neq0,S_{20}\neq0,W_0\neq0,W_2\neq0$.
\end{itemize}

\modifiedBlue{State points within} a stable bulk phase
\modifiedBlue{maximize $-\Omega[\rho]$ so that for the pressure $p$ one has
$p=-\frac{1}{\mathcal{V}}\Omega[\rho^\text{eq}]\geq-\frac{1}{\mathcal{V}}\Omega[\rho]$.}
At \modifiedGreen{phase coexistence distinct sets}
of order parameters give rise to the same value of the
\modifiedMagenta{reduced} pressure:
\begin{align}
 & p^*(T,\mu,d)
 :=-\frac{\beta\Omega[\rho^\text{eq}]}{\mathcal{V}}\nonumber\\
 &= n_0+
    \frac{1}{4\mathcal{V}_d}\Int{\mathcal{V}_d}{3}{r}n^\text{eq}(\vec{r})
    \left[\zeta_0(\vec{r})+S_2^\text{eq}(\vec{r})\zeta_2(\vec{r})\right]\nonumber\\
 & -n_0\frac{\partial_{Q_0}\mathcal{J}(Q_0)}{2\mathcal{V}_d}
    \Int{\mathcal{V}_d}{3}{r'}\Int{\mathcal{S}}{2}{\omega'}
    \bar\rho(\vec{r}',\vec\omega')\times\nonumber\\
 &  \Int{\mathcal{V}}{3}{r''}\Int{\mathcal{S}}{2}{\omega''}
    \bar\rho(\vec{r}'',\vec\omega'')f_M(|\vec{r}'-\vec{r}''|,\vec\omega',\vec\omega''),
 \label{eq:pressure}
\end{align}
where $\zeta_l$, $l=0,2$, are the coefficients \modifiedGreen{in} the
\modifiedBlue{expansion of the} effective one-particle potential $\beta\psi$
\modifiedBlue{(Eq.~(\ref{eq:Eff1Potential})) in terms of Legendre polynomials.
The derivation of Eq.~(\ref{eq:pressure}) is provided in Appendix~\ref{sec:appendix:pressure}.}
The equilibrium value of $d$ maximizes
$p^*(T,\mu,d)$ for fixed temperature and chemical potential,
provided its value is larger than for any isotropic or nematic
phase for the same \modifiedBlue{state} $(T,\mu)$:
\begin{equation}
  \left.\frac{\partial p^*(T,\mu,d)}{\partial d}\right|_{T,~\mu}=0.
 \label{eq:EquiLayerSpacing}
\end{equation}
Under these conditions a smectic phase with layer spacing $d$
is the stable phase.

\modifiedRed{
\subsubsection{\label{sec:theory:DFT:Crystallization} Crystallization}
\modifiedGreen{As already mentioned in Sec.}~\ref{sec:theory:DFT:Formalism},
the formalism, presented so far, captures isotropic, nematic, and smectic-A phases.
However, for sufficiently \modifiedGreen{low temperatures and sufficiently high densities}
one expects crystallization to \modifiedGreen{occur}.
As will be discussed in Sec.~\ref{sec:discussion},
the DFT formalism presented in Sec.~\ref{sec:theory:DFT:Formalism}
predicts \modifiedGreen{distinct} variants of smectic-A phases to be stable at large packing fractions
(compare the phase diagrams in \modifiedOrange{
Figs.~\ref{fig:phasediagrams1}, \ref{fig:phasediagrams2}, and \ref{fig:phasediagrams3}}).
In order to assess the stability of those smectic-A-type phases with respect to crystallization,
we follow an approach similar to that used in investigations of melting and freezing 
in colloidal suspensions (see, e.g., Ref.~\cite{Loewen1994} for a review).
\modifiedGreen{To this end we consider} an expansion of the grand potential functional
$\beta\Omega[\rho]$ in terms of number density profiles $\rho$ around
the value $\rho_N$ of a uniform nematic phase.
Hence, the reference fluid is homogenous but shows orientational order.
For \modifiedGreen{simplicity},
we \modifiedGreen{take} all particles 
\modifiedGreen{to be} perfectly aligned with the director $\vec{\hat n}$,
which, without loss of generality, points into the $z$-direction.
Thus, the value of the grand potential around the homogenous
reference density $\rho_N$ of the nematic fluid is given by the following expansion:
\begin{equation}
 \begin{split}
    \beta\Omega[\rho]&=
    \beta\Omega[\rho_N]
   +\Int{\mathcal{V}}{3}{r}\rho(\vec{r})\ln\left(\frac{\rho(\vec{r})}{\rho_N}\right)\\
  &-\frac{1}{2}\Int{\mathcal{V}}{3}{r}\Int{\mathcal{V}}{3}{r'}
    c^{(2)}\left(\vec{r}-\vec{r'}\right)\Delta\rho(\vec{r})\Delta\rho(\vec{r'})\\
  &+\mathcal{O}(\Delta\rho^3),
 \label{eq:CrystOmega}
 \end{split}
\end{equation}
where $c^{(2)}\left(\vec{r}-\vec{r'}\right)$ is the (two-particle) direct correlation function
and $\Delta\rho(\vec{r}):=\rho(\vec{r})-\rho_N$ gives the deviation of the density at position $\vec{r}$
from the homogeneous density $\rho_N$.
\modifiedGreen{We} note,
that considering a perfectly aligned system allows us to disregard the orientational 
degrees of freedom in Eq.~(\ref{eq:CrystOmega}).
In order to proceed we perform the following substitution:
\begin{equation}
 \begin{split}
  &-\frac{1}{2}\Int{\mathcal{V}}{3}{r}\Int{\mathcal{V}}{3}{r'}
   c^{(2)}\left(\vec{r}-\vec{r'}\right)\Delta\rho(\vec{r})\Delta\rho(\vec{r'})
  +\mathcal{O}(\Delta\rho^3)\\
=:&
  -\frac{1}{2}\Int{\mathcal{V}}{3}{r}\Int{\mathcal{V}}{3}{r'}
   \bar c^{(2)}\left(\vec{r}-\vec{r'}\right)\Delta\rho(\vec{r})\Delta\rho(\vec{r'}),
 \label{eq:effective_DirCorFunc}
 \end{split}
\end{equation}
where the second order term, involving the direct correlation
function $c^{(2)}$, and the higher order terms of Eq.~(\ref{eq:CrystOmega}) 
are replaced by an effective description of the direct correlation function
$\bar c^{(2)}$.
The \modifiedGreen{motivation for} using
an effective direct correlation function $\bar c^{(2)}$ (Eq.~(\ref{eq:effective_DirCorFunc}))
is to avoid evaluating terms $\propto\mathcal{O}(\Delta\rho^3)$ in Eq.~(\ref{eq:CrystOmega}).
However, simply truncating the series at second order and using the direct correlation function
$c^{(2)}\left(\vec{r}-\vec{r'}\right):=
-\frac{\delta^2\mathcal{F}[\rho]}{\delta\rho(\vec{r})\delta\rho(\vec{r}')}$
from Eqs.~(\ref{eq:F_WDA}), (\ref{eq:Eff1Pot_PL}), (\ref{eq:Eff1Pot_ERPA}),
and (\ref{eq:Eff1Potential}) leads to unphysical results
(in particular one observes stable columnar phases,
which in the present case of calamitic mesogenes~\modifiedOrange{\cite{DeGennes1974}}
\modifiedGreen{appear to be} an artifact),
due to the absence of the higher order terms.
It turns out that using a second order approach in the spirit of Onsager~\cite{Onsager1949}
\modifiedGreen{in order} to incorporate the hard-core interactions cures this \modifiedGreen{defect}.
We \modifiedGreen{emphasize}, that this approach is rather simplistic and
not intended to yield quantitatively precise results.
However, it allows \modifiedGreen{one} to estimate the onset of crystallization consistently
with our DFT approach \modifiedGreen{described in} Sec.~\ref{sec:theory:DFT:Formalism},
\modifiedGreen{because} the Parsons-Lee approach \modifiedGreen{used} (Eq.~(\ref{eq:Eff1Pot_PL}))
can be understood as a modification of the Onsager functional.
Thus we choose the following form of the direct correlation function,
in order to keep the effective description consistent with the formalism 
of Sec.~\ref{sec:theory:DFT:Formalism}:
\begin{equation}
 \begin{split}
  &\bar c^{(2)}\left(\vec{r}-\vec{r'}\right)=
   -f_M(\vec{r}-\vec{r'},\vec{\hat z},\vec{\hat z})\,+\\
  &(1+f_M(\vec{r}-\vec{r'},\vec{\hat z},\vec{\hat z}))
   \beta U(\vec{r}-\vec{r'},\vec{\hat z},\vec{\hat z}).
 \label{eq:effective_DirCorFunc_2}
 \end{split}
\end{equation}
The crystalline density profile \modifiedGreen{will} be described by a
\modifiedGreen{superposition} of Gaussians~\cite{Loewen1994},
which are \modifiedGreen{centered} at the sites $\vec{R}=\vec{R}_{||}+\vec{R}_\perp$
of a \modifiedGreen{three-dimensional} hexagonal lattice $\mathcal{R}$:
\begin{equation}
 \begin{split}
   \rho(\vec{r})=&\frac{\alpha_\perp}{\pi}\sqrt{\frac{\alpha_{||}}{\pi}}
   \sum_{\vec{R}\in\mathcal{R}}
   \exp\left(-\alpha_\perp(\vec{r}_\perp-\vec{R}_\perp)^2\right)\times\\
  &\exp\left(-\alpha_{||}(\vec{r}_{||}-\vec{R}_{||})^2\right),
 \label{eq:HexGaussians}
 \end{split}
\end{equation}
where $\vec{r}_{||}$ and $\vec{R}_{||}$ are the projection
of the position $\vec{r}$ and of the lattice site \modifiedGreen{vector} $\vec{R}$,
respectively, \modifiedGreen{onto} the $z$-direction, while $\vec{r}_\perp$ and $\vec{R}_\perp$
are the respective projections \modifiedGreen{onto} the $x$-$y$-plane.
The Gaussians are described by two parameters:
$1/(2\alpha_{||})$ is the mean-square displacement in
$z$-direction, while $1/\alpha_\perp$ is the mean-square displacement
in lateral direction
(\modifiedGreen{perpendicular to the $z$-direction} and parallel to the $x$-$y$-plane).
\modifiedGreen{We note,
that the definitions of the mean-square displacements $1/(2\alpha_{||})$ and $1/\alpha_\perp$
differ by a factor of $1/2$, due to the different \modifiedMagenta{dimensionality} of the respective
Gaussian contributions, which is one-dimensional for $1/(2\alpha_{||})$ and
two-dimensional for $1/\alpha_\perp$.}
The hexagonal lattice $\mathcal{R}$ is defined by its
primitive vectors $\vec{a}_1=a(\sqrt{3}\vec{\hat x}+\vec{\hat y})/2$,
$\vec{a}_2=a(\vec{\hat y}-\sqrt{3}\vec{\hat x})/2$,
and $\vec{a}_3=L\vec{\hat z}$.
The lateral nearest neighbor spacing $a$ is related to the volume $V_c$ of
the elementary cell via $V_c=\sqrt{3}a^2L/2$.
Note, that we choose the height of the elementary cell
to be equal to the particle length $L$,
which leads to $d=L$ in case of a \modifiedGreen{smectic-A} phase.
Our choice of the density profile allows \modifiedGreen{us} to represent
the following four types of bulk phases:
\begin{itemize}
 \item nematic fluid: $\alpha_{||}=\alpha_\perp=0$,
 \item smectic-A fluid: $\alpha_{||}>0,\alpha_\perp=0$,
 \item hexagonal columnar phase: $\alpha_{||}=0,\alpha_\perp>0$,
 \item hexagonal crystal: $\alpha_{||}>0,\alpha_\perp>0$.
\end{itemize}
The motivation for choosing a three-dimensional hexagonal lattice structure
is, on one hand, that the smectic-A phase as well as a crystalline structure
can be recaptured by tuning the parameters $\alpha_{||}$ and $\alpha_\perp$ accordingly (see above).
On the other hand, \modifiedMagenta{because} the particles are taken to be perfectly aligned with the $z$-direction,
their cross-sections parallel to the $x$-$y$-plane are \modifiedMagenta{circles}.
Therefore a hexagonal \modifiedMagenta{structure} perpendicular to the $x$-$y$-plane
\modifiedMagenta{appears to be a plausible candidate}.
In order to calculate $\beta\Omega[\rho]$ in Eq.~(\ref{eq:CrystOmega}),
we have to evaluate Eq.~(\ref{eq:effective_DirCorFunc}),
which can be written as 
\begin{equation}
 \begin{split}
  -\frac{1}{2}\Int{\mathcal{V}}{3}{r}\Int{\mathcal{V}}{3}{r'}
   \bar c^{(2)}\left(\vec{r}-\vec{r'}\right)
   \Delta\rho(\vec{r})\Delta\rho(\vec{r'})=\\
  -\frac{1}{2}\rho_N^2\mathcal{V}
   \sum_{\vec{G}\in\mathcal{G}\setminus\{0\}}
   \hat{\bar c}^{(2)}(\vec{G})
   \exp\left(-\frac{\vec{G}_\perp^2}{2\alpha_\perp}
             -\frac{\vec{G}_{||} ^2}{2\alpha_{||}}\right),
 \label{eq:effective_DirCorFunc_3}
 \end{split}
\end{equation}
where $\vec{G}=\vec{G}_{||}+\vec{G}_\perp$ denotes a site of the reciprocal lattice $\mathcal{G}$
of $\mathcal{R}$ and $\hat{\bar c}^{(2)}(\vec{G})$ is the Fourier transform 
of the direct correlation function (Eq.~(\ref{eq:effective_DirCorFunc_2})).
In Eq.~(\ref{eq:effective_DirCorFunc_3}) we used
the Fourier representation of $\Delta\rho(\vec{r})$:
\begin{equation}
 \Delta\rho(\vec{r})=\rho_N\sum_{\vec{G}\in\mathcal{G}\setminus\{0\}}
 \exp\left(i\,\vec{G}\cdot\vec{r}
           -\frac{\vec{G}_\perp^2}{4\alpha_\perp}
           -\frac{\vec{G}_{||} ^2}{4\alpha_{||}}
     \right).
\end{equation}
Note, that the mean density of the inhomogeneous fluid
described by Eq.~(\ref{eq:HexGaussians}) is equal 
to the density $\rho_N$ of the homogenous (nematic) reference fluid.
In order to assess the stability of the four aforementioned types of phases
for a given \modifiedGreen{reduced temperature $T^*=kT/\epsilon_0$,
where $\epsilon_0$ is the interaction strength of the 
Gay-Berne potential $U_\text{GB}$ (see Eq.~(\ref{eq:Pairpot_GB_epsilon})),}
and density $\rho_N$, i.e., for a given point in the phase diagrams
\modifiedMagenta{shown in Figs.~\ref{fig:phasediagrams1}, \ref{fig:phasediagrams2},
and \ref{fig:phasediagrams3},} the difference of the
grand potential density $\beta\Omega[\rho]/\mathcal{V}$
(Eq.~(\ref{eq:CrystOmega}) with Eqs.~(\ref{eq:effective_DirCorFunc})
and (\ref{eq:effective_DirCorFunc_3}))
from the value $\beta\Omega[\rho_N]/\mathcal{V}$
of the homogeneous nematic reference fluid 
is evaluated for $\alpha_\perp\geq0$ and $\alpha_{||}\geq0$:
\begin{equation}
 \begin{split}
  & \frac{\Delta\beta\Omega}{\mathcal{V}}:=\frac{\beta\Omega[\rho]-\beta\Omega[\rho_N]}{\mathcal{V}}=
    \frac{1}{\mathcal{V}}
    \Int{\mathcal{V}}{3}{r}\rho(\vec{r})\ln\left(\frac{\rho(\vec{r})}{\rho_N}\right)\\
  &-\frac{1}{2}\rho_N^2
   \sum_{\vec{G}\in\mathcal{G}\setminus\{0\}}
   \hat{\bar c}^{(2)}(\vec{G})
   \exp\left(-\frac{\vec{G}_\perp^2}{2\alpha_\perp}
             -\frac{\vec{G}_{||} ^2}{2\alpha_{||}}\right).
 \label{eq:OmegaDiff_Crystallization}
 \end{split}
\end{equation}
}

In order to illustrate, how the onset of crystallization is determined,
we consider the following set of pair potential parameters:
$L/R=4,\epsilon_R/\epsilon_L=2,D/R=0.9,\lambda_D/R=5$, and $\gamma/(R\epsilon_0)=0.045$.
\modifiedGreen{With this we} evaluate numerically Eq.~(\ref{eq:OmegaDiff_Crystallization})
for a set of four thermodynamic state points
with packing fraction $\eta_N=0.42$ and \modifiedGreen{reduced}
temperatures $T^*\in\{0.8,0.82,0.85,0.87\}$.
The values of $\Delta\beta\Omega/\mathcal{V}$ for $\alpha_\perp R^2\in[0,120]$
and $\alpha_{||}R^2\in[0,12]$ are shown in Fig.~\ref{fig:OmegaDiff_Crystallization}.
For $T^*=0.87$ and $0.85$ the smectic-A phase is stable \modifiedGreen{with respect to} crystallization, while
for $T^*=0.8$ it becomes \modifiedGreen{unstable with respect to a hexagonal crystalline phase}.
$T^*=0.82$ is close to coexistence of the smectic-A phase and the hexagonal crystal,
\modifiedGreen{because in this case} the grand potential $\beta\Omega[\rho]$
exhibits two almost equally deep local minima corresponding to these two phases.
Repeating \modifiedGreen{this} procedure for \modifiedGreen{various}
packing fractions $\eta_N$ allows \modifiedGreen{one} to detect the
\modifiedGreen{phase} transition from a stable \modifiedGreen{smectic-A} phase
\modifiedGreen{to} a stable crystal.

\modifiedOrange{Alternatively,} the location of the melting 
of the hexagonal lattice structure in lateral direction
can be \modifiedOrange{estimated by invoking}
a Lindemann criterion~\cite{Lindemann1910,Gilvarry1956,Zheng1998}.
It states that \modifiedOrange{if} the scaled root mean square displacement
\modifiedOrange{$1/(a\sqrt{\alpha_\perp})$} of the (lateral) hexagonal lattice
with lattice spacing $\frac{a}{R}=\sqrt{\frac{\pi}{3\sqrt{3}\eta_0}}$
and packing fraction $\eta_0$ \modifiedOrange{exceeds} a certain threshold value
$\delta$ (the \modifiedOrange{so-called} critical Lindemann parameter)
the lattice vibrations are sufficiently strong
to destroy the (lateral) lattice structure.
Evaluating \modifiedOrange{$1/(a\sqrt{\alpha_\perp})$} from
the minimum of $\Delta\beta\Omega/\mathcal{V}$ (Eq.~(\ref{eq:OmegaDiff_Crystallization}))
corresponding to a three-dimensional hexagonal lattice structure
(Fig.~\ref{fig:OmegaDiff_Crystallization})
along the (pink) melting curves in Figs.~\ref{fig:phasediagrams2} and \ref{fig:phasediagrams3}
yields for $\eta_0\lesssim0.4$ a lateral root mean square displacement 
\modifiedOrange{$1/(a\sqrt{\alpha_\perp})\gtrsim0.1$} and for $\eta_0\gtrsim0.4$
a lateral root mean square displacement \modifiedOrange{$1/(a\sqrt{\alpha_\perp})\lesssim0.1$}.
Thus, for the widely used, \modifiedOrange{common}
critical Lindemann parameter $\delta\approx0.1$
the (pink) melting curves shown in
Figs.~\ref{fig:phasediagrams2} and \ref{fig:phasediagrams3}
lie below (above) \modifiedOrange{those} respective melting curves,
\modifiedOrange{which have been} obtained by applying
the Lindemann criterion, for packing fractions $\eta_0$ larger (smaller) than $0.4$.
Hence the Lindemann criterion $\delta\approx0.1$ leads to the (pink) melting curves
in Figs.~\ref{fig:phasediagrams2} and \ref{fig:phasediagrams3} only for
$\eta_0\approx0.4$; \modifiedOrange{otherwise}
the critical Lindemann parameter has to be considered as
(monotonically decreasing) function of the packing fraction:
$\delta(\eta_0)\in[0.06,0.2]$ for $\eta_0\in[0.3,0.49]$.
This result leads us to the conclusion that the Lindemann criterion,
assuming a constant critical Lindemann parameter $\delta=\text{const.}$,
is not applicable here.

\subsection{\label{sec:theory:GCMC} Grand canonical Monte Carlo simulation}

\modifiedBlue{We have carried out}
grand canonical Monte Carlo \modifiedBlue{(MC)} simulations,
\modifiedGreen{based on} the molecular model introduced in Sec.~\ref{sec:theory:model}.
The simulations are performed in a cubic simulation box
of side length $V^{1/3}/R\in\{12.75,13.2,15.0\}$,
employing periodic boundary conditions.
Standard Metropolis importance sampling of the grand canonical 
Boltzmann distribution $P(\chi)\propto\exp(\beta\mu N[\chi]-\beta H[\chi])$
with the chemical potential $\mu$, the total number of ILC molecules $N[\chi]$,
and the Hamiltonian
\begin{equation}
  \beta H[\chi]=
  \sum_{\substack{i,j\\j>i}}^N
  \beta U(\vec{r}_{ij},\vec\omega_i,\vec\omega_j)
  \Theta(R_\text{cut}-|\vec{r}_{ij}|)
 \label{Hamiltonian}
\end{equation}
\modifiedBlue{\modifiedGreen{which governs} the set $\{\chi\}$ of all configurations.}
The pair interaction $U(\vec{r}_{ij},\vec\omega_i,\vec\omega_j)$
(Eq.~(\ref{eq:Pairpot})) is truncated at \modifiedBlue{the}
cut-off distance $|\vec{r}_{ij}|=R_\text{cut}<V^{1/3}/2$.
Each simulation run consists of \modifiedBlue{$N^\text{mon}=0.2-1\times10^6$}
Monte Carlo moves,
\modifiedBlue{from which we monitor} the observables of interest (see below).
\modifiedBlue{In addition,} between two consecutive monitoring moves
ca.\ \modifiedBlue{$N^\text{relax}=500$} relaxation moves are included
in order to reduce correlations between successive \modifiedBlue{(monitored)} configurations
along the \modifiedBlue{MC} trajectory.
\modifiedBlue{Thus, a simulation consists of
$N^\text{tot}=N^\text{mon}\times N^\text{relax}=1-5\times10^8$ simulation moves in total.}
Each Monte Carlo move can be either a translation and rotation of one particle
(randomly chosen with probability $P_\text{t\&r}$),
an insertion of one particle of orientation $\vec\omega$ at position $\vec{r}$
(chosen with probability $(1-P_\text{t\&r})/2$),
or a removal of one particle (chosen with probability $(1-P_\text{t\&r})/2$).
In \modifiedBlue{the} case of \emph{t}ranslation and \emph{r}otation,
the trial orientation is chosen randomly within 
\modifiedBlue{the interval $0\leq\theta\leq\theta_\text{max}$}
around the orientation \modifiedBlue{of the particle under consideration}.
The trial translational displacement is done within a cube-like volume $v$
around the position of the particle \modifiedBlue{under consideration}.
\MODIFIED{In order to optimize the acceptance rate of the trial configurations along the MC trajectory
the displacement volume $v$, the maximum polar angle $\theta_\text{max}$,
and the probability $P_\text{t\&r}$ have been adapted accordingly.
We note, that the initial configuration for each simulation is isotropic,
which allows the system to \MODIfied{freely} form any kind of structure.}

The spatial arrangement of \modifiedBlue{the} particles can be investigated via the
\modifiedBlue{local number} density
\begin{equation}
  \rho^\text{loc}(\vec{r}):=l^{-3}\left\langle N^\text{loc}(\vec{r},[\chi]) \right\rangle,
 \label{eq:LocalDens}
\end{equation}
where $N^\text{loc}(\vec{r},[\chi])$ is the number of particles
for a given configuration $\chi$ in a cube-like partial volume $l^3$
of the simulation box located at position $\vec{r}$;
$\langle\cdots\rangle$ denotes the thermal average.
\modifiedGreen{Upon} monitoring the local density on a simple cubic lattice
of sample points within the \modifiedBlue{simulation box of} volume $V$
the structure of the fluid is inferred. 

The degree of orientational order can be characterized
by considering the local orientational order parameter 
\begin{equation}
  S_2^{\text{loc}}(\vec{r}):=
  \frac{3}{2}
  \bigg\langle
  \frac{\sum_{i=1}^{N^\text{loc}}(\vec\omega_i\cdot\vec{\hat n}[\chi])^2}
  {N^\text{loc}(\vec{r},[\chi])}
  \bigg\rangle-\frac{1}{2},
 \label{eq:LocalS2}
\end{equation}
where, \modifiedBlue{for a given configuration $\chi$,
$\vec\omega_i\cdot\vec{\hat n}$} is the projection of \modifiedBlue{the}
long axis $\vec\omega_i$ of the $i$-th particle
\modifiedBlue{onto} the global director $\vec{\hat n}[\chi]$.
Here, ``global'' means that all particles within the simulation box $V$ are considered,
while ``local'' means that only particles in the relevant partial volume $l^3$ are considered.
The director \modifiedGreen{$\vec{\hat n}[\chi]$ corresponding to configuration $\chi$}
is obtained by calculating the eigenvector corresponding to the largest eigenvalue
of the orientational ordering matrix
(\modifiedBlue{i.e., the tensor} order parameter)~\cite{DeGennes1974}
\modifiedGreen{for the \modifiedMagenta{considered} configuration $\chi$}:
\begin{equation}
  Q_{ab}[\chi]:=\frac{3}{2N[\chi]}\sum_{i=1}^N(\vec\omega_i)_a(\vec\omega_i)_b-\frac{\delta_{ab}}{2},
 \label{eq:OrderingMatrix}
\end{equation}
where $(\vec\omega_i)_a$ denotes the $a$-th component of vector $\vec\omega_i$.
For $S_2^{\text{loc}}(\vec{r})\lesssim1$ particles located at $\vec{r}$
are predominantly aligned with the director $\vec{\hat n}$,
while for $S_2^{\text{loc}}(\vec{r})\gtrsim-0.5$ the particles are predominantly perpendicular
to the director.
For \modifiedBlue{$|S_2^{\text{loc}}(\vec{r})|\ll1$},
particles at $\vec{r}$ do not \modifiedGreen{exhibit} orientational ordering.

\section{\label{sec:discussion}Results and Discussion}

In this section we discuss the phase diagrams for \modifiedBlue{various} kinds
of ILCs, \modifiedBlue{characterized} by the set of parameters
describing \modifiedBlue{their} pair potential (Eq.~(\ref{eq:Pairpot})).
First, \modifiedBlue{we study} the phase behavior \modifiedBlue{by}
using the DFT framework presented in Sec.~\ref{sec:theory:DFT}.
\modifiedBlue{After} having discussed the theoretical predictions
of the present DFT approach, we confirm the \modifiedBlue{corresponding}
qualitative \modifiedBlue{features of} the phase behavior
via Monte Carlo simulations.
For convenience \modifiedBlue{we introduce}
the reduced temperature $T^*:=kT/\epsilon_0$,
\modifiedGreen{where $\epsilon_0$ is the interaction strength of the 
Gay-Berne potential $U_\text{GB}$ (see Eq.~(\ref{eq:Pairpot_GB_epsilon})),}
and the reduced chemical potential $\mu^*:=\beta\mu-\ln(4\pi\Lambda^3)$;
and $\eta_0=n_0\,LR^2\pi/6$ denotes the mean packing fraction.

\subsection{\label{sec:discussion:PhaseDiagrams}Phase diagrams}

\subsubsection{\label{sec:discussion:PhaseDiagrams:Comparison}
Comparison between \modifiedBlue{ordinary} liquid crystals and ILCs}
\begin{figure}[!t]
 \includegraphics[width=0.5\textwidth]{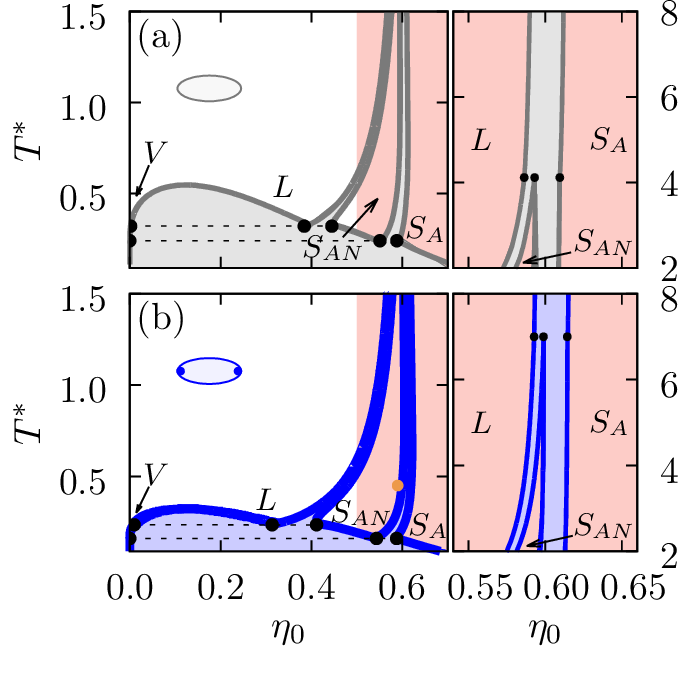}
  \caption{
  Panel (a) shows the phase diagram for ordinary liquid crystals with
  $L/R=2$ \modifiedBlue{and} $\epsilon_R/\epsilon_L=2$.
  \modifiedBlue{Panel} (b) \modifiedBlue{corresponds to} ionic liquid crystals described by 
  $L/R=2,\epsilon_R/\epsilon_L=2,D/R=0.9,\lambda_D/R=5$\modifiedBlue{, and} $\gamma/(R\epsilon_0)=0.0045$.
  \modifiedBlue{
  The black dots connected by a dashed line in the left panels indicate
  three-phase coexistence of the vapor ($V$), the liquid ($L$), and the narrow smectic $S_{AN}$ phase,
  and three-phase coexistence of the vapor,
  \modifiedGreen{the $S_{AN}$}, and the ordinary smectic $S_A$ phase, \modifiedGreen{respectively}.} 
  The black dots \modifiedBlue{in the right panels indicate the location of 
  $L$-$S_{AN}$-$S_A$ three-phase coexistence
  \modifiedGreen{(here the connection by dashed lines is omitted)} which occurs
  at the triple point temperature $T_t^*\approx4.11$ for the ordinary liquid crystals
  and at $T_t^*\approx7.0$ for the ILC fluid.}
  The orange dot ($\textcolor[RGB]{238,154,073}{\bullet}$)
  denotes the \modifiedBlue{state} point $(T^*=0.45,\mu^*=20)$ in the ILC phase diagram
  \modifiedBlue{for which, cf., Fig.~\ref{fig:structure1} provides the corresponding
  order parameter profiles.}
  \modifiedRed{
  The \modifiedGreen{salmon-colored} area represents the region $\eta_0\geq0.5$ of the phase diagram
  for which the lateral spacing in between neighboring particles
  on a hexagonal lattice becomes less than $10\%$ of the particle diameter $R$,
  i.e., $a/R\leq1.1$. \modifiedGreen{Thus, the particles are densely packed and previous simulations
  suggest the \modifiedMagenta{occurrence}
  of crystallization in this high density regime~\cite{DeMiquel2002}.}
  \modifiedGreen{According to the left panels the left bottom corner of the $S_{AN}$
  phase appears to be stable against crystallization.}
  }
  }
 \label{fig:phasediagrams1}
\end{figure}
\begin{figure}[!t]
 \includegraphics[width=0.45\textwidth]{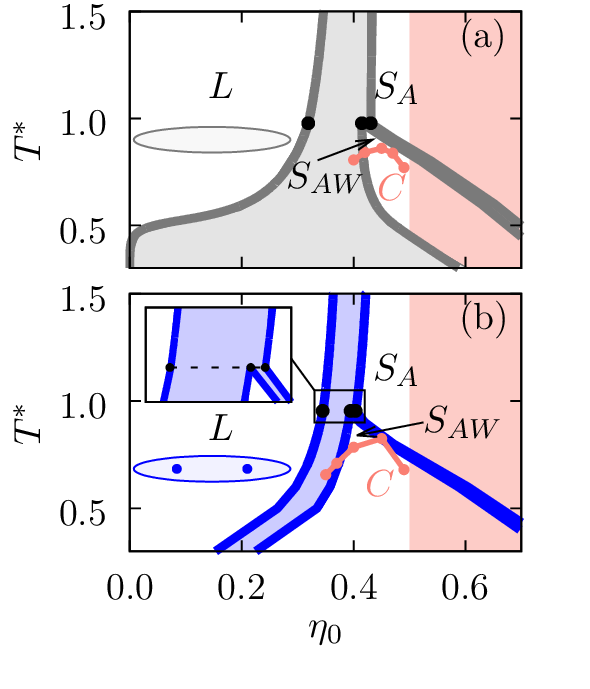}
  \caption{
  Panel (a) shows the phase diagram for ordinary liquid crystals with
  $L/R=4,\epsilon_R/\epsilon_L=2$ and panel (b) for ionic liquid crystals described by 
  $L/R=4,\epsilon_R/\epsilon_L=2,D/R=0.9,\lambda_D/R=5$\modifiedBlue{, and} $\gamma/(R\epsilon_0)=0.045$.
  The black dots indicate the location of $L$-$S_{AW}$-$S_A$ three-phase coexistence
  \modifiedBlue{
  and the inset in panel (b) gives an enlarged view of the vicinity of the triple point
  \modifiedGreen{of} the ILC fluid.}
   \modifiedRed{
   The pink curve \modifiedGreen{indicates} the
   \modifiedMagenta{\emph{onset}} of crystallization ($C$),
   \modifiedGreen{obtained by} the method described in Sec.~\ref{sec:theory:DFT:Crystallization}.
   \modifiedGreen{
   We note, that this approach does not allow
   \modifiedMagenta{one} to analyze two-phase regions
   involving the phase $C$, because it compares the grand potential minima associated
   with smectic-A and crystalline phases for given $(\eta_0,T^*)$.
   }
   Like in Fig.~\ref{fig:phasediagrams1},
   the \modifiedGreen{salmon-colored}
   area represents the region $\eta_0\geq0.5$ of the phase diagram
   for which the lateral spacing in between neighboring particles
   on a hexagonal lattice becomes less than $10\%$ of the particle diameter $R$, i.e., $a/R\leq1.1$.
   \modifiedGreen{Hence, the particles are densely packed and previous simulations
   report the occurrence of a solid phase in this density regime~\cite{DeMiquel2002}.}
   }
  }
 \label{fig:phasediagrams2}
\end{figure}
\begin{figure}[!t]
 \includegraphics[width=0.5\textwidth]{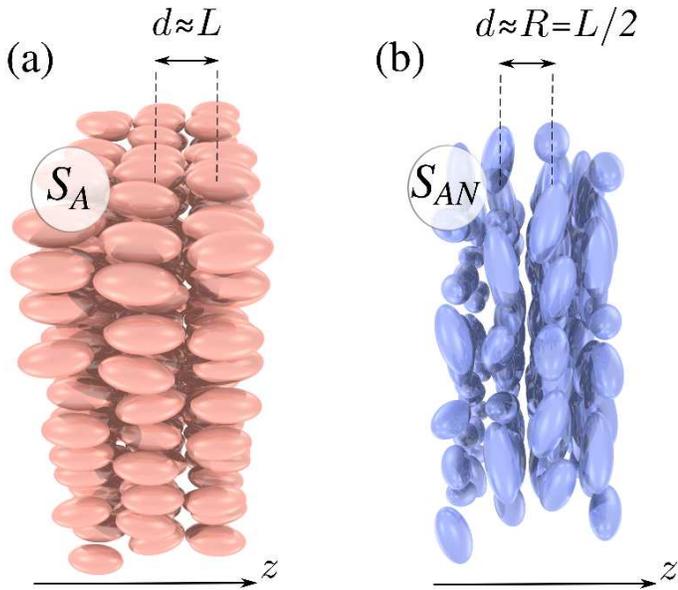}
  \caption{
  \modifiedGreen{
  Smectic configurations of particles with $L/R=2$.
  Panel (a) depicts three layers of 
  an ordinary smectic $S_A$ phase for a system composed of
  particles of length-to-breadth ratio $L/R=2$.
  The particles are mostly aligned with the layer normal (i.e., the $z$-direction),
  which leads to a layer spacing \modifiedMagenta{comparable to}
  the size of the particle length $L$,
  as expected for an ordinary $S_A$ phase.
  Panel (b) \modifiedMagenta{represents}
  a typical configuration of the $S_{AN}$ phase for $L/R=2$.
  Four smectic layers are shown in which
  the particles are oriented mainly perpendicular to the layer normal
  \MODIFIED{(i.e., the $z$-direction)}.
  However, they do not show a preferred orientation in the $x$-$y$-plane.
  This behavior leads to a layer spacing $d$ which is 
  \modifiedMagenta{comparable to} the size 
  of the particle diameter $R$.
  \MODIFIED{In order to clearly visualize the smectic layers of the $S_{AN}$ phase,
  the diameter of the blue particles in panel (b) \MODIfied{is reduced slightly}.
  This leads to a small gap in between the smectic layers of this illustration.}
  \modifiedMagenta{
  We note, that the smectic-A phases shown in panel (a) and (b)
  could not be observed in simulations and \modifiedOrange{thus}
  the depicted configurations \modifiedOrange{are not snapshots but}
  have been composed} artificially for illustration purposes.}
  }
 \label{fig:compare_SA_SAN}
\end{figure}

The phase behavior of ILCs and ordinary liquid crystals is studied
by considering their respective phase diagrams in the 
\modifiedBlue{$(T^*,\eta_0)$ plane}.
In Fig.~\ref{fig:phasediagrams1}(a) uncharged liquid crystals
of length-to-breadth ratio $L/R=2$ and
with Gay-Berne anisotropy parameter $\epsilon_R/\epsilon_L=2$ are considered.
\modifiedBlue{In} Fig.~\ref{fig:phasediagrams1}(b)
\modifiedBlue{the phase behavior of} ionic liquid crystals \modifiedBlue{is shown},
described by $L/R=2,\epsilon_R/\epsilon_L=2,D/R=0.9,\lambda_D/R=5,$
\modifiedBlue{and $\gamma/(R\epsilon_0)=0.0045$}.
In both cases, at low packing fractions and at low temperatures,
we observe \modifiedBlue{the} coexistence of a dilute and a dense isotropic phase,
which we refer to as liquid\,($L$)-vapor\,($V$) coexistence.
One finds that the critical temperature is lowered for the ILC fluid,
which is a well-known observation for ionic systems~\cite{Fisher1994};
here it is induced by the enhanced repulsion \modifiedBlue{between}
the ILC molecules.
\modifiedBlue{Although low} critical temperatures
\modifiedBlue{are} a general \modifiedBlue{feature of} Coulombic systems,
the \modifiedBlue{precise} location of the critical point is very sensitive
to the \modifiedBlue{details of the} model and \modifiedBlue{the}
method \modifiedBlue{used}~\cite{Fisher1994}.
\modifiedBlue{For both types of fluids,}
increasing the mean packing fraction $\eta_0$ \modifiedBlue{leads to}
a first-order phase transition to a smectic phase,
\modifiedBlue{in agreement with the corresponding results in Ref.~\cite{Kondrat_et_al2010}}.
Remarkably, at sufficiently low temperatures, 
before forming an ordinary smectic-A structure ($S_A$),
a smectic phase \modifiedBlue{appears} in which the particles are oriented predominantly
perpendicular to the director
\modifiedGreen{
of the smectic phase, i.e., the $z$-direction along which the periodically oscillating 
density occurs.}
Since this behavior leads to a layer spacing which is comparable to the 
\modifiedBlue{diameter} $R$ of the particles
and therefore is \emph{narrower} (\modifiedMagenta{\emph{N}})
than \modifiedBlue{in} an ordinary $S_A$ phase, \modifiedBlue{in which}
the layer spacing is comparable to the length $L$ of the particles,
we refer to this smectic structure as \modifiedBlue{the} $S_{AN}$ phase.
\modifiedGreen{(\modifiedOrange{Figure}~\ref{fig:compare_SA_SAN}
provides a comparison of the structure 
of both types of smectic phases, $S_{AN}$ and $S_A$,
for particles with length-to-breadth ratio $L/R=2$.)}
However, at \modifiedBlue{high} temperatures a first-order
phase transition \modifiedBlue{occurs} directly from the liquid ($L$) to the $S_A$ phase.
The low- and the high-temperature regimes are separated by a triple point,
indicated by the black dots in the respective plot of Fig.~\ref{fig:phasediagrams1}, 
at which the liquid ($L$), the narrow smectic ($S_{AN}$),
and the ordinary smectic phase ($S_A$) \modifiedBlue{coexist}.
For the ionic liquid crystal
the triple point temperature ($T_t^*\approx7.0$) is significantly \modifiedBlue{higher}
than for the ordinary uncharged liquid crystal ($T_t^*\approx4.11$).
\modifiedBlue{Thus for ILCs the orientationally less-ordered smectic phase $S_{AN}$
remains stable at temperatures which are higher than for the ordinary liquid crystals.}
For \modifiedBlue{large $T^*$} the $L$-$S_A$ coexistence curves coincide 
for liquid crystals and ILCs, \modifiedBlue{because in the high-temperature regime
the same hard-core repulsion is the dominant interaction.}

\modifiedRed{In the context of (uncharged) ordinary liquid crystals,
to our knowledge \modifiedGreen{the $S_{AN}$ phase}
has not been reported previously.
Since particles with length-to-breadth ratio $L/R=2$
\modifiedGreen{and Gay-Berne anisotropy parameter $\epsilon_R/\epsilon_L=2$
exhibit a rather isotropic pair potential $U(\vec{r}_{12},\vec\omega_1,\vec\omega_2)$},
it is very likely
that the occurrence of liquid-crystalline phases in
such a system is an artifact of the DFT method 
\modifiedMagenta{described in} Sec.~\ref{sec:theory:DFT:Formalism},
which is unable to capture the formation of genuine crystalline structures.
For a hexagonal lattice structure the lateral lattice spacing
$\frac{a}{R}=\sqrt{\frac{\pi}{3\sqrt{3}\eta_0}}$
(see Sec.~\ref{sec:theory:DFT:Crystallization}) takes a value of 
\modifiedGreen{$a/R\approx1.1$ for $\eta_0\approx0.5$.}
Since this means that the free space \modifiedGreen{$(a-R)/R$} in lateral direction
in between neighboring particles on the hexagonal lattice is less than 10\%
of their diameter $R$, \modifiedGreen{the particles are densely packed in the high density region
$\eta_0\geq0.5$ and previous simulations~\cite{DeMiquel2002} on systems
of pure \modifiedMagenta{(i.e., uncharged)} Gay-Berne particles of length-to-breadth ratio
$L/R=3$ report the occurrence of a solid phase
for number densities $n_0\gtrsim0.32\,R^{-3}$
(denoted as $\rho$ in \modifiedMagenta{Ref.}~\cite{DeMiquel2002}) which correspond to
$\eta_0=n_0LR^2\pi/6\gtrsim0.5$ for $L/R=3$.}
Thus, as is shown by the \modifiedGreen{salmon-colored} area in Fig.~\ref{fig:phasediagrams1}
the \modifiedGreen{thermodynamically} 
stable state points of the liquid crystalline phases $S_A$ and $S_{AN}$
\modifiedGreen{lie} almost completely inside this \modifiedGreen{(expected)} crystallization region.
We note, that the occurrence of two different types of ``smectic'' phases
(i.e., $S_A$ and $S_{AN}$)
within the DFT approach of Sec.~\ref{sec:theory:DFT:Formalism} can be a hint on the presence
of \modifiedMagenta{actually} different types of crystalline phases in such systems,
which are distinguishable either
by their lattice structure or by the degree of orientational ordering of the particles
on the lattice sites.
Within this interpretation of the phase diagrams in Fig.~\ref{fig:phasediagrams1}
the $S_A$ phase would be the \modifiedGreen{analogue}
of a crystalline phase with additional orientational ordering, while
the $S_{AN}$ phase mimics a crystalline phase with a lower degree of orientational ordering
(\modifiedGreen{i.e., a} plastic crystal)
}

Figure~\ref{fig:phasediagrams2} \modifiedBlue{provides another}
comparison \modifiedBlue{between} (a) uncharged liquid crystal molecules and
(b) ILC molecules with $D/R=0.9,\lambda_D/R=5,\gamma/(R\epsilon_0)=0.045$;
both \modifiedBlue{types of molecules share the same}
length-to-breadth ratio $L/R=4$ and \modifiedBlue{the ratio} 
$\epsilon_R/\epsilon_L=2$.
\modifiedBlue{These particles are twice as elongated as those
in Fig.~\ref{fig:phasediagrams1}.}
\modifiedBlue{In this case there is} no $L$-$V$ coexistence; however,
for the uncharged liquid crystal \modifiedGreen{(a)} \modifiedBlue{it is still metastable,}
giving rise to a shoulder-like shape of the 
left hand side of the liquid-smectic two-phase region
indicated by the gray-colored area in Fig.~\ref{fig:phasediagrams2}(a).
For the ILC fluid the liquid-smectic two-phase region
(\modifiedBlue{light-}blue-colored area in Fig.~\ref{fig:phasediagrams2}(b))
is narrower compared to \modifiedBlue{its counterpart for the ordinary}
liquid crystals.
\modifiedBlue{At low temperatures this gives} rise to
stability of smectic structures,
\modifiedRed{with respect to the isotropic liquid phase,}
already at smaller
mean packing fractions $\eta_0$.
\modifiedBlue{This is caused by} the presence of the additional electrostatic
repulsion which imposes an energetic penalty on a homogeneous liquid
already at packing fractions
\modifiedBlue{which are smaller than the corresponding ones for}
ordinary liquid crystal fluids. 
Similar to the previous case of the shorter particles,
two distinct types of smectic structures can be observed.
At sufficiently low temperatures, 
before forming an ordinary smectic-A structure ($S_A$)
\modifiedBlue{upon increasing $\eta_0$},
a smectic phase is observed \modifiedBlue{the layer spacing of which}
is considerably larger \modifiedBlue{than in} the high-temperature $S_A$ phase.
Remarkably, it shows an alternating structure \modifiedBlue{in which}
a majority of \modifiedBlue{the} particles 
within the smectic layers \modifiedBlue{is} oriented predominantly parallel to the director
and a minority of \modifiedBlue{the} particles \modifiedBlue{is}
located in between the layers with \modifiedBlue{an orientation which is}
predominantly perpendicular to the director.
Since to our knowledge such a bulk structure has not \modifiedBlue{yet}
been observed in the context of smectic phases,
we \modifiedBlue{shall} refer to this novel structure as
\modifiedBlue{the} $S_{AW}$ phase,
emphasizing the extraordinarily \emph{wide} ({\scriptsize\emph{W}}) layer spacing.
Again three-phase coexistence \modifiedBlue{occurs as}
indicated by black dots in the respective plots.
\modifiedBlue{It} marks the transition to
the high-temperature regime in which a first-order phase transition directly from
the liquid to the $S_A$ phase \modifiedBlue{takes place}.
In both cases \modifiedBlue{(Figs.~\ref{fig:phasediagrams2}(a) and (b))}
the triple point temperature is about $T_t^*\approx1.0$.
\modifiedBlue{We note} that the $S_{AW}$ phase
has not been observed for ordinary liquid crystals,
because commonly \modifiedBlue{at low temperatures}
Gay-Berne fluids exhibit crystalline phases,
\modifiedBlue{as shown by previous studies~\cite{DeMiquel2004}.}
\modifiedRed{
In order to estimate the onset of crystallization
in these systems, \modifiedGreen{we have calculated the corresponding} coexistence curves,
\modifiedGreen{shown} as pink curves in Fig.~\ref{fig:phasediagrams2},
\modifiedGreen{for} a smectic-A phase $S_A$ and a hexagonal lattice structure $C$,
\modifiedGreen{by} using the method of Sec.~\ref{sec:theory:DFT:Crystallization}.
It turns out that the onset of crystallization appears 
close to the $S_A$-$S_{AW}$ transition for both cases in Fig.~\ref{fig:phasediagrams2}.
This result suggests that at most in a small thermodynamic
\modifiedGreen{pocket} the $S_{AW}$ phase remains stable
\modifiedGreen{against crystallization}.
Considering the simplicity of the method \modifiedGreen{used}
(see Sec.~\ref{sec:theory:DFT:Crystallization}),
which does not allow \modifiedGreen{one} to precisely determine the onset of crystallization,
the stability of the $S_{AW}$ for those two cases
\modifiedGreen{(a) and (b)} seems to be an artifact of the approximations used.
Thus, one cannot expect a genuine $S_{AW}$ phase to occur for 
the two cases \modifiedGreen{considered} in Fig.~\ref{fig:phasediagrams2},
which is in agreement with previous findings.
}
Nevertheless the $S_{AW}$ phase can be stable for an ILC fluid,
\modifiedBlue{because} the presence of the charges is capable
\modifiedBlue{to alter} the bulk phase diagram significantly.
In the next section we \modifiedBlue{shall} discuss
the \modifiedBlue{influence} of the location of the charges
on the phase diagram and
we \modifiedBlue{shall} demonstrate that
one can enhance the stability of the $S_{AW}$ phase
at \modifiedBlue{higher} temperatures by 
positioning the charges at the \modifiedBlue{tips} of the particles.
Finally, Monte Carlo simulations \modifiedBlue{for}
such kind of ILC fluids will be presented.
The simulation results show that the $S_{AW}$ phase is
\modifiedBlue{indeed} observable for ionic liquid crystal fluids. 
\modifiedBlue{
Finally we note that
in order to study the onset of crystallization 
quantitatively on a more precise level,
one should consider a free energy functional 
which accounts for positional correlations more carefully 
than the present DFT approach (see Sec.~\ref{sec:theory:DFT}).
Treating the hardcore interactions of the anisotropic particles
within \emph{fundamental measure theory}~\cite{Rosenfeld1994,
Hansen-Goos2009,Hansen-Goos2010,Wittmann2015,Wittmann2016}
is an appropriate \modifiedGreen{and promising approach}.
}

\subsubsection{Dependence on the location of the charges}
\begin{figure}[!t]
 \includegraphics[width=0.45\textwidth]{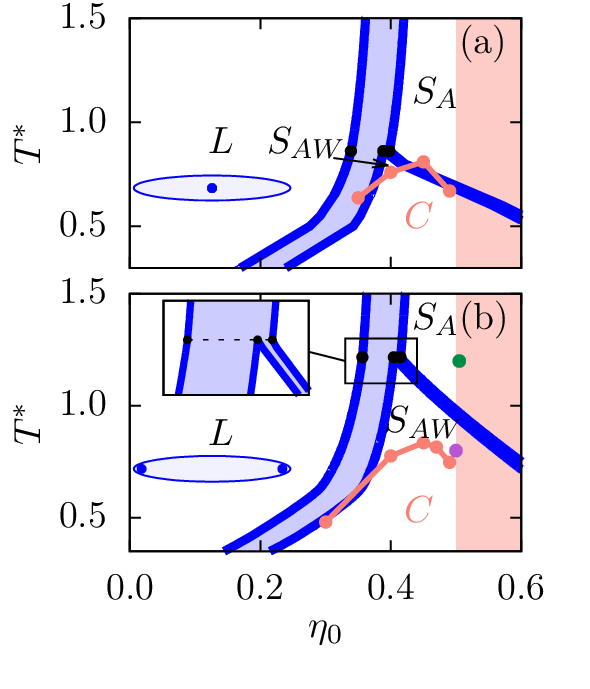}
  \caption{
  Phase diagrams for ILCs with
  (a) $L/R=4,\epsilon_R/\epsilon_L=2,D/R=0,\lambda_D/R=5$\modifiedBlue{, and} $\gamma/(R\epsilon_0)=0.045$
  and
  (b) $L/R=4,\epsilon_R/\epsilon_L=2,D/R=1.8,\lambda_D/R=5$\modifiedBlue{, and} $\gamma/(R\epsilon_0)=0.045$.
  The colored dots denote the \modifiedBlue{state} points \modifiedGreen{$(T^*,\mu^*)=(0.8,20)$}
  ($\textcolor[RGB]{186,085,211}{\bullet}$, see\modifiedGreen{, cf.,} Fig.~\ref{fig:structure3})
  and $(1.2,18)$
  ($\textcolor[RGB]{000,139,069}{\bullet}$, see\modifiedGreen{, cf.,} Fig.~\ref{fig:structure2}),
  while the black dots indicate $L$-$S_{AW}$-$S_A$ three-phase coexistence.
  The inset in panel (b) gives an enlarged view of the vicinity of the triple point.
   \modifiedRed{
   Like in Fig.~\ref{fig:phasediagrams2}, the pink curve \modifiedGreen{indicates}
   the onset of crystallization and the \modifiedGreen{salmon-colored} area
   represents the region $\eta_0\geq0.5$ of the phase diagram
   for which the lateral spacing in between neighboring particles on a hexagonal lattice
   becomes less than $10\%$ of the particle diameter $R$, i.e., $a/R\leq1.1$.
   \modifiedGreen{Hence, the particles are densely packed and previous simulations
   report the occurrence of a solid phase in this density regime~\cite{DeMiquel2002}.}
   }
  }
 \label{fig:phasediagrams3}
\end{figure}
\begin{figure}[!t]
 \includegraphics[width=0.5\textwidth]{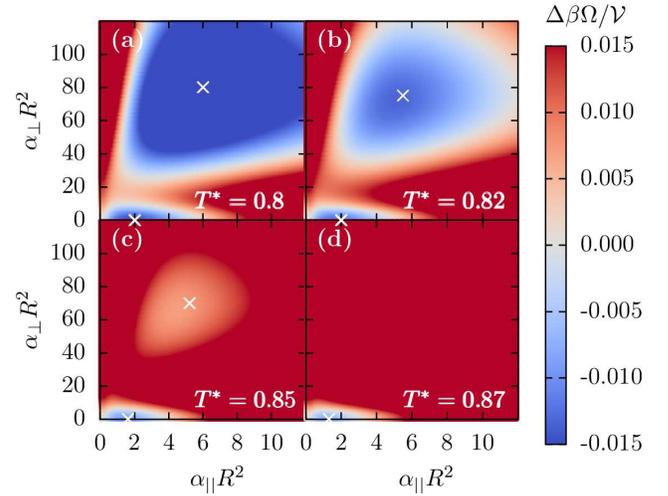}
  \caption{
  \modifiedRed{
  \modifiedGreen{
  Grand potential density $\Delta\beta\Omega/\mathcal{V}$ (Eq.~(\ref{eq:OmegaDiff_Crystallization}))
  of spatially non-uniform structures (crystalline or smectic, Eq.~(\ref{eq:HexGaussians}))
  in the range $\alpha_\perp R^2\in[0,120]$ and $\alpha_{||}R^2\in[0,12]$
  relative to that for a spatially uniform nematic fluid of the same density at 
  packing fraction $\eta_N=0.42$ and temperatures
  $T^*=0.8,0.82,0.85,$ \modifiedMagenta{and} $0.87$.
  Crosses ($\times$) denote local minima of $\Delta\beta\Omega/\mathcal{V}$. 
  By construction \modifiedMagenta{one has}
  $\Delta\beta\Omega/\mathcal{V}=0$ for $\alpha_{||}=\alpha_\perp=0$.}
  For $T^*=0.8$ \modifiedMagenta{(panel (a))} the global minimum of $\Delta\beta\Omega/\mathcal{V}$
  is at $(\alpha_{||}R^2\approx6,\alpha_\perp R^2\approx80)$ which is a hexagonal crystal structure.
  \modifiedMagenta{The} $S_A$ phase with $(\alpha_{||}R^2\approx2,\alpha_\perp R^2=0)$ is metastable.
  By increasing temperature one finds coexistence of the $S_A$ phase and of the crystal
  to occur close to $T^*=0.82$ \modifiedMagenta{(panel (b))} and for larger temperature $T^*=0.85,0.87$
  \modifiedMagenta{(panels (c) and (d))} the $S_A$ phase becomes stable.
  }
  }
 \label{fig:OmegaDiff_Crystallization}
\end{figure}

\modifiedBlue{Here} we investigate the dependence of the phase behavior of ILCs 
on the position $D$ of the charges of the particles for
$L/R=4$, $\epsilon_R/\epsilon_L=2$,
$\lambda_D/R=5$, $\gamma/(R\epsilon_0)=0.045$.
Figure~\ref{fig:phasediagrams3}(a) shows the case of the two charges 
\modifiedBlue{merged} in the geometrical center of \modifiedBlue{the} molecule,
\modifiedBlue{i.e.,} $D/R=0$. \modifiedBlue{In Fig.~\ref{fig:phasediagrams3}(b)}
the two charges are located \modifiedBlue{near} the \modifiedBlue{tips},
\modifiedBlue{i.e.,} $D/R=1.8$. For $D/R=0$ the phase diagram coincides
almost quantitatively with the \modifiedBlue{corresponding}
phase diagram \modifiedBlue{in Fig.~\ref{fig:phasediagrams2}(b)} for $D/R=0.9$,
besides a slight change in the location of the $S_{AW}$-$S_A$ two-phase region.
Thus, the change in the pair potential by moving the charges
from the center to the moderate distance $D/R=0.9$ 
\modifiedBlue{turns out} to be insufficient
for a significant change of the phase behavior. 
However, moving the charges to the \modifiedBlue{tips} of the particles
changes the shape of the pair potential \modifiedBlue{significantly}
\modifiedBlue{(Figs.~\ref{fig:pairpot}(c) and (d))},
and \modifiedBlue{this leads indeed to} a considerable change
in the phase behavior.
Figure~\ref{fig:phasediagrams3}(b) shows that
for ILC molecules with charges at their \modifiedBlue{tips}
\modifiedBlue{($D/R=1.8$ and $L/R=4$)}
the $L$-$S_{AW}$-$S_A$ triple point 
(\modifiedBlue{see} the inset of Fig.~\ref{fig:phasediagrams3}(b)
\modifiedBlue{providing} an enlarged view of the vicinity of the triple point)
is shifted to \modifiedBlue{a} higher temperature $T_t^*\approx1.22$.
\modifiedBlue{Thus} the low-temperature smectic phase $S_{AW}$ becomes stable at
temperatures, \modifiedBlue{which are higher than in the cases in
Figs.~\ref{fig:phasediagrams2} and \ref{fig:phasediagrams3}(a).}
Again, we estimate the location of the onset of crystallization \modifiedGreen{by}
employing the method of Sec.~\ref{sec:theory:DFT:Crystallization}.
The \modifiedGreen{corresponding} results (pink curves in Fig.~\ref{fig:phasediagrams3})
show that, in \modifiedGreen{the} case of the charges being located right at the tips
(panel (b)) the stable region of the $S_{AW}$ phase is enhanced compared
\modifiedGreen{with} the other cases (Figs.~\ref{fig:phasediagrams2} and \ref{fig:phasediagrams3}(a)),
due to the higher $L$-$S_{AW}$-$S_A$ triple point temperature.
Hence, an $S_{AW}$ phase is expected to indeed occur for long thin particles
with charges located at the tips (Fig.~\ref{fig:phasediagrams3}(b)),
whereas it is preempted by crystallization otherwise
(Figs.~\ref{fig:phasediagrams2} and \ref{fig:phasediagrams3}(a)).
\MODIFIED{
If the charges are localized at the tips of the molecules\MODIfied{,} the smectic phase $S_{AW}$
with wide layer spacing is stabilized in the intermediate temperature regime\MODIfied{, i.e.,}
in between the high temperature ordinary smectic $S_{A}$ phase and
crystalline structures $C$ at low \MODIfied{temperatures}
(at intermediate densities), \MODIfied{which is} due to the effective electrostatic repulsion
of neighboring smectic layers. However, in the other cases, i.e., 
\MODIfied{if} the charges are localized close to the center of mass or
\MODIfied{if} there are no charges at all, the ordinary smectic phase $S_A$ with 
\MODIfied{densely} packed smectic layers ($d\approx L$) is entropically preferred over
the wide smectic phase $S_{AW}$ at intermediate temperatures (and intermediate packing fractions).
\MODIfied{However, in the present case}, the $S_{AW}$ phase is
\MODIfied{more stable than} the ordinary smectic $S_A$ phase
only at temperatures below the freezing transition
\MODIfied{where the actually stable phase is the crystalline one}.
}

\begin{figure}[!t]
 \includegraphics[width=0.5\textwidth]{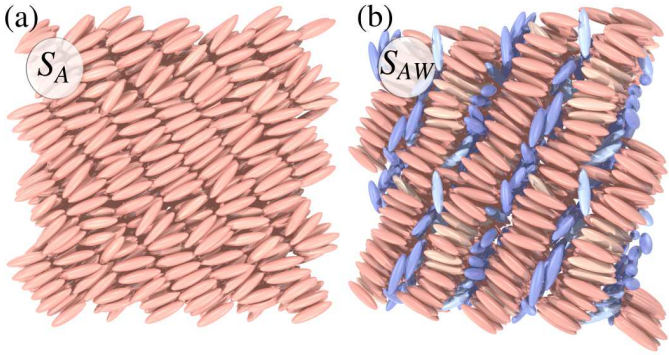}
  \caption{
  \modifiedGreen{Smectic configurations of ILCs with $L/R=4$.}
  Panel (a) shows a 
  \modifiedBlue{configuration appearing during} a simulation performed at temperature $T^*=0.6$;
  the chemical potential $\mu^*=0.9$ is tuned such that $\eta_0\approx0.389$.
  Panel (b) depicts a \modifiedBlue{configuration} for \modifiedGreen{$(T^*,\mu^*)=(0.5,-2.6)$} 
  giving rise to $\eta_0\approx0.324$.
  \modifiedBlue{For both (a) and (b) the parameters of the}
  pair potential \modifiedBlue{are given} by
  $L/R=4,\epsilon_R/\epsilon_L=3,D/R=1.8,\lambda_D/R=5,\gamma/(R\epsilon_0)=0.045$, \modifiedBlue{and}
  $R_\text{cut}/R=6$.
  At \modifiedBlue{the higher} temperature $T^*=0.6$, one \modifiedBlue{finds} the
  \modifiedBlue{ordinary} smectic \modifiedBlue{$S_A$} phase,
  while for the lower temperature $T^*=0.5$ the novel $S_{AW}$ phase is observed.
  \modifiedBlue{The latter is}
  characterized by an alternating structure of particles
  \modifiedBlue{such that within} the smectic layers
  \modifiedBlue{the particles are} oriented parallel 
  to the layer normal (\modifiedBlue{pale pink} particles) \modifiedBlue{whereas the}
  particles in between the layers \modifiedBlue{are oriented perpendicularly} to it
  \modifiedMagenta{but without lateral orientational order}
  (\modifiedBlue{blue} particles).
  }
 \label{fig:sim_snapshots}
\end{figure}

\modifiedBlue{We have studied the latter case}
of ILCs with \modifiedBlue{the} charges at their \modifiedBlue{tips}
also \modifiedBlue{by using grand canonical}
Monte Carlo simulations. In Fig.~\ref{fig:sim_snapshots}
two \modifiedBlue{configurations are shown which appear during}
simulations performed \modifiedBlue{for $(T^*,\mu^*)=(0.6,0.9)$ in panel (a)},
and \modifiedBlue{for $(T^*,\mu^*)=(0.5,-2.6)$ in panel (b)}.
Here the pair potential is described by
$L/R=4$, $\epsilon_R/\epsilon_L=3$, $D/R=1.8$, $\lambda_D/R=5$,
$\gamma/(R\epsilon_0)=0.045$, \modifiedBlue{and $R_\text{cut}/R=6$}.
The chemical potentials are chosen to be sufficiently large,
such that in both cases the system forms a smectic structure.
In panel (a), one observes an ordinary $S_A$ phase
\modifiedBlue{according to which the}
particles are located in the smectic layers
with a preferred orientation parallel to the director $\vec{\hat n}$,
\modifiedGreen{i.e., the layer normal}.
\modifiedBlue{Instead, at the lower temperature $T^*=0.5$}
panel (b) shows a different structure.
Here, an increased layer spacing is observed.
\modifiedBlue{The} space in between the layers
is populated by \modifiedBlue{numerous} particles
\modifiedBlue{which are preferentially oriented}
perpendicular to the layer normal.
\modifiedBlue{This} is the same periodic structure
which \modifiedBlue{we have found} 
within our DFT approach for the low-temperature $S_{AW}$ phase
(compare Fig.~\ref{fig:phasediagrams3}(b)).
\modifiedBlue{Furthermore}, in agreement with the present theory,
increasing the \modifiedBlue{rescaled} chemical potential $\mu^*$ at low but fixed temperature $T^*$,
at sufficiently large packing fraction $\eta_0$ \modifiedBlue{one finds}
a transition from the $S_{AW}$ phase to the $S_A$ phase.
By increasing the chemical potential $\mu^*$ the packing fraction is also increased
and ultimately a dense packing of smectic layers, 
\modifiedBlue{corresponding to} the $S_A$ phase, is preferred over the
\modifiedBlue{smectic $S_{AW}$ phase} with wide layer spacing.
(See also the discussion of our simulational results in the next section.)

It is worth mentioning that a similar kind of structure has been reported for a system of
hard discs interacting via an additional
anisotropic Yukawa potential~\cite{Jabbari-Farouji2013,Jabbari-Farouji2014}.
In this canonical Monte Carlo study a structure called \emph{intergrowth texture}
has been observed which shows a periodic structure of two alternating layers of particles.
The directors of both layers are perpendicular to each other.
Nevertheless, unlike the $S_{AW}$ phase,
the particles \modifiedBlue{within} each layer of an intergrowth texture are not localized.
Thus they do not \modifiedBlue{exhibit positional order}
in any direction and \modifiedBlue{cannot} be categorized as a smectic structure.
In contrast to monodisperse systems, like in the present study, 
alternating smectic layer structures have already been observed in binary mixtures
of particles with different geometries~\cite{Koda1994,vanRoij1996,Cinacchi2004,Martinez-Raton2005}.
For such systems the alternating layer structure is driven
by segregation of the two particles species.
\modifiedBlue{It is worth mentioning, that} due to fluctuations,
even in the common $S_A$ phase there is a non-vanishing probability
to \modifiedBlue{find} particles in between the smectic layers with perpendicular orientation
(see, e.g., Ref.~\cite{vanRoij1995}).

\modifiedBlue{
Finally, we note that for instance particles with a electric quadrupole are known
to form smectic $S_C$ phases, in which the director $\vec{\hat n}$ is tilted
\modifiedGreen{with respect to the} normal of the smectic layers (see, e.g., Ref.~\cite{Neal1998}).
Such kind of liquid crystals are of particular interest for 
technological applications \modifiedGreen{such as} fast electro-optic displays, because those 
materials can be ferroelectric~\cite{Meyer1975}.
}
%

\subsection{\label{sec:discussion:SmecticStructure}Variety of smectic structures}

\modifiedBlue{We have illustrated how the phase behavior} of ionic liquid crystals
\modifiedBlue{varies as function of the parameters characterizing the pair potential.}
In particular, the occurring smectic phases show distinct layer spacings.
In order to analyze the structure of the \modifiedBlue{various} smectic bulk phases in more detail,
we discuss the density profiles in terms of the \modifiedBlue{local} packing fraction
$\eta(z)=n(z)\,LR^2\pi/6$ and the \modifiedBlue{spatially varying}
orientational order parameter $S_2(z)$ (compare Sec.~\ref{sec:theory:DFT:Formalism}).
\begin{figure}[!t]
 \includegraphics[width=0.45\textwidth]{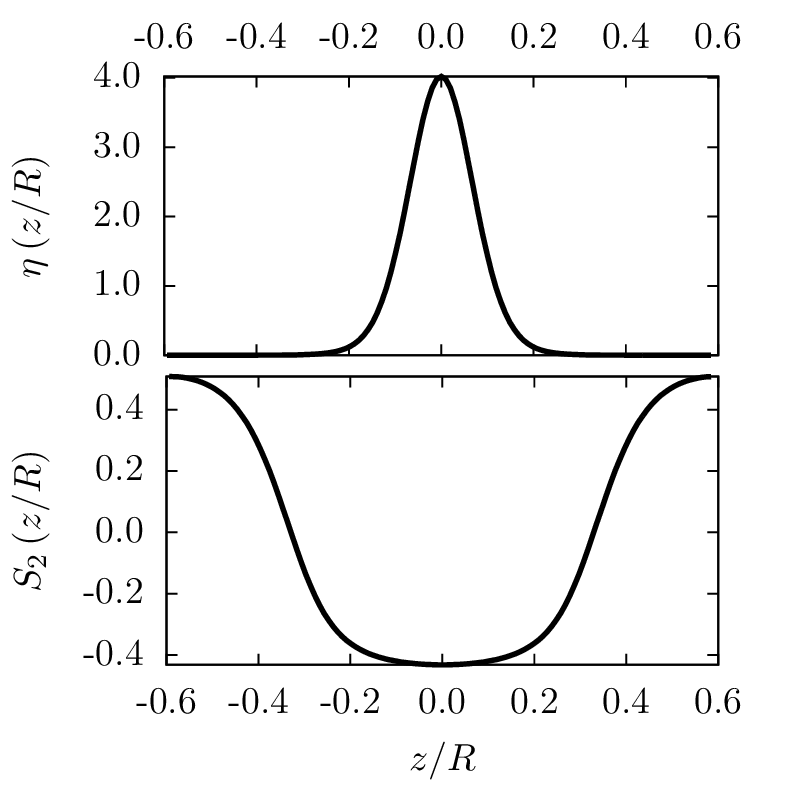}
  \caption{
  \modifiedBlue{Local packing fraction $\eta(z/R)=n(z/R)LR^2\pi/6$
  and orientational order parameter
  $S_2(z/R)=\Int{\mathcal{S}}{2}{\omega}P_2(\cos(\theta))\rho(z/R,\theta)/n(z/R)$
  with the total number density $n(z/R)=\Int{\mathcal{S}}{2}{\omega}\rho(z/R,\theta)$}
  \modifiedGreen{displayed} within one period $d/R\approx1.2$ \modifiedGreen{for}
  \modifiedBlue{the state} point \modifiedBlue{$(T^*,\mu^*)=(0.45,20)$ indicated in}
  Fig.~\ref{fig:phasediagrams1}(a) (\modifiedBlue{orange dot} $\textcolor[RGB]{238,154,073}{\bullet}$) for
  $L/R=2, \epsilon_R/\epsilon_L=2, D/R=0.9, \lambda_D/R=5$\modifiedBlue{, and} $\gamma/(R\epsilon_0)=0.0045$.
  The smectic layer spacing $d$ is \modifiedBlue{smaller} than the length
  \modifiedBlue{of the particles}.
  \modifiedBlue{At the center of the smectic layers the ILC molecules are}
  oriented mainly perpendicular to the \modifiedGreen{layer normal} $\vec{\hat n}$
  \modifiedBlue{as one can infer from}
  the negative value of the orientational order parameter $S_2(z=0)\approx-0.4$
  ($S_{AN}$ phase).
  }
 \label{fig:structure1}
\end{figure}
First, \modifiedBlue{we} consider the smectic phase $S_{AN}$ observed for
$L/R=2, \epsilon_R/\epsilon_L=2, D/R=0.9, \lambda_D/R=5$
\modifiedBlue{and, $\gamma/(R\epsilon_0)=0.0045$}
(compare Fig.~\ref{fig:phasediagrams1}(b)).
In Fig.~\ref{fig:structure1} the relevant profiles $\eta(z)$ and $S_2(z)$ are plotted
for the \modifiedBlue{state} point \modifiedBlue{$(T^*,\mu^*)=(0.45,20.0)$}
\modifiedBlue{indicated in} Fig.~\ref{fig:phasediagrams1}(b)
(\modifiedGreen{orange dot} $\textcolor[RGB]{238,154,073}{\bullet}$).
The smectic layer spacing is $d/R\approx1.2$; $S_2(z)\approx-0.4$ at
\modifiedBlue{$|z/d|\ll1$} shows that \modifiedBlue{within the smectic layers}
the particles are oriented predominantly perpendicular to the \modifiedGreen{layer normal}.
This \modifiedBlue{finding} is \modifiedBlue{plausible because}
$d/R\approx1.2$ is much smaller than the length of the particles $L/R=2$
\modifiedBlue{which enforces the particles to tilt towards the smectic layers}.
\modifiedBlue{The packing fraction profile} $\eta(z)$ in Fig.~\ref{fig:structure1}
\modifiedBlue{tells} that the particles
are strongly localized \modifiedBlue{within} the layers.  
\modifiedBlue{The layer spacing does not vary significantly as function of temperature
and \modifiedGreen{of} \modifiedMagenta{the} chemical potential \modifiedGreen{within}
the thermodynamic region of a stable smectic phase}
\modifiedRed{(\modifiedGreen{according} to the DFT
method \modifiedGreen{introduced in} Sec.~\ref{sec:theory:DFT:Formalism}).}
\modifiedBlue{In the smectic $S_{AN}$ phase 
the \modifiedGreen{layer normal $\vec{\hat n}$} still points into the $z$-direction, so that
the particles do not have a preferred lateral orientation,
but they avoid an orientation parallel to the director.}
This behavior seems to be \modifiedGreen{caused by} the small length-to-breadth ratio $L/R=2$
and the small value of the anisotropy parameter $\epsilon_R/\epsilon_L=2$,
\modifiedBlue{which renders} \modifiedGreen{these} particles \modifiedBlue{relatively} isotropic.
\modifiedBlue{This} is even more pronounced
in \modifiedBlue{the} case of the ILC fluid \modifiedBlue{shown} in Fig.~\ref{fig:phasediagrams1}(b)
due to the additional electrostatic repulsion\modifiedBlue{, which} leads to a higher
$L$-$S_{AW}$-$S_A$ triple point temperature.
\modifiedBlue{For both smectic phases, $S_A$ and $S_{AN}$,}
the layer spacing does not \modifiedBlue{vary} significantly 
\modifiedBlue{as function of} temperature and chemical potential
within the \modifiedBlue{thermodynamic} region of \modifiedBlue{a stable}
smectic phase.
\modifiedRed{
Again, we emphasize that for the shorter particles,
described by $L/R=2$ and $\epsilon_R/\epsilon_L=2$,
the stability of the liquid-crystalline phases $S_{A}$ and $S_{AN}$
is very likely to be an artifact of the
\modifiedGreen{method employed (see Sec.~\ref{sec:theory:DFT:Formalism})}.
The transition from an isotropic liquid phase to those mesophases
occurs at large densities \modifiedGreen{for} which one already expects crystalline structures 
to emerge (see the discussion \modifiedGreen{in} Sec.~\ref{sec:discussion:PhaseDiagrams:Comparison}).
}

\begin{figure}[!t]
 \includegraphics[width=0.45\textwidth]{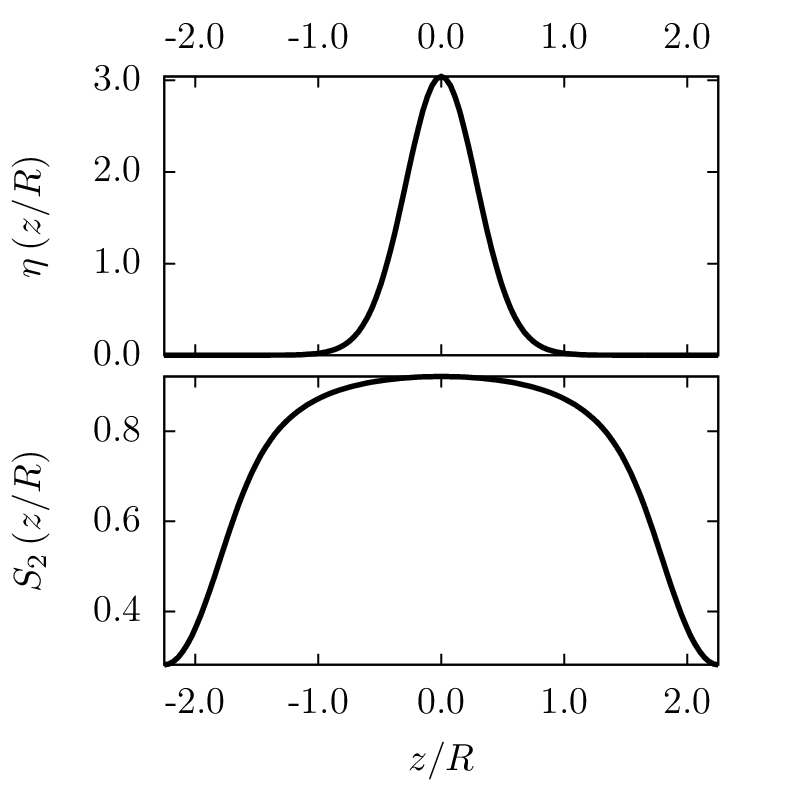}
  \caption{
  \modifiedBlue{Local packing fraction $\eta(z/R)$ and
  scalar orientational order parameter $S_2(z/R)$}
  within one period $d/R\approx4.5$ at \modifiedBlue{the state} point
  \modifiedGreen{$(T^*,\mu^*)=(1.2,18)$}
  in Fig.~\ref{fig:phasediagrams3}(b) (\modifiedBlue{green dot} $\textcolor[RGB]{000,139,069}{\bullet}$) for
  $L/R=4, \epsilon_R/\epsilon_L=2, D/R=1.8, \lambda_D/R=5$\modifiedBlue{, and} $\gamma/(R\epsilon_0)=0.045$.
  The ILC molecules are strongly \modifiedBlue{localized}
  in the \modifiedBlue{center of the} smectic layers \modifiedBlue{where they are} oriented mainly
  parallel to the \modifiedGreen{layer normal} $\vec{\hat n}$
  \modifiedBlue{as} indicated by the large positive value
  of the orientational order parameter \modifiedBlue{$S_2(z=0)\approx0.9$} (ordinary $S_A$ phase).
  }
 \label{fig:structure2}
\end{figure}
\begin{figure}[!t]
 \includegraphics[width=0.45\textwidth]{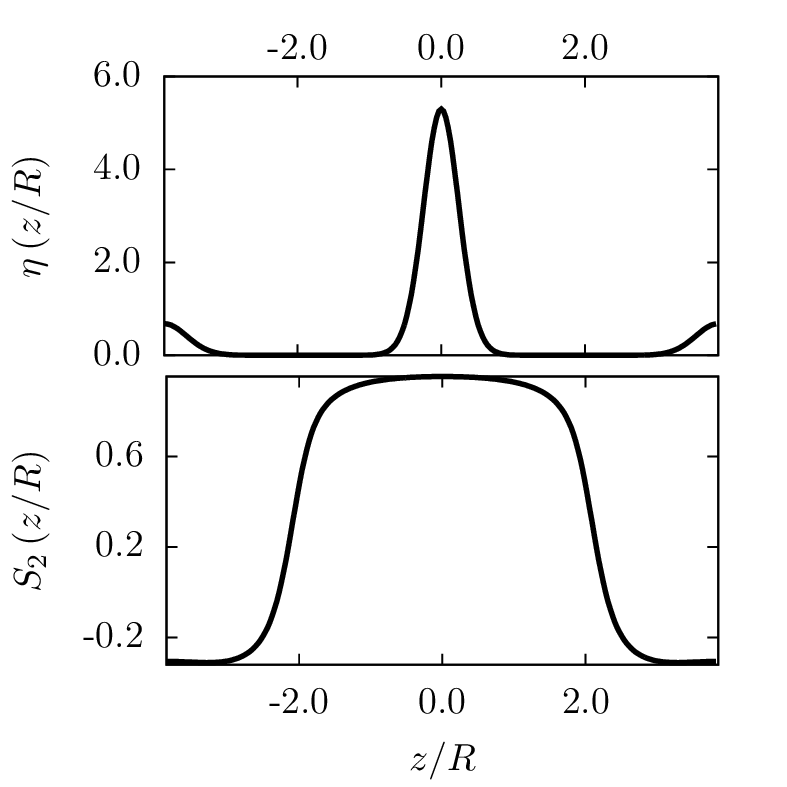}
  \caption{
  \modifiedBlue{Local packing fraction $\eta(z/R)$ and
  scalar orientational order parameter $S_2(z/R)$}
  within one period $d/R\approx7.7$ at the \modifiedBlue{state} point
  \modifiedGreen{$(T^*,\mu^*)=(0.8,20)$}
  in Fig.~\ref{fig:phasediagrams3}(b)
  (\modifiedBlue{magenta dot} $\textcolor[RGB]{186,085,211}{\bullet}$) for
  $L/R=4, \epsilon_R/\epsilon_L=2, D/R=1.8, \lambda_D/R=5$\modifiedBlue{, and} $\gamma/(R\epsilon_0)=0.045$.
  One observes an alternating structure \modifiedBlue{with the} majority of \modifiedBlue{the}
  particles \modifiedBlue{being} located \modifiedBlue{at the center of}
  the layers ($z\approx0$) with \modifiedBlue{an} orientation parallel to the layer normal,
  while a significant minority of particles is located in between the layers
  \modifiedBlue{(i.e., $|z|/R\approx d/(2R)\approx3.85$)}
  with preferentially perpendicular orientation ($S_{AW}$ phase).
  }
 \label{fig:structure3}
\end{figure}

Now, \modifiedBlue{we} turn to the ILC molecules described by \modifiedBlue{the parameter set}
$L/R=4, \epsilon_R/\epsilon_L=2, D/R=1.8, \lambda_D/R=5$,
\modifiedBlue{and $\gamma/(R\epsilon_0)=0.045$},
\modifiedBlue{the phase diagram of which} is shown in Fig.~\ref{fig:phasediagrams3}(b).
\modifiedGreen{In Fig.~\ref{fig:phasediagrams3}(b),}
at \modifiedBlue{the state} point \modifiedBlue{$(T^*,\mu^*)=(1.2,18)$}
(\modifiedBlue{green dot} $\textcolor[RGB]{000,139,069}{\bullet}$) the $S_A$ phase is stable with
a layer spacing $d/R\approx4.5$.
The profiles of $\eta(z)$ and $S_2(z)$ are shown in Fig.~\ref{fig:structure2}.
As expected for an ordinary smectic-A phase,
the ILC molecules are located in the layers
with \modifiedGreen{an orientation} predominantly parallel
to the \modifiedGreen{layer normal $\vec{\hat n}$}.
In contrast to the shorter particles \modifiedGreen{discussed in} Fig.~\ref{fig:structure1},
here the layer spacing $d/R\approx4.5$
is \modifiedGreen{comparable with} the size of
the length $L/R=4$ \modifiedBlue{of the particles} and thus there is enough space for the particles
to be aligned with the \modifiedGreen{layer normal $\vec{\hat n}$}.

At low temperatures and large packing fractions,
one \modifiedBlue{finds} the novel wide smectic $S_{AW}$ phase.
For \modifiedBlue{$(T^*,\mu^*)=(0.8,20)$}
(\modifiedBlue{magenta dot} $\textcolor[RGB]{186,085,211}{\bullet}$ in Fig.~\ref{fig:phasediagrams3}(b))
this structure is shown in Fig.~\ref{fig:structure3}.
The equilibrium layer spacing is $d/R\approx7.7$,
which is significantly larger than \modifiedBlue{the one} for the high-temperature
\modifiedBlue{ordinary} smectic $S_A$ phase
(compare Fig.~\ref{fig:structure2}).
\modifiedBlue{The wide smectic $S_{AW}$ phase} shows an increased number of particles
localized in between the layers, \modifiedBlue{i.e., around} $|z|\approx d/2$.
They are oriented preferentially perpendicular to the layer normal
\modifiedGreen{$\vec{\hat n}$}
\MODIFIED{(with no orientational ordering \MODIfied{within} the $x$-$y$-plane)},
while particles in the layers, \modifiedBlue{i.e., for $|z/d|\ll1$},
are predominantly aligned with the \modifiedGreen{normal $\vec{\hat n}$},
like in the $S_A$ phase.

\begin{figure}[!t]
 \includegraphics[width=0.40\textwidth]{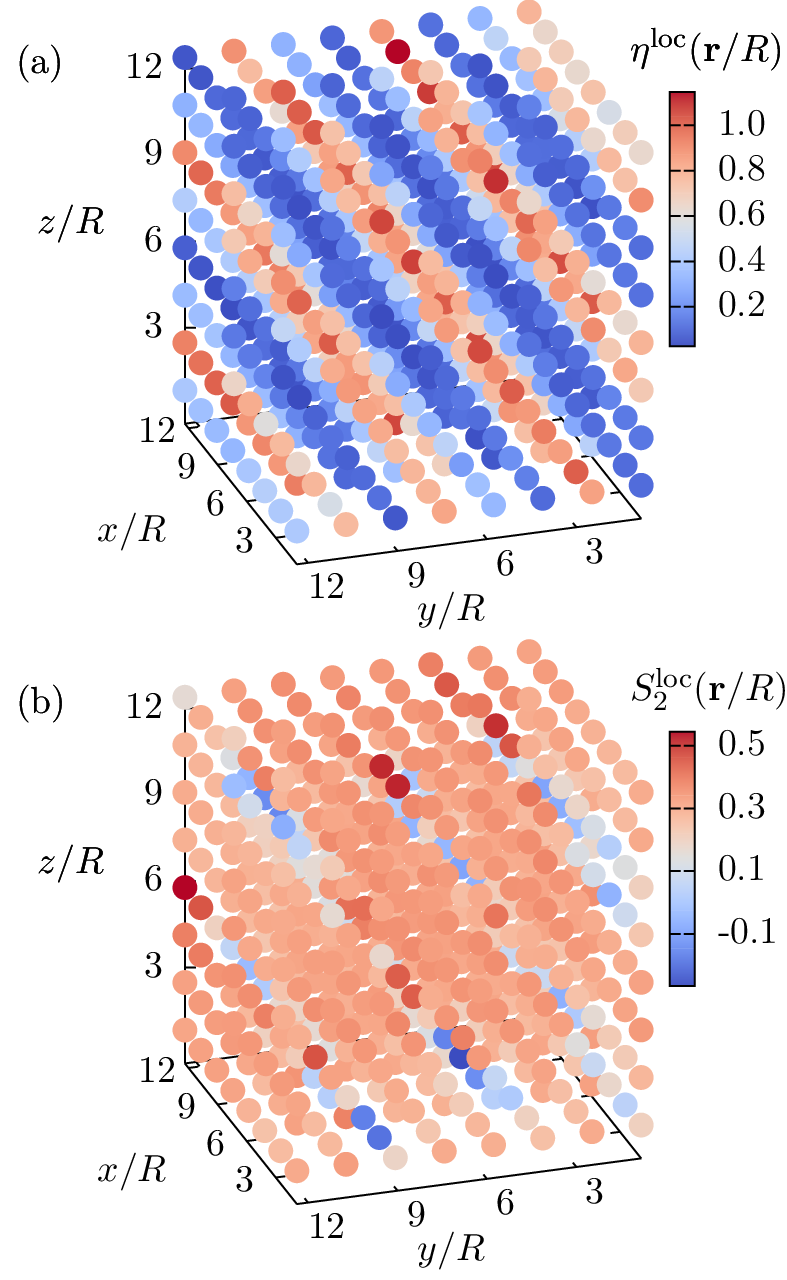}
  \caption{
  \modifiedBlue{
  Local packing fraction $\eta^\text{loc}(\vec{r}/R)$ (a) and
  scalar orientational order parameter $S_2^\text{loc}(\vec{r}/R)$ (b)}
  obtained \modifiedBlue{from grand canonical} Monte Carlo simulations
  on a periodic cubic box of side length $V^{1/3}/R=13.2$ at temperature $T^*=0.5$
  and chemical potential $\mu^*=-1.2$, giving rise to a global
  mean packing fraction $\eta_0\approx0.436$ in the simulation box.
  \modifiedBlue{Each colored dot represents a site of a 
  simple cubic lattice on which we monitor $\eta^\text{loc}(\vec{r}/R)$
  \modifiedGreen{and} $S_2^\text{loc}(\vec{r}/R)$, \modifiedGreen{respectively,}
  along the MC trajectory.
  The color-coding of the dots can be inferred from the respective color key.} 
  \modifiedBlue{The parameters of the pair potential are given by}
  $L/R=4, \epsilon_R/\epsilon_L=3, D/R=1.8, \lambda_D/R=5, \gamma/(R\epsilon_0)=0.045$\modifiedBlue{, and}
  $R_\text{cut}/R=6$.
  The ILC molecules are \modifiedGreen{concentrated}
  in the smectic layers \modifiedGreen{which are} oriented mainly
  parallel to the \modifiedGreen{layer normal,
  which can be inferred from the positive value of $S_2^\text{loc}(\vec{r}/R)>0.3$
  for positions $\vec{r}$ which correspond to large $\eta^\text{loc}(\vec{r}/R)>0.6$; this}
  \modifiedBlue{corresponds} to an ordinary smectic $S_A$ phase.
  \modifiedGreen{Some sample points $\vec{r}$ yield a negative value of $S_2(\vec{r}/R)<0$,
  corresponding to particles which eventually moved out of the smectic layers
  and then \modifiedMagenta{turned} perpendicular.}
  \modifiedBlue{Here, in contrast to the DFT approach,
  the layer normal does not necessarily point into the $z$-direction,
  but it is tilted towards one of the edges of the simulation box.}
  }
 \label{fig:mc_structure1}
\end{figure}
\begin{figure}[!t]
 \includegraphics[width=0.40\textwidth]{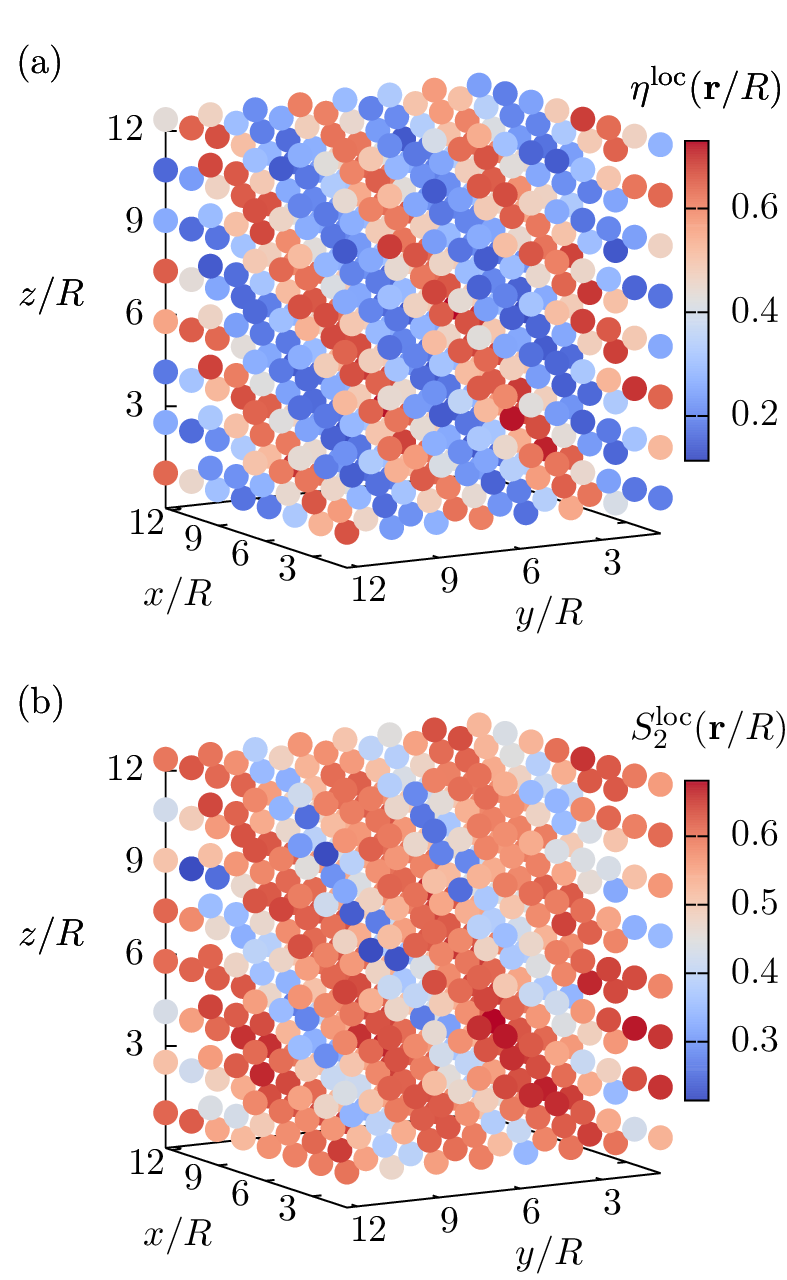}
  \caption{
  \modifiedBlue{Same as Fig.~\ref{fig:mc_structure1} but for
  $T^*=0.6$ and $\mu^*=1.7$ which corresponds to $\eta_0\approx0.394$.
  This state point belongs to an ordinary smectic $S_A$ phase,}
  \modifiedGreen{which can be inferred from the positive value of $S_2^\text{loc}(\vec{r}/R)>0.4$
  for positions $\vec{r}$ which correspond to large $\eta^\text{loc}(\vec{r}/R)>0.6$.
  The temperature $T^*=0.6$ is higher than for the state point $(0.5,-1.2)$
  \modifiedMagenta{discussed} in Fig.~\ref{fig:mc_structure1}.
  \modifiedMagenta{Thus} the particles in the smectic layers are less localized
  leading to smaller maximum values of the
  local packing fraction $\eta_\text{max}^\text{loc}(\vec{r}/R)\approx0.7$
  (compare the maxima $\eta_\text{max}^\text{loc}(\vec{r}/R)\approx1.1$
  in Fig.~\ref{fig:mc_structure1}).}
  \modifiedBlue{The layer normal points towards one of the diagonals of the simulation box.}
  }
 \label{fig:mc_structure2}
\end{figure}
\begin{figure}[!t]
 \includegraphics[width=0.40\textwidth]{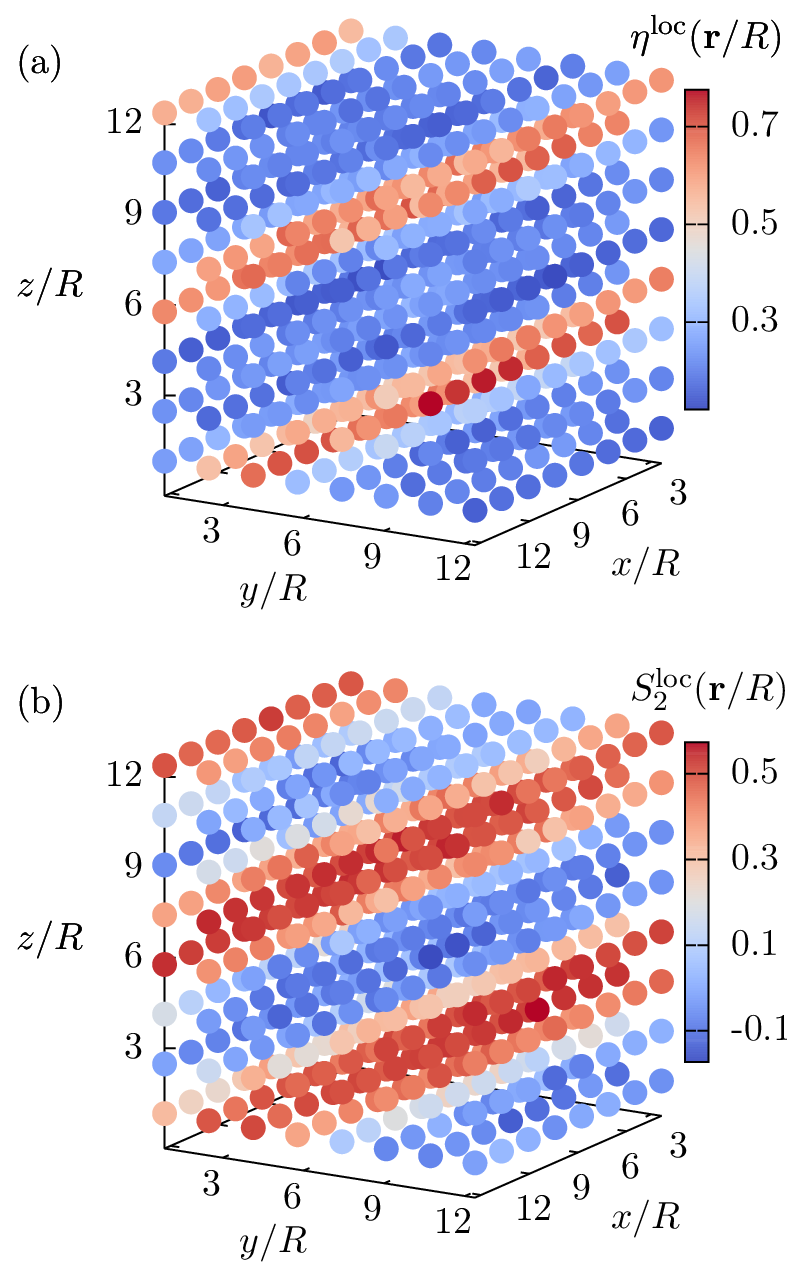}
  \caption{
  \modifiedBlue{Same as Figs.~\ref{fig:mc_structure1} and \ref{fig:mc_structure2}
  but for $T^*=0.5$ and $\mu^*=-2.6$ which corresponds to $\eta_0\approx0.324$.
  For this state point} one observes an alternating structure
  of a majority of particles, which are located in the layers
  with \modifiedBlue{their orientations} parallel to the layer normal,
  while a significant minority of \modifiedBlue{the} particles is located
  in between the layers with perpendicular orientation ($S_{AW}$ phase).
  \modifiedBlue{The layer normal points towards one of the edges of the simulation box.}
  }
 \label{fig:mc_structure3}
\end{figure}
So far, we \modifiedBlue{have} discussed the structural properties
of the \modifiedBlue{various} ILC smectic phases, \modifiedBlue{as}
\modifiedBlue{predicted by the present density functional theory.}
\modifiedBlue{For comparison,}
in Figs.~\ref{fig:mc_structure1}, \ref{fig:mc_structure2}, and \ref{fig:mc_structure3}
the local packing fraction $\eta^\text{loc}(\vec{r})=\rho^\text{loc}(\vec{r})\,LR^2\pi/6$
and the local orientational order parameter $S_2^\text{loc}$,
\modifiedBlue{as} obtained by Monte Carlo simulations of an ILC fluid with
$L/R=4,\epsilon_R/\epsilon_L=3,\lambda_D=5,D/R=1.8,\gamma/(R\epsilon_0)=0.045$,
\modifiedBlue{and $R_\text{cut}/R=6$},
are shown on a simple cubic lattice of sample points
within the cubic simulation box of side length $V^{1/3}/R=13.2$.
Figures~\ref{fig:mc_structure1} and \ref{fig:mc_structure2} clearly
show an ordinary \modifiedBlue{smectic} $S_A$ phase, characterized by a periodic structure
\modifiedBlue{in which} the particles are located \modifiedGreen{inside} the smectic layers,
indicated by the red data points for large \modifiedBlue{values of $\eta^\text{loc}(\vec{r})$}
and a predominant alignment of particles along the layer normal,
indicated by the large positive value of the local orientational order parameter 
$S_2^\text{loc}(\vec{r})\geq0.5$ for nearly all sample points $\vec{r}$.
The data points of Fig.~\ref{fig:mc_structure1} are obtained for
\modifiedBlue{$(T^*,\mu^*)=(0.5,-1.2)$, i.e., $\eta_0\approx0.436$}
and the data of Fig.~\ref{fig:mc_structure2} \modifiedBlue{corresponds to}
\modifiedBlue{$(T^*,\mu^*)=(0.6,1.7)$, i.e., $\eta_0\approx0.394$}.
However, if the temperature is sufficiently low and the chemical potential
is \modifiedBlue{chosen} such that the mean packing fraction is not too large,
e.g., \modifiedBlue{$(T^*,\mu^*)=(0.5,-2.6)$ so that $\eta_0\approx0.324$},
one observes the novel smectic structure $S_{AW}$ \modifiedBlue{as}
shown in Fig.~\ref{fig:mc_structure3}.
The alternating orientation of particles gives rise to the alternating pattern
of blue ($S_2^\text{loc}(\vec{r})<0$) and red ($S_2^\text{loc}(\vec{r})>0$)
data points \modifiedGreen{for} the orientational order parameter along the layer normal
\modifiedBlue{(Fig.~\ref{fig:mc_structure3}(b))}.
\modifiedBlue{For these simulation results} the layer normal
and the $z$-direction are not parallel,
\modifiedBlue{because} the \modifiedBlue{start} configuration is isotropic,
which \MODIFIED{in principle} allows the system to form any structure \modifiedBlue{without bias}.
(Without cost of free energy the sample can be rotated so that the layer normal
is parallel to the $z$ axis.)
\MODIFIED{
However, the layer normal tends to \MODIfied{be parallel to} one of the
diagonals of the simulation box (compare Figs.~\ref{fig:mc_structure1}-\ref{fig:mc_structure3}),
which \MODIfied{is likely to} be related to the cubic geometry
of the simulation box and \MODIfied{thus appears to be a finite-size effect}.
}
In agreement with the present \modifiedBlue{DFT approach}, the simulations \modifiedBlue{tell}
that for the $S_A$ phase the smectic layer spacing is \modifiedBlue{of} the size of
the particle length $L/R=4$ while for the $S_{AW}$ phase
\modifiedBlue{$d$} is significantly \modifiedBlue{larger.}
\modifiedBlue{In this phase there are}
small (local) maxima of the local packing fraction $\eta^\text{loc}(\vec{r})$
in between the layers
(indicated \modifiedBlue{in Fig.~\ref{fig:mc_structure3}} by the light blue dots
being surrounded by dark blue dots).
Note that \modifiedBlue{although} some sample points
in Fig.~\ref{fig:mc_structure1} show a negative value
of the local orientational order parameter $S_2^\text{loc}(\vec{r})<0$ in between the smectic layers,
this does not indicate a realization of the smectic \modifiedBlue{$S_{AW}$} phase,
but is due to the well-known observation,
that \modifiedBlue{in course of the simulation} some particles move out of the smectic layers
and \modifiedBlue{then} turn perpendicular,
\modifiedBlue{because there is only a narrow gap}
in between the smectic layers of \modifiedBlue{an} ordinary $S_A$ phase
(see, e.g., \modifiedBlue{Ref.}~\cite{vanRoij1995}).

The present \modifiedBlue{DFT predicts} that the triple point temperature for an ILC fluid
can be increased \modifiedBlue{relative to the corresponding one for} an ordinary liquid crystal,
\modifiedBlue{provided} the location $D$ of the charges, their interaction strength $\gamma$,
and the screening length $\lambda_D$ \modifiedBlue{are chosen suitably}
(compare Figs.~\ref{fig:phasediagrams2} and \ref{fig:phasediagrams3}).
Therefore \modifiedBlue{the} $S_{AW}$ phase can \modifiedBlue{occur}
for \modifiedBlue{ILCs (see Fig.~\ref{fig:sim_snapshots}(b))} 
if the $S_{AW}$-$S_A$ coexistence \modifiedBlue{curve} is
shifted above the melting transition.
\modifiedRed{
In contrast, \modifiedGreen{for} ordinary liquid crystals (see Fig.~\ref{fig:phasediagrams2}(a))
the formation of the $S_{AW}$ phase 
is preempted by crystallization, which is in agreement with
the findings of previous studies, e.g., Ref.~\cite{DeMiquel2004}.
}

It is worth mentioning, that the \modifiedBlue{so-called} \emph{intergrowth texture} structure
observed in \modifiedBlue{Refs.}~\cite{Jabbari-Farouji2013,Jabbari-Farouji2014} can also
be interpreted as an ionic liquid crystal phenomenon,
\modifiedBlue{because} there hard discs \modifiedBlue{have been considered}
\modifiedBlue{which interact} via an additional, anisotropic Yukawa potential.

\subsection{\label{sec:discussion:d_vs_kT}Temperature dependence of the layer spacing}
\begin{figure}[!t]
 \includegraphics[width=0.50\textwidth]{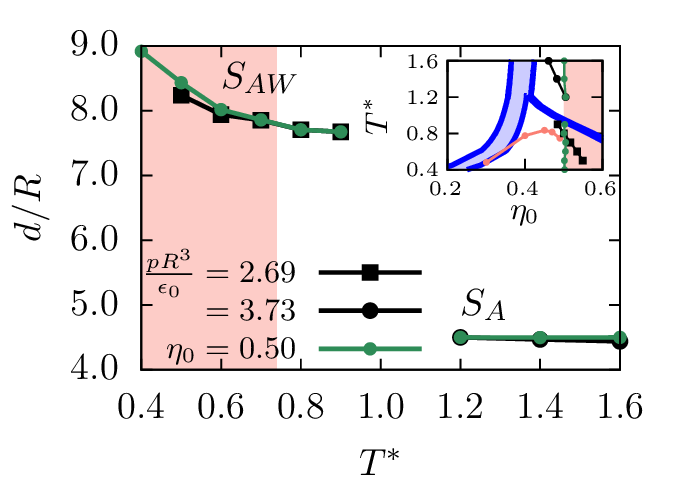}
  \caption{
  Equilibrium layer spacing \modifiedBlue{$d/R$},
  \modifiedBlue{as} obtained within the present DFT approach,
  \modifiedBlue{as function of} temperature $T^*$
  \modifiedBlue{for the parameter set
  $L/R=4,\epsilon_R/\epsilon_L=2,D/R=1.8,\lambda_D/R=5$, and $\gamma/(R\epsilon_0)=0.045$
  of the pair potential.}
  For $T^*>1.0$ the \modifiedBlue{ordinary smectic $S_A$} phase is stable
  (\modifiedBlue{see} the inset, which shows the considered thermodynamic paths).
  \modifiedBlue{For the considered paths, the layer spacing depends only}
  \modifiedBlue{weakly} on temperature.
  \modifiedMagenta{The pressure $p:=\Omega[\rho^\text{eq}]/\mathcal{V}$ is obtained by
  multiplying Eq.~(\ref{eq:pressure}) by $kT$.}
  For the low-temperature wide \modifiedBlue{smectic} $S_{AW}$ phase
  \modifiedBlue{the dependence on temperature is more pronounced. This is}
  due to the free space in between the layers and the electrostatic repulsion,
  which becomes more \modifiedBlue{effective upon} decreasing temperature.
  \modifiedGreen{However, this effect is prominent only in that region
  of the $S_{AW}$ phase where it is metastable with respect to
  crystallization, i.e., for $T^*\lesssim0.74$ (salmon-colored area).
  The pink curve in the inset indicates the onset of crystallization.}
  }
 \label{fig:d_vs_kT}
\end{figure}
The high-temperature phase $S_A$ and
the low-temperature phase $S_{AW}$ \modifiedBlue{exhibit} distinct
\modifiedBlue{structural properties (Figs.~\ref{fig:structure2} and \ref{fig:structure3})}.
\modifiedBlue{In particular} the \modifiedGreen{size} of the layer spacing \modifiedBlue{differs}.
\modifiedBlue{Our analysis reveals} that for the $S_A$ phase,
\modifiedBlue{in which} the layer spacing is \modifiedBlue{about}
the size of \modifiedBlue{the length $L$ of the particles,
the layer thickness varies only weakly as function of temperature (see Fig.~\ref{fig:d_vs_kT}).}
\modifiedBlue{Along two thermodynamic paths} within the domain of
the stable $S_A$ phase -- one at fixed mean packing fraction $\eta_0\approx0.5$
(green \modifiedGreen{dotted} vertical path in the \modifiedBlue{corresponding} phase diagram
\modifiedBlue{shown in the inset of Fig.~\ref{fig:d_vs_kT};
compare Fig.~\ref{fig:phasediagrams3}(b))}
and the other one at fixed pressure \modifiedMagenta{
$p=-\Omega[\rho^\text{eq}]/\mathcal{V}=3.73\,\epsilon_0/R^3$
(black dotted path)} -- the layer spacing
does not change much and takes a value of about $d/R\approx4.4-4.5$,
\modifiedBlue{which is a common finding for phases \modifiedGreen{of the $S_A$-tpe}.}
\modifiedRed{
Interestingly, for both paths (black squared path with \modifiedMagenta{$p=2.69\,\epsilon_0/R^3$}
and green dotted path with $\eta_0\approx0.5$; $T^*\leq0.9$)
the low-temperature wide smectic phase $S_{AW}$,
which, compared with the $S_A$ phase, exhibits 
an increased layer spacing (compare Figs.~\ref{fig:structure2} and \ref{fig:structure3}), 
also does not show a considerable temperature dependence of the layer spacing 
within its \modifiedGreen{region of thermodynamic stability}
(see \modifiedGreen{white background in} Fig.~\ref{fig:d_vs_kT} for $T^*\in[0.74,0.9]$).
However, within the region of the $S_{AW}$ phase
\modifiedGreen{being metastable} with respect to crystallization,
i.e., \modifiedGreen{for} $T^*\lesssim0.74$ in Fig.~\ref{fig:d_vs_kT}
(\modifiedGreen{salmon-colored} area), for both paths \modifiedGreen{there is}
a pronounced temperature dependence of the layer spacing.
}
Since the smectic layers of the $S_{AW}$ phase are not as densely packed
as the layers of an ordinary $S_A$ phase,
the free space in between the layers allows for 
\modifiedBlue{a certain softness of} the layer spacing.
The increase of the layer spacing with decreasing temperature can be understood intuitively,
\modifiedBlue{because} upon lowering temperature the electrostatic repulsion becomes more
\modifiedBlue{efficient so that} the smectic layers widen.
\modifiedRed{
Nevertheless, since this behavior is only observable within
the metastable region of the $S_{AW}$ phase,
we conclude that one expects \modifiedGreen{only} a weak temperature dependence
of the layer spacing for the $S_{AW}$ phase, 
\modifiedGreen{analogous} to the $S_A$ phase.
}

For the shorter particles with $L/R=2$, \modifiedBlue{we have found both for}
the narrow $S_{AN}$ phase as well as for the ordinary $S_A$ phase
a very weak dependence of the layer spacing on temperature,
like \modifiedBlue{in Fig.~\ref{fig:d_vs_kT}} for the high-temperature smectic $S_A$ phase.

\section{\label{sec:conclusions}Conclusions and Summary}

Ionic liquid crystals have been investigated by means of density functional theory
(Sec.~\ref{sec:theory:DFT}) and \modifiedBlue{grand canonical} Monte Carlo simulations
(Sec.~\ref{sec:theory:GCMC}).
\modifiedBlue{To this end} a coarse-grained description of the ILC molecules
\modifiedBlue{(Fig.~\ref{fig:ellipsoids})}
as rigid ellipsoids interacting via \modifiedBlue{a} molecular pair potential
$U(\vec{r}_{12},\vec\omega_1,\vec\omega_2)$
\modifiedBlue{(Eq.~(\ref{eq:Pairpot}) and Fig.~\ref{fig:pairpot})}
has been employed
\modifiedBlue{(see Sec.~\ref{sec:theory:model})}.

\modifiedGreen{This} \modifiedMagenta{study demonstrates} that
ILC fluids show a \modifiedBlue{rich} phenomenology 
\modifiedBlue{concerning} their phase behavior and
\modifiedBlue{their} structural bulk properties.
\modifiedBlue{Beyond} the qualitative differences in the phase behavior of ordinary 
liquid crystals and ILCs,
we \modifiedBlue{have} examined in detail the dependence of the thermal and structural properties of
ILC fluids on the length-to-breadth ratio $L/R$
and \modifiedBlue{on} the distance $D$ of the charges from the geometrical center of the molecules.
This analysis \modifiedBlue{leads} to the following main conclusions:
\begin{itemize}
 \item[(1)]
 Comparing ordinary (uncharged) liquid crystals and ILCs,
 within the present DFT approach
 a lowering of the liquid-vapor critical point
 \modifiedBlue{of the latter} is observed
 \modifiedBlue{(see Fig.~\ref{fig:phasediagrams1})}.
 Additionally, \modifiedBlue{for ILCs} the liquid-smectic two-phase region
 becomes narrower, giving rise to a stable smectic structure 
 at smaller packing fractions $\eta_0$
 \modifiedBlue{(see Fig.~\ref{fig:phasediagrams2})}.
 \item[(2)]
 For the shorter particles \modifiedBlue{with} length-to-breadth ratio $L/R=2$
 \modifiedBlue{there is} an ordinary $S_A$ phase at high temperatures and
 large mean packing fractions. \modifiedBlue{At} low temperatures and
 intermediate mean packing fractions \modifiedBlue{there is} a distinct smectic \modifiedBlue{$S_{AN}$}
 structure in which the particles are \modifiedBlue{oriented} parallel to the smectic layers, 
 \modifiedBlue{i.e.,} perpendicular to the \modifiedGreen{layer normal $\vec{\hat n}$}
 \modifiedBlue{(see Fig.~\ref{fig:structure1})},
 and thus do not show a preferred orientation.
 \modifiedGreen{(\modifiedMagenta{Figure}~\ref{fig:compare_SA_SAN} provides a comparison of the structure 
 of both types of smectic phases, $S_{AN}$ and $S_A$,
 for particles with length-to-breadth ratio $L/R=2$.)}
 This behavior seems to be related to the small length-to-breadth ratio $L/R=2$
 and \modifiedBlue{to} the small value of the anisotropy parameter $\epsilon_R/\epsilon_L=2$
 \modifiedBlue{of the underlying Gay-Berne pair potential.}
 \modifiedBlue{This renders} the particles rather isotropic,
 which is even more pronounced in \modifiedGreen{the} case
 of the ILC fluid (Fig.~\ref{fig:phasediagrams1}(b))
 due to the additional electrostatic repulsion, and \modifiedBlue{thus}
 leads to \modifiedBlue{a} higher $L$-$S_{AN}$-$S_A$ triple point temperature.
 \modifiedRed{
 However, considering the large packing fractions $\eta_0\geq0.5$ for which
 the liquid-crystalline phases are predicted \modifiedGreen{to occur} in these systems,
 \modifiedGreen{we have found} that for \modifiedGreen{a} hexagonal lattice structure
 this leads to a lateral lattice spacing of $a/R\leq1.1$
 (see Secs.~\ref{sec:theory:DFT:Crystallization}
 and \ref{sec:discussion:PhaseDiagrams:Comparison}).
 \modifiedGreen{Hence, the particles are densely packed in this density regime
 and previous simulations~\cite{DeMiquel2002} on systems
 of Gay-Berne particles of length-to-breadth ratio
 $L/R=3$ report the onset of crystallization within \modifiedMagenta{that} regime}.
 \modifiedGreen{On this basis, at least in parts, the thermodynamic}
 stability of the liquid-crystalline phases $S_{A}$ and $S_{AN}$
 can be expected to be an artifact of the
 \modifiedGreen{method employed (see Sec.~\ref{sec:theory:DFT:Formalism})},
 which cannot capture crystalline phases.
 The prediction of the liquid-crystalline phases $S_A$ and $S_{AN}$,
 which show an \modifiedGreen{periodically varying} density profile in $z$-direction,
 can be considered as a hint on the presence of \modifiedGreen{various} types of crystalline phases
 at large densities in these systems.
 The $S_A$ phase can be interpreted as an \modifiedGreen{analogue} to a
 crystalline phase with additional orientational ordering, while
 the $S_{AN}$ phase mimics a crystalline phase
 with a lower degree of orientational ordering
 (\modifiedGreen{i.e.,} plastic crystals).
 }
 \item[(3)]
 For the longer particles of length-to-breadth ratio $L/R=4$,
 besides the isotropic liquid ($L$) and the ordinary smectic \modifiedBlue{$S_A$} phase
 \modifiedBlue{(Fig.~\ref{fig:structure2})},
 at low temperatures and sufficiently large packing fractions
 the novel $S_{AW}$ phase \modifiedBlue{(Fig.~\ref{fig:structure3})} \modifiedBlue{occurs}
 (see Figs.~\ref{fig:phasediagrams2} and \ref{fig:phasediagrams3}).
 It is characterized by a considerably larger layer spacing than \modifiedGreen{in}
 the ordinary $S_A$ phase
 \modifiedBlue{(compare Figs.~\ref{fig:structure2} and \ref{fig:structure3})}.
 While \modifiedBlue{the} majority of particles
 \modifiedBlue{is} oriented mostly parallel to the \modifiedGreen{layer normal},
 \modifiedBlue{as} indicated by a large value of the orientational order parameter $S_2(z=0)>0.8$
 within the smectic layers, a significant \modifiedBlue{number} of particles is located in between
 the smectic layers. Those particles tend to be perpendicular
 to the \modifiedGreen{layer normal}, giving rise to $S_2(|z|\lesssim d/2)<0$. 
 \item[(4)]
 \modifiedBlue{Concerning} the phase behavior of ILCs
 \modifiedBlue{as function of} the location of the charges in the molecules,
 \modifiedBlue{we have} found that for \modifiedBlue{the parameter set} 
 $L/R=4,\lambda_D/R=5,\epsilon_R/\epsilon_L=2,\gamma/(R\epsilon_0)=0.045$
 positioning the charges at an intermediate distance \modifiedBlue{$D/R\leq0.9$}
 from the \modifiedGreen{geometric} center
 does not alter the phase behavior much
 \modifiedBlue{as compared to positioning the charges in the center}
 (\modifiedBlue{see} Figs.~\ref{fig:phasediagrams2}(b) and \ref{fig:phasediagrams3}(a)).
 However if the charges are located almost at the \modifiedBlue{tips} of the molecules
 ($L/R=4,D/R=1.8$, \modifiedBlue{Fig.~\ref{fig:phasediagrams3}(b)})
 \modifiedBlue{there is a} significant change in the phase behavior.
 \modifiedGreen{The coexistence of the phases $S_{AW}$ and $S_A$}
 is shifted towards higher \modifiedGreen{temperatures}.
 This shift for $D/R=1.8$ stabilizes the $S_{AW}$ phase in a temperature regime 
 below the ordinary $S_A$ phase \modifiedBlue{but} above the melting curve,
 \modifiedRed{
 unlike the other cases studied
 (Figs.~\ref{fig:phasediagrams2} and \ref{fig:phasediagrams3}(b))
 for which our analysis,
 using the method \modifiedMagenta{discussed in} Sec.~\ref{sec:theory:DFT:Crystallization},
 yields that the $S_{AW}$ phase is expected to be preempted by crystallization
 (compare the pink curves in Figs.~\ref{fig:phasediagrams2} and \ref{fig:phasediagrams3}
 \modifiedGreen{which are obtained by the procedure
 outlined in Fig.~\ref{fig:OmegaDiff_Crystallization}}
 ).
 }
 Accordingly, we \modifiedBlue{have observed} the $S_{AW}$ phase, within \modifiedBlue{our}
 present \modifiedBlue{grand canonical} Monte Carlo simulations, for an ILC fluid
 \modifiedBlue{the charges of which are located at the tips} of the molecules.
 In qualitative agreement with \modifiedBlue{DFT}, the simulational results yield
 an ordinary smectic \modifiedBlue{$S_A$} phase at high temperatures and large packing fractions
 \modifiedBlue{(see Figs.~\ref{fig:mc_structure1} and \ref{fig:mc_structure2})};
 the layer spacing is of the size of the particles and nearly all
 of them are located within the smectic layers aligned with the \modifiedGreen{layer normal}
 (see Fig.~\ref{fig:sim_snapshots}(a)).
 At lower temperatures the novel \modifiedBlue{$S_{AW}$} smectic phase with wide layer spacings
 \modifiedBlue{occurs, such that} a considerable fraction of particles is located in between
 the smectic layers with mainly perpendicular orientation
 with respect to the \modifiedGreen{layer normal}
 (see Figs.~\ref{fig:sim_snapshots}(b) and \ref{fig:mc_structure3}).
 \item[(5)]
 Analyzing the dependence of the smectic layer spacing on temperature
 for \modifiedBlue{the parameter set}
 $L/R=4,\epsilon_R/\epsilon_L=2,\lambda_D=5,D/R=1.8,\gamma/(R\epsilon_0)=0.045$
 \modifiedBlue{reveals distinct behaviors of the smectic $S_A$ and $S_{AW}$ phases
 (see Fig.~\ref{fig:d_vs_kT})}:
 While the \modifiedGreen{layer spacing of the}
 ordinary high-temperature smectic \modifiedBlue{$S_A$} phase
 does not \modifiedBlue{vary notably as function of} temperature,
 \modifiedBlue{which is a common finding for ordinary $S_A$ phases,}
 increasing layer spacings for decreasing temperatures
 \modifiedRed{
 can be observed for the low-temperature smectic phase $S_{AW}$.
 \modifiedGreen{This can be understood in terms of}
 the free space in between the smectic layers \modifiedGreen{which} gives rise to
 \modifiedGreen{a certain} softness in the layer spacing.
 Due to the enhanced \modifiedGreen{effective} electrostatic repulsion at lower temperatures,
 the layers tend to widen upon lowering the temperature.
 However, this behavior is prominent only in the metastable region
 of the $S_{AW}$ phase, while within the stable region
 of the $S_{AW}$ phase, in analogy to the high-temperature
 $S_A$ phase, \modifiedGreen{there is no pronounced} temperature dependence of 
 the layer spacing.
 }
 
 Like the high-temperature $S_A$ phase \modifiedBlue{for long particles}
 \modifiedGreen{(see Fig.~\ref{fig:d_vs_kT}),}
 the layer spacing of the smectic \modifiedBlue{$S_A$ and $S_{AN}$} phases,
 observed for shorter particles with $L/R=2$, \modifiedGreen{does not exhibit}
 a considerable \modifiedGreen{temperature dependence}.
\end{itemize}
\modifiedBlue{We point out} that the theoretical framework \modifiedGreen{presented here}
is also applicable \modifiedBlue{for studying interfaces such as the}
free interfaces of coexisting bulk phases or 
\modifiedBlue{the interface of an} ionic liquid crystal 
\modifiedBlue{in contact with an electrode.}
In particular 
\modifiedBlue{it will be interesting to study the interfacial features of
these materials as they emerge from} the interplay of ionic and liquid-crystalline properties.
\modifiedBlue{We also} stress the importance of the 
choice of the projected density \modifiedBlue{distribution} $\bar\rho(\vec{r},\vec\omega)$
(Eq.~\ref{eq:WeightedDensity}) \modifiedBlue{with respect to} our theoretical approach,
\modifiedBlue{because} the incorporation of
\modifiedGreen{second-order} Fourier modes is \modifiedBlue{indispensable}
\modifiedBlue{for capturing} the novel wide smectic $S_{AW}$ phase.

A \modifiedBlue{natural} extension of the \modifiedGreen{study} presented here \modifiedBlue{is}
to consider a density functional of binary fluid \modifiedBlue{mixtures}
allowing for an explicit \modifiedBlue{description} of the counterions.
However, \modifiedBlue{we do} not expect the phase behavior to be
crucially affected by this higher degree of sophistication,
\modifiedBlue{because} studies using multicomponent integral equations~\cite{Harnau2000,Harnau2002}
showed that the counterions, which are smaller than the ILC molecules,
give rise to an effective screening between the latter,
rendering the \modifiedBlue{use} of a screened Coulomb potential,
like in Eq.~(\ref{eq:PairPot_ES}), a reasonable approach.
\MODIFIED{
However, a more realistic description could be obtained by \MODIfied{determining}
the Debye screening length $\lambda_D$ \MODIfied{in accordance with} Eq.~(\ref{eq:Debyelength}),
instead of treating it as a control parameter with \MODIfied{a}
fixed value $\lambda_D/R=5$. \MODIfied{Within this approach},
the full range of the Debye screening length in the \MODIfied{various}
density and temperature regimes could be incorporated
into the model, which could lead to interesting new phase \MODIfied{behaviors}
and structural phenomena.
While for dilute electrolyte solutions one typically finds $\lambda_D/R\gg1$, 
\MODIfied{in dense ionic liquids} the Debye screening length $\lambda_D$
can become smaller than the particle diameter $R$.
Thus, the value $\lambda_D/R=5$ used throughout this study lays in between
those two limiting cases.
}

\modifiedRed{The \modifiedGreen{description} of
the reference hard-core system within \modifiedGreen{an approach more sophisticated}
than Eq.~(\ref{eq:effective_DirCorFunc_2}),
\modifiedGreen{such as \modifiedMagenta{fundamental measure theory}},
would allow for a more reliable \modifiedGreen{calculation} of the transition towards crystalline
phases. We consider this \modifiedGreen{as} a necessary step in order to accurately predict
the \modifiedGreen{extent} of $S_{AW}$ stability \modifiedGreen{at low} temperatures.}

\modifiedMagenta{Induced by an external electric field,
qualitatively new phenomena might occur.}

\begin{acknowledgments}
We thank M.\ P.\ Allen and D.\ Frenkel for valuable comments.
\end{acknowledgments}




\appendix

\section{\label{sec:appendix:derivation}Derivation of Eq.~(\ref{eq:DensityGenericForm})}

\modifiedBlue{As explained below Eq.~(\ref{eq:ExpansionCoeffs2}), 
for bulk phases \modifiedMagenta{one has} 
$Q_i=\text{const}$. Accordingly, Eq.~(\ref{eq:Eff1Pot_PL}) reduces to}
\begin{align}
  &\beta\psi_\text{PL}(\vec{r},\vec\omega,[\bar\rho]) =
  -\mathcal{J}(Q_0)\times
  \nonumber\\
  &\Int{\mathcal{V}}{3}{r'}\Int{\mathcal{S}}{2}{\omega'}
   \bar\rho(\vec{r}',\vec\omega')
   f_M(\vec{r}-\vec{r}',\vec\omega,\vec\omega').
\label{eq:appendix1}
\end{align}
Using the definition \modifiedBlue{in Eq.~(\ref{eq:WeightedDensity})}
of the projected density $\bar\rho(\vec{r},\vec\omega)$,
\modifiedBlue{with} $Q_i=\text{const}$ for bulk phases,
we obtain \modifiedBlue{six} terms in the integrand of Eq.~(\ref{eq:appendix1}),
one for each \modifiedBlue{$Q_i,~i=0,\cdots,5$}.
Changing the integration variable
from $\vec{r}'$ to $\vec{\tilde r}=\vec{r}'-\vec{r}$ yields
\begin{align}
   \beta\psi_\text{PL}(\vec{r},\vec\omega,[\bar\rho]) &=\nonumber\\
   \modifiedBlue{-\frac{\mathcal{J}(Q_0)}{4\pi}}\bigg[~
   Q_0
   &\Int{\mathcal{V}}{3}{\tilde r}\Int{\mathcal{S}}{2}{\omega'}
   f_M(\vec{\tilde r},\vec\omega,\vec\omega')
   \nonumber\\
   +Q_1\cos(2\pi z/d)
   &\Int{\mathcal{V}}{3}{\tilde r}\Int{\mathcal{S}}{2}{\omega'}
   f_M(\vec{\tilde r},\vec\omega,\vec\omega')\cos(2\pi\tilde z/d)
  \nonumber\\
   +Q_2\cos(4\pi z/d)
   &\Int{\mathcal{V}}{3}{\tilde r}\Int{\mathcal{S}}{2}{\omega'}
   f_M(\vec{\tilde r},\vec\omega,\vec\omega')\cos(4\pi\tilde z/d)
  \nonumber\\
   +Q_3
   &\Int{\mathcal{V}}{3}{\tilde r}\Int{\mathcal{S}}{2}{\omega'}
   f_M(\vec{\tilde r},\vec\omega,\vec\omega')\modifiedBlue{5P_2(\cos(\theta'))}
  \nonumber\\
   +Q_4\cos(2\pi z/d)
   &\Int{\mathcal{V}}{3}{\tilde r}\Int{\mathcal{S}}{2}{\omega'}
   f_M(\vec{\tilde r},\vec\omega,\vec\omega')\nonumber\\
   &\times \modifiedBlue{5P_2(\cos(\theta'))}\cos(2\pi\tilde z/d)\nonumber\\
  \nonumber\\
   +Q_5\cos(4\pi z/d)
   &\Int{\mathcal{V}}{3}{\tilde r}\Int{\mathcal{S}}{2}{\omega'}
   f_M(\vec{\tilde r},\vec\omega,\vec\omega')\nonumber\\
   &\times \modifiedBlue{5P_2(\cos(\theta'))}\cos(4\pi\tilde z/d)~\bigg],
\label{eq:appendix2}
\end{align}
where we have used the relation $\cos(x+y)=\cos(x)\cos(y)-\sin(x)\sin(y)$
and that the integration domains $\mathcal{V}$
(associated with $\vec{r}'$) and $\mathcal{\tilde V}$ (associated with $\vec{\tilde r}$)
\modifiedMagenta{become equal and approach the three-dimensional space $\mathbb{R}^3$.
We note that unlike dipolar fluids~\cite{Groh1994}, due to the absence of
\modifiedOrange{long-ranged} interactions
caused by the screening of the charges (see Eq.~(\ref{eq:PairPot_ES})),
here the free energy functional does not depend on the sample shape.
Thus, \modifiedOrange{asymptotically}
the replacement of $\mathcal{\tilde V}$ by $\mathcal{V}$ is valid.}
The \modifiedBlue{integrals} involving \modifiedBlue{terms proportional to $\sin(y)$}
vanish \modifiedBlue{because} the \MODIFIED{Mayer f-function $f_M$} is even in $\vec{\tilde r}$.
Note that \modifiedBlue{via $f_M(\vec{\tilde r},\vec\omega,\vec\omega')$}
the integrals in Eq.~(\ref{eq:appendix2})
still carry a non-trivial dependence on the polar angle $\theta$.
However, the effective one-particle potential $\beta\psi[\bar\rho]$
\modifiedBlue{follows from integrating Eq.~(\ref{eq:appendix2})
over $\vec\omega$ and inserting this into Eq.~(\ref{eq:Eff1Potential})},
\modifiedBlue{rendering the corresponding} 
Legendre expansion coefficients $\zeta_l(\vec{r})$.
Using Eqs.~(\ref{eq:appendix2}) and (\ref{eq:Eff1Potential}),
we find that $\beta\psi[\bar\rho]$ has the same dependence on $z$
and $\theta$ \modifiedBlue{as} the projected density. Note,
that for the contribution $\beta\psi_\text{ERPA}[\bar\rho]$
\modifiedBlue{to} the effective one-particle potential,
due to interactions beyond the contact distance
$|\vec{r}_{12}|\geq R\sigma$ (Eq.~(\ref{eq:Eff1Pot_ERPA})),
one obtains the same result \modifiedBlue{concerning}
the spatial and \modifiedBlue{the} orientational dependence, \modifiedBlue{because}
$(1+f_M(\vec{\tilde r},\vec\omega,\vec\omega'))\beta U(\vec{\tilde r},\vec\omega,\vec\omega')$
is an even function \modifiedBlue{of} $\vec{\tilde r}$, too.

The \modifiedBlue{dependence on} $\vec{r}$ and $\vec\omega$ of the
integral in Eq.~(\ref{eq:Calc_c1}) \modifiedBlue{involves, inter alia,} 
the functional derivative
\modifiedBlue{
$\frac{\delta\bar\rho (\vec{r}'',\,\vec\omega'',[\rho])}{\delta\rho(\vec{r},\,\vec\omega)}$.
}
Again using the definition of the projected density
\modifiedBlue{(Eqs.~(\ref{eq:WeightedDensity})-(\ref{eq:ExpansionCoeffs2})) one finds}
\modifiedBlue{
\begin{align}
  &\frac{\delta\bar\rho (\vec{r}'',\vec\omega'',[\rho])}{\delta\rho(\vec{r},\vec\omega)}
  =\frac{1}{4\pi}\Theta(d/2-|z-z''|)\bigg[1+\nonumber\\
  &2\cos(2\pi z/d)\cos(2\pi z''/d)+
   2\cos(4\pi z/d)\cos(4\pi z''/d)+\nonumber\\
  &5P_2(\cos(\theta))P_2(\cos(\theta''))\big(1+
   2\cos(2\pi z/d)\cos(2\pi z''/d)+\nonumber\\
  &2\cos(4\pi z/d)\cos(4\pi z''/d)\big)\bigg].
 \label{eq:appendix3}
\end{align}
}
Thus, the second \modifiedBlue{summand in} Eq.~(\ref{eq:Calc_c1}) shares the same
\modifiedBlue{type of} dependence 
on $z$ and $\theta$ like the first \modifiedBlue{summand.}
Finally, the equilibrium profile \modifiedBlue{follows from solving} Eq.~(\ref{eq:ELG}),
which indeed \modifiedBlue{exhibits} the generic form given by Eq.~(\ref{eq:DensityGenericForm}).
Note, that \modifiedBlue{concerning the} bulk phases the Heaviside step function $\Theta(d/2-|z-z'|)$
\modifiedBlue{acts} only as to confine the spatial integration domain to \modifiedBlue{a single}
periodic cell, but does not \modifiedBlue{generate} a further dependence on the position $z$,
\modifiedBlue{because the bulk phases are considered to have a periodic structure only.}

It is worth mentioning that the same \modifiedBlue{line of argument} holds for the solution
\modifiedBlue{following from} the modified one-particle direct correlation function $\tilde c^{(1)}$
\modifiedBlue{in} Eq.~(\ref{eq:Calc_c1_Approx}), \modifiedBlue{because} here the first term
is again given by the effective one-particle potential $\beta\psi[\bar\rho]$
and the \modifiedBlue{second} term is constant
for \modifiedBlue{periodic} bulk phases \modifiedBlue{(see above)}.
Thus, \modifiedBlue{this} solution \modifiedBlue{has} the same
\modifiedBlue{functional} form as Eq.~(\ref{eq:DensityGenericForm}).

\section{\label{sec:appendix:comparison}
Comparison \modifiedBlue{between} the exact and the approximate solution of the Euler-Lagrange equation}

\modifiedBlue{We consider} an ionic liquid crystal with
$L/R=4,D/R=1.8,\lambda_D/R=5,\epsilon_R/\epsilon_L=2$\modifiedBlue{, and} $\gamma/(R\epsilon_0)=0.045$
(see Sec.~\ref{sec:theory:model}).
The \modifiedGreen{corresponding} phase diagram, obtained by the modified Euler-Lagrange equation
\modifiedGreen{(Eqs.~(\ref{eq:ELG}) and (\ref{eq:Calc_c1_Approx}))},
is shown in Fig.~\ref{fig:phasediagrams3}(b).
In order to compare the exact solution and the solution of the modified
Euler-Lagrange equation, we consider 
\modifiedBlue{three thermodynamic state} points $(T^*,\mu^*)$ in the phase diagram.
For $\modifiedBlue{(T^*,\mu^*)=(0.8,20)}$, the modified Euler-Lagrange equation yields a stable,
wide smectic phase $S_{AW}$ with
$\eta_0\approx0.5002,\tilde W_0:=W_0\,LR^2\pi/6\approx0.7059,S_{20}\approx0.3646,W_2\approx0.7605$
\modifiedGreen{(Eq.~(\ref{eq:OrderParameters}))},
and smectic layer spacing $d/R\approx7.71$.
For the same \modifiedBlue{state} point the exact solution 
\modifiedBlue{also belongs to} a wide smectic \modifiedBlue{$S_{AW}$} phase
with $\eta_0\approx0.5002,\tilde W_0\approx0.7058,S_{20}\approx0.3657,W_2\approx0.7603$
and smectic layer spacing $d/R\approx7.70$.
If we now choose a \modifiedBlue{state} point at \modifiedBlue{a} higher temperature
$\modifiedBlue{(T^*,\mu^*)=(1.2,18)}$,
which \modifiedBlue{corresponds} to the same mean packing fraction \modifiedBlue{$\eta_0$}
(hence, the two considered \modifiedBlue{state} points \modifiedBlue{lie on} a vertical line in the 
phase diagram \modifiedBlue{shown in} Fig.~\ref{fig:phasediagrams3}(b)),
one finds that
\modifiedBlue{for these values the solution of the modified Euler-Lagrange equation
belongs to the} high-temperature smectic \modifiedBlue{$S_A$} phase with
$d/R\approx4.50,\eta_0\approx0.5045,\tilde W_0\approx0.9207,S_{20}\approx0.7418,W_2\approx0.2680$,
as the exact solution \modifiedBlue{does} (\modifiedBlue{with} 
$d/R\approx4.52,\eta_0\approx0.5049,\tilde W_0\approx0.9218,S_{20}\approx0.7373,W_2\approx0.2774$).
Choosing the \modifiedBlue{state} point 
\modifiedBlue{$(T^*,\mu^*)=(0.8,29)$} increases the mean packing fraction
\modifiedBlue{to $\eta_0\approx0.59$}
for \modifiedBlue{which}
both schemes \modifiedBlue{predict} a transition from the $S_{AW}$ phase to the $S_A$ phase
(compare Table~\ref{tab:op_comparison}).
We conclude that \modifiedBlue{although} the exact location of the phase transition
between the observed bulk phases might be slightly shifted,
both minimization schemes give rise to the same qualitative phase behavior;
for the \modifiedBlue{considered} cases good agreement even on a quantitative level 
\modifiedBlue{has been} found.
Finally, Table~\ref{tab:op_comparison} summarizes the results for the order parameters
and the smectic layer spacing $d$ for the stable smectic phases 
\modifiedBlue{as predicted by} both solutions at the considered
\modifiedBlue{state} points $\modifiedBlue{(T^*,\mu^*)=(0.8,20)}$, $(0.8,29)$, and $(1.2,18)$.
\begin{table}
 \begin{adjustbox}{center}
 \begin{tabular}{|m{1.3cm}|c|m{1cm}|c|c|c|c|c|}
 \hline
 method  &$(T^*,\mu^*)$ & stable phase & $d/R$  & $\eta_0$ & $\tilde W_0$ & $S_{20}$ & $W_2$\\\hline\hline
 ~~~~~I  &$(0.8,20)$    &  $~~~S_{AW}$ & $7.70$ & $0.5002$ & $0.7058$     & $0.3657$ & $0.7603$  \\\hline
 ~~~~~I  &$(0.8,29)$    &  $~~~S_A$    & $4.55$ & $0.5858$ & $1.1138$     & $0.8063$ & $0.2307$  \\\hline
 ~~~~~I  &$(1.2,18)$    &  $~~~S_A$    & $4.52$ & $0.5049$ & $0.9218$     & $0.7373$ & $0.2774$  \\\hline
 \end{tabular}
 \end{adjustbox}
 \vspace*{0.05cm}
 \begin{adjustbox}{center} 
 \begin{tabular}{|m{1.3cm}|c|m{1cm}|c|c|c|c|c|}
 \hline
 ~~~~\,II&$(0.8,20)$    &  $~~~S_{AW}$ & $7.71$ & $0.5002$ & $0.7059$     & $0.3646$  & $0.7605$ \\\hline
 ~~~~\,II&$(0.8,29)$    &  $~~~S_A$    & $4.60$ & $0.5862$ & $1.1147$     & $0.7885$  & $0.2645$ \\\hline
 ~~~~\,II&$(1.2,18)$    &  $~~~S_A$    & $4.50$ & $0.5045$ & $0.9207$     & $0.7418$  & $0.2680$ \\\hline
 \end{tabular}
 \end{adjustbox}
 \caption{\label{tab:op_comparison}
 Comparison of the results for the exact solution of the Euler-Lagrange equation
 (method I, \modifiedGreen{Eqs.~(\ref{eq:ELG})\,-\,(\ref{eq:Calc_c1})})
 and the solution \modifiedBlue{obtained from}
 the modified one-particle direct correlation function $\tilde c^{(1)}$
 \modifiedBlue{given by} Eq.~(\ref{eq:Calc_c1_Approx}) (method II)
 \modifiedBlue{for three thermodynamic state points $(T^*,\mu^*)$.
 We compare the results of both methods for the layer spacing $d/R$,
 the mean packing fraction $\eta_0$, the first Fourier mode \modifiedGreen{$\tilde W_0=W_0LR^2\pi/6$}
 of the local packing fraction $\eta(z)$, the mean scalar orientational order parameter $S_{20}$,
 and the first Fourier mode $W_2$ of the scalar orientational order parameter profile $S_2(z)$
 \modifiedGreen{(see Eq.~(\ref{eq:OrderParameters}))}.}
 }
\end{table}
%

\modifiedBlue{
\section{\label{sec:appendix:pressure}Derivation of Eq.~(\ref{eq:pressure})}
In order to evaluate the \modifiedGreen{reduced} pressure $p^*=-\beta\Omega[\rho^\text{eq}]/V$,
the grand potential functional $\beta\Omega[\rho]$ (see Eq.~(\ref{eq:Omega})) is evaluated for
the solution \modifiedGreen{$\rho^\text{eq}(\vec{r},\vec\omega)$}
of the modified Euler-Lagrange equation,
\modifiedGreen{which solves Eq.~(\ref{eq:ELG}) with the modified
one-particle direct correlation function \modifiedMagenta{$\tilde c^{(1)}$}
given by Eq.~(\ref{eq:Calc_c1_Approx}):}
\begin{align}
 p^*=&-\frac{\beta\Omega[\rho^\text{eq}]}{\mathcal{V}}
    = -\frac{\beta\mathcal{F}[\rho^\text{eq}]}{\mathcal{V}}
      -\frac{1}{\mathcal{V}}\Int{\mathcal{V}}{3}{r}\Int{\mathcal{S}}{2}{\omega}
       \rho^\text{eq}(\vec{r},\vec\omega)\times\nonumber\\
     & \bigg[\tilde c^1(\vec{r},\vec\omega,[\rho^\text{eq}])-1\bigg]=
       \frac{1}{\mathcal{V}}\Int{\mathcal{V}}{3}{r}\Int{\mathcal{S}}{2}{\omega}
       \rho^\text{eq}(\vec{r},\vec\omega)\times\nonumber\\
     & \bigg[
       \frac{1}{2}\beta\psi(\vec{r},\vec\omega,\bar\rho[\rho^\text{eq}])+1
      -\frac{\partial_{Q_0}\mathcal{J}(Q_0)}{2\mathcal{V}_d}\times\nonumber\\
     & \Int{\mathcal{V}}{3}{r'}\Int{\mathcal{S}}{2}{\omega'}
       \bar\rho(\vec{r}',\vec\omega')\Theta(d/2-|z-z'|)\times\nonumber\\
     & \Int{\mathcal{V}}{3}{r''}\Int{\mathcal{S}}{2}{\omega''}
       \bar\rho(\vec{r}'',\vec\omega'')f_M(|\vec{r}'-\vec{r}''|,\vec\omega',\vec\omega'')
       \bigg].
 \label{eq:appendix4}
\end{align}
In the third step of Eq.~(\ref{eq:appendix4}),
we used the definition of the excess free energy functional (Eq.~(\ref{eq:F_WDA})).
\modifiedGreen{Since} the last two terms of Eq.~(\ref{eq:appendix4})
depend on the position $\vec{r}$ and the orientation
$\vec\omega$ \modifiedGreen{only} via $\rho^\text{eq}(\vec{r},\vec\omega)$,
the integrals over $\vec{r}$ and $\vec\omega$ can be \modifiedGreen{carried out:}
\begin{align}
 p^*=& \frac{1}{2\mathcal{V}_d}\Int{\mathcal{V}_d}{3}{r}\Int{\mathcal{S}}{2}{\omega}
       \rho^\text{eq}(\vec{r},\vec\omega)
       \beta\psi(\vec{r},\vec\omega,\bar\rho[\rho^\text{eq}])+\nonumber\\
     &+n_0-n_0\frac{\partial_{Q_0}\mathcal{J}(Q_0)}{2\mathcal{V}_d}
       \Int{\mathcal{V}_d}{3}{r'}\Int{\mathcal{S}}{2}{\omega'}
       \bar\rho(\vec{r}',\vec\omega')\times\nonumber\\
     &~\Int{\mathcal{V}}{3}{r''}\Int{\mathcal{S}}{2}{\omega''}
       \bar\rho(\vec{r}'',\vec\omega'')f_M(|\vec{r}'-\vec{r}''|,\vec\omega',\vec\omega''),
 \label{eq:appendix5}
\end{align}
where we used
$n_0=\Int{\mathcal{V}}{3}{r}\Int{\mathcal{S}}{2}{\omega}\rho^\text{eq}(\vec{r},\vec\omega)$
and \modifiedMagenta{that} in the first term 
\modifiedGreen{the entire system of volume $\mathcal{V}$
can be considered to be composed of a set of periodic cells of volume
$\mathcal{V}_d$ (see Sec.~\ref{sec:theory:DFT} below Eq.~(\ref{eq:ExpansionCoeffs2})).}
Finally, in order to simplify the first term of Eq.~(\ref{eq:appendix5})
we again write the equilibrium density profile
$\rho^\text{eq}(\vec{r},\vec\omega)=n^\text{eq}(\vec{r})f^\text{eq}(\vec{r},\vec\omega)$
as the product of the total number density $n(\vec{r})$
and the orientational distribution function $f(\vec{r},\vec\omega)$
and use the definition of the effective one-particle potential $\beta\psi(\vec{r},\vec\omega)$
(\modifiedGreen{given as} a Legendre polynomial \modifiedGreen{series} up to second order with the 
expansion coefficients $\zeta_l(\vec{r})$, $l=0,2$, see Eq.~(\ref{eq:Eff1Potential})).
\modifiedGreen{This leads to}
\begin{align}
 p^*=& n_0+
       \frac{1}{4\mathcal{V}_d}\Int{\mathcal{V}_d}{3}{r}n^\text{eq}(\vec{r})
       \left[\zeta_0(\vec{r})+S_2^\text{eq}(\vec{r})\zeta_2(\vec{r})\right]\nonumber\\
     &-n_0\frac{\partial_{Q_0}\mathcal{J}(Q_0)}{2\mathcal{V}_d}
       \Int{\mathcal{V}_d}{3}{r'}\Int{\mathcal{S}}{2}{\omega'}
       \bar\rho(\vec{r}',\vec\omega')\times\nonumber\\
     &~\Int{\mathcal{V}}{3}{r''}\Int{\mathcal{S}}{2}{\omega''}
       \bar\rho(\vec{r}'',\vec\omega'')f_M(|\vec{r}'-\vec{r}''|,\vec\omega',\vec\omega''),
 \label{eq:appendix6}
\end{align}
which \modifiedGreen{agrees with Eq.~(\ref{eq:pressure}).}
}

\bibliography{literature}
\end{document}